\documentclass[a4paper,fleqn,usenatbib]{mnras}
\usepackage{newtxtext,newtxmath}
\usepackage[T1]{fontenc}
\usepackage{ae,aecompl}

\usepackage{graphicx}	% Including figure files
\usepackage{amsmath}	% Advanced maths commands
\usepackage{amssymb}	% Extra maths symbols
\usepackage{xspace}
\usepackage{multirow}
\usepackage{anyfontsize}
\usepackage{pdflscape}
\newcommand{\kms}{kms$^{-1}$\xspace}	% km/s
\newcommand{\ta}{TWA~6\xspace}
\newcommand{\tb}{TWA~8A\xspace}
\newcommand{\prot}{\ensuremath{P_{\text{rot}}}\xspace}
\newcommand{\sv}{Stokes~\emph{V}\xspace}
\newcommand{\si}{Stokes~\emph{I}\xspace}
\newcommand{\teff}{\ensuremath{T_{\text{eff}}}\xspace}
\newcommand{\logg}{\ensuremath{\log{g}}\xspace}
\newcommand{\rsun}{\text{R}\ensuremath{_{\sun}}\xspace}
\newcommand{\rstar}{\ensuremath{R_{\star}}\xspace}
\newcommand{\msun}{\text{M}\textsubscript{\sun}\xspace}
\newcommand{\lsun}{\text{L}\textsubscript{\sun}\xspace}
\newcommand{\mjup}{\text{M}\textsubscript{Jup}\xspace}
\newcommand{\vrad}{\ensuremath{v_{\text{rad}}}\xspace}
\newcommand{\vsini}{\ensuremath{v\sin{i}}\xspace}
\newcommand{\vmic}{\ensuremath{v_{\text{mic}}}\xspace}
\newcommand{\radd}{$\text{rad~d}^{-1}$\xspace}

\newcommand{\msunyr}{\text{M}\textsubscript{\sun}\text{yr}$^{-1}$\xspace}
\newcommand{\ha}{H$\alpha$\xspace}
\newcommand{\hb}{H$\beta$\xspace}
\newcommand{\pc}{per~cent\xspace}

\title[The weak-line T Tauri stars TWA 6 and TWA 8A]{Magnetic topologies of young suns: \\The weak-line T Tauri stars TWA 6 and TWA 8A}

\author[C. A. Hill et al.]{C. A. Hill$^{1}$\thanks{E-mail: chill@irap.omp.eu}, C. P. Folsom$^{1}$, J.-F. Donati$^{1}$, G. J. Herczeg$^{2}$, G. A. J. Hussain$^{3,1}$, \newauthor S. H. P. Alencar$^{4}$, S. G. Gregory$^{5}$ and the MaTYSSE collaboration. \\
$^{1}$IRAP, Universit\'{e} de Toulouse, CNRS, UPS, CNES, 14 Avenue Edouard Belin, Toulouse, F-31400, France \\
$^{2}$Kavli Institute for Astronomy and Astrophysics, Peking University, Yi He Yuan Lu 5, Haidian Qu, Beijing 100871, China\\
$^{3}$ESO, Karl-Schwarzschild-Str. 2, D-85748 Garching, Germany \\
$^{4}$Departamento de F\`{i}sica - ICEx-UFMG, Av. Ant\^{o}nio Carlos, 6627, 30270--901 Belo Horizonte, MG, Brazil \\
$^{5}$University of St Andrews, St Andrews, KY16 9SS, UK\\}

\date{Accepted XXX. Received YYY; in original form ZZZ}

\pubyear{2018}

\begin{document}
\label{firstpage}
\pagerange{\pageref{firstpage}--\pageref{lastpage}}
\maketitle

\begin{abstract}
We present a spectropolarimetric study of two weak-line T Tauri stars (wTTSs), \ta and \tb, as part of the MaTYSSE (Magnetic Topologies of Young Stars and the Survival of close-in giant Exoplanets) program. Both stars display significant Zeeman signatures that we have modelled using Zeeman Doppler Imaging (ZDI). The magnetic field of \ta is split equally between poloidal and toroidal components, with the largest fraction of energy in higher-order modes, with a total unsigned flux of 840~G, and a poloidal component tilted $35\degr$ from the rotation axis. \tb has a 70~\pc poloidal field, with most of the energy in higher-order modes, with an unsigned flux of 1.4~kG (with a magnetic filling factor of 0.2), and a poloidal field tilted $20\degr$ from the rotation axis. Spectral fitting of the very strong field in \tb (in individual lines, simultaneously for Stokes $I$ and $V$) yielded a mean magnetic field strength of $6.0\pm0.5$~kG. The higher field strengths recovered from spectral fitting suggests that a significant proportion of magnetic energy lies in small-scale fields that are unresolved by ZDI. So far, wTTSs in MaTYSSE appear to show that the poloidal-field axisymmetry correlates with the magnetic field strength. Moreover, it appears that classical T Tauri stars (cTTSs) and wTTSs are mostly poloidal and axisymmetric when mostly convective and cooler than $\sim4300$~K, with hotter stars being less axisymmetric and poloidal, regardless of internal structure.
\end{abstract}

%%%%%%%%%%%%%%%%%%%%%%%%%%%%%%%%%%%%%%%%%%%%%%%%%%

%%%%%%%%%%%%%%%%% BODY OF PAPER %%%%%%%%%%%%%%%%%%

\section{Introduction}
During the first few hundred thousand years of low-mass star formation, class-I pre-main sequence (PMS) stars accrete significant amounts of material from their surrounding dusty envelopes. After around 0.5~Myr, these protostars emerge from their dusty cocoons and are termed classical T Tauri stars (cTTSs / class-II PMS stars) if they are still accreting from their surrounding discs, or weak-line T Tauri stars (wTTSs / class-III PMS stars) if they have exhausted the gas from the inner disc cavity. During the PMS phase, stellar magnetic fields have their largest impact on the evolution of the star. These fields control accretion processes and trigger outflows/jets \citep{bouvier2007}, dictate the star's angular momentum evolution by enforced spin-down through star-disc coupling (e.g., \citealt{davies2014}), and alter disc dynamics and planet formation \citep{baruteau2014}. Moreover, as PMS stars are gravitationally contracting towards the MS, the change in stellar structure from fully to partly convective is expected to alter the stellar dynamo mechanism and the resulting magnetic field topology.

Previous work through the MaPP (Magnetic Protostars and Planets) survey revealed that the large-scale topologies of 11 cTTSs remained relatively simple and mainly poloidal when the host star is still fully or largely convective, but become much more complex when the host star turns mostly radiative \citep{gregory2012,donati2013}. This survey concluded that these fields likely originated from a dynamo, varying over time-scales of a few years \citep{donati2011, donati2012, donati2013}, and resembling those of mature stars with comparable internal structure \citep{morin2008b}.

The nature of the magnetic fields of wTTSs and how they depend on fundamental parameters is less well known. These evolutionary phases are the initial conditions in which disc-less PMS stars initiate their unleashed spin up towards the zero-age main sequence (ZAMS). Hence, it is crucial to characterize their magnetic fields and how they depend on mass, temperature, age and rotation. To this end, we are performing a spectropolarimetric study of around 30 wTTSs through the MaTYSSE (Magnetic Topologies of Young Stars and the Survival of close-in giant Exoplanets) programme, mainly allocated on ESPaDoNS at the Canada-France-Hawaii Telescope (CFHT), complemented by observations with NARVAL on the Telescope Bernard Lyot, and with HARPS on the ESO 3.6-m Telescope. By using Zeeman Doppler Imaging (ZDI) to characterize the magnetic fields of wTTSs, we are able to test stellar dynamo theories and models of low-mass star formation. Moreover, by filtering out the activity-related jitter from the radial velocity (RV) curves, we are able to potentially detect hot Jupiters (hJs; see \citealt{donati2016a}), and thus verify whether core accretion and migration is the most likely mechanism for forming close-in giant planets \citep[e.g.,][]{alibert2005}.

Here, we present our detailed analysis of the wTTSs \ta and \tb as part of the MaTYSSE survey. Both targets are members of the TW Hydrae association, which, at an age of $10\pm3$~Myr \citep{bell2015}, is in transition between the T Tauri and the post T Tauri phase, and thus provides a very interesting period in which to study the properties of the member stars as they spin-up towards the ZAMS. Our phase-resolved spectropolarimetric observations are documented in Section~\ref{sec:observations}, with the stellar and disc properties presented in Section~\ref{sec:prop}. We discuss the spectral energy distributions, several emission lines, and the accretion status of both stars in Section~\ref{sec:sed}.
 In Section~\ref{sec:tomography} we present our results after applying our tomographic modelling technique to the data. In Section~\ref{sec:specfit} we present our results of our spectral fitting to the \si and \sv spectra, and in Section~\ref{sec:rv} we discuss our analysis of the filtered RV curves. Finally, we discuss and summarize our results and their implications for low-mass star and planet formation in Section~\ref{sec:discussion}.

\section{Observations}
\label{sec:observations}
Spectropolarimetric observations of \ta were taken in February 2014, with observations of \tb taken in March and April 2015, both using ESPaDOnS at the 3.6-m~CFHT. Spectra from ESPaDOnS span the entire optical domain (from 370--1000~nm) at a resolution of 65,000 (i.e., a resolved velocity element of 4.6~\kms) over the full wavelength range, in both circular or linear polarization \citep{donati2003}. 

A total of 22 circularly-polarized (\sv) and unpolarized (\si) spectra were collected for \ta over a timespan of 16 nights, corresponding to around 29.6 rotation cycles (where $\prot =0.5409$~d, \citealt{kiraga2012}). Time sampling was fairly regular, with the longest gap of 6 nights occurring towards the end of the run. For \tb, 15 spectra were collected with regular time sampling over a 15 night timespan, corresponding to around 3.2 rotation cycles (where \prot = 4.638~d, \citealt{kiraga2012}).

All polarization spectra consist of four individual sub-exposures (each lasting 406~s for \ta, and 1115~s for \tb), taken in different polarimeter configurations to allow the removal of all spurious polarization signatures at first order. All raw frames were processed using the \textsc{Libre ESpRIT} software package, which performs bias subtraction, flat fielding, wavelength calibration, and optimal extraction of (un)polarized \'{e}chelle spectra, as described in the previous papers of the series (\citealt{donati1997}, also see \citealt{donati2010b, donati2011, donati2014}), to which the reader is referred for more information. The peak signal-to-noise ratios (S/N, per 2.6~\kms velocity bin) achieved on the collected spectra range between 111--197 (median 164) for \ta, and 209--369 (median 340) for \tb, depending on weather/seeing conditions. All spectra are automatically corrected for spectral shifts resulting from instrumental effects (e.g., mechanical flexures, temperature or pressure variations) using atmospheric telluric lines as a reference. This procedure provides spectra with a relative RV precision of better than 0.030~\kms \citep[e.g.][]{moutou2007,donati2008a}. A journal of all observations is presented in Table~\ref{tab:obs} for both stars.

\begin{table}
\centering
\caption{Journal of ESPaDOnS observations of \ta (first 22 rows) and \tb (last 15 rows), each consisting of a sequence of 4 subexposures lasting 406~s and 1115~s for \ta and \tb, respectively. Columns 1--4 list (i) the UT date of the observation, (ii) the corresponding UT time at mid exposure, (iii) the Barycentric Julian Date (BJD), and (iv) the peak signal-to-noise ratio (per 2.6~\kms velocity bin) of each observation. Columns 5 and 6 respectively list the S/N in \si LSD profiles (per 1.8~\kms velocity bin), and the rms noise level (relative to the unpolarized continuum level $I_{\text{c}}$ and per 1.8~\kms velocity bin) in the \sv LSD profiles. Column~6 indicates the rotational cycle associated with each exposure, using the ephemerides given in Equation~\ref{eq:ephemeris}.}
\label{tab:obs}	
\setlength\tabcolsep{3pt} % default value: 6pt
\begin{tabular}{lcccccc}
\hline													
\multicolumn{1}{c}{Date} 	&	 UT 	&	BJD	&	S/N	&	S/N$_{\textsc{lsd}}$	&	 $\sigma_{\textsc{lsd}}$ 	&	Cycle	\\
(2014) &	 (hh:mm:ss) 	&	(2456693.9+)	&	 	&		&	 (0.01\%) 	&		\\
\hline													
Feb 04	&	11:15:00	&	0.07239	&	160	&	1796	&	5.6	&	0.151	\\
Feb 04	&	12:17:52	&	0.11605	&	184	&	2260	&	4.5	&	0.232	\\
Feb 07	&	10:28:34	&	3.04027	&	134	&	1619	&	6.2	&	5.638	\\
Feb 07	&	11:30:04	&	3.08299	&	131	&	1559	&	6.4	&	5.717	\\
Feb 07	&	13:04:25	&	3.14851	&	158	&	1849	&	5.4	&	5.838	\\
Feb 09	&	09:27:53	&	4.99822	&	168	&	1990	&	5.1	&	9.258	\\
Feb 09	&	10:46:58	&	5.05313	&	169	&	1973	&	5.1	&	9.359	\\
Feb 09	&	11:48:45	&	5.09605	&	132	&	1639	&	6.1	&	9.439	\\
Feb 10	&	11:08:25	&	6.06807	&	163	&	1914	&	5.3	&	11.236	\\
Feb 10	&	12:10:38	&	6.11128	&	178	&	2126	&	4.7	&	11.316	\\
Feb 11	&	09:37:38	&	7.00507	&	183	&	2273	&	4.4	&	12.968	\\
Feb 11	&	11:05:43	&	7.06624	&	197	&	2461	&	4.1	&	13.081	\\
Feb 11	&	11:54:04	&	7.09981	&	169	&	1922	&	5.2	&	13.143	\\
Feb 12	&	09:21:45	&	7.99407	&	164	&	1933	&	5.2	&	14.797	\\
Feb 12	&	11:22:19	&	8.07780	&	159	&	1869	&	5.4	&	14.951	\\
Feb 12	&	12:49:30	&	8.13834	&	160	&	1855	&	5.4	&	15.063	\\
Feb 13	&	10:35:31	&	9.04533	&	111	&	1514	&	6.6	&	16.740	\\
Feb 13	&	13:06:11	&	9.14997	&	180	&	2094	&	4.8	&	16.934	\\
Feb 19	&	09:23:25	&	14.99545	&	160	&	1847	&	5.4	&	27.741	\\
Feb 19	&	10:51:16	&	15.05645	&	181	&	2223	&	4.5	&	27.853	\\
Feb 19	&	12:29:16	&	15.12451	&	192	&	2387	&	4.2	&	27.979	\\
Feb 20	&	12:06:25	&	16.10867	&	137	&	1345	&	7.5	&	29.799	\\
\hline
(2015) & & (2457107.9+) &	& & &		\\													
\hline
Mar 25	&	11:43:05	&	0.06756	&	338	&	3847	&	2.6	&	0.020	\\
Mar 26	&	11:04:43	&	1.04090	&	343	&	3812	&	2.6	&	0.230	\\
Mar 27	&	11:40:03	&	2.06545	&	340	&	3863	&	2.6	&	0.451	\\
Mar 28	&	11:30:18	&	3.05868	&	341	&	3841	&	2.6	&	0.665	\\
Mar 29	&	12:16:00	&	4.09040	&	302	&	3348	&	3.0	&	0.887	\\
Mar 30	&	11:49:48	&	5.07220	&	369	&	4244	&	2.4	&	1.099	\\
Mar 31	&	08:28:33	&	5.93245	&	357	&	4071	&	2.5	&	1.285	\\
Apr 01	&	08:26:55	&	6.93130	&	349	&	3960	&	2.5	&	1.500	\\
Apr 03	&	11:25:50	&	9.05552	&	253	&	2670	&	3.8	&	1.958	\\
Apr 04	&	11:34:15	&	10.06136	&	355	&	4091	&	2.5	&	2.175	\\
Apr 05	&	08:42:10	&	10.94184	&	332	&	3764	&	2.7	&	2.365	\\
Apr 06	&	08:30:36	&	11.93379	&	353	&	4013	&	2.5	&	2.579	\\
Apr 08	&	09:13:10	&	13.97061	&	202	&	1754	&	5.7	&	3.018	\\
Apr 09	&	07:18:42	&	14.90036	&	253	&	2819	&	3.6	&	3.218	\\
Apr 09	&	08:45:35	&	14.95143	&	309	&	3621	&	2.8	&	3.229	\\
\hline													
\end{tabular}
\end{table}

\section{Stellar and disc properties}
\label{sec:prop}
Both stars are part of the TW~Hya association \citep[TWA, e.g., ][]{jayawardhana1999,webb1999,donaldson2016}, one of the closest young star associations at a distance of $\simeq 50$~pc \citep[e.g.,][]{zuckerman2004}. Furthermore, at an age of $10\pm3$~Myr \citep{bell2015}, TWA is at a crucial evolutionary phase where star-disk interactions have ceased, and where the T Tauri stars are rapidly spinning up as they continue their gravitational contraction towards the main sequence \citep[e.g.,][]{rebull2004}.

Both stars are classed as T~Tauri due to strong \ion{Li}{i}~6708~\AA\ absorption \citep[e.g., ][]{webb1999}, with mean equivalent widths (EW) of around 0.45~\AA\ (20~\kms) and 0.38~\AA\ (17~\kms) for \ta and \tb, respectively (slightly lower than the 0.56~\AA\ and 0.53~\AA\ found by \citealt{torres2003}). Furthermore, our spectra show that the strength of \ion{Li}{i}~6708~\AA\ absorption does not vary significantly for either star, indicating a lack of veiling (in agreement with \citealt{herczeg2014}), and confirming their status as wTTSs (see Section~\ref{sec:sed} and \ref{sec:acc} for further discussion). Moreover, both stars show very regular periodic light-curves that do not appear like those of cTTS, further supporting their non-accreting status. 

For \ta, we adopt the photometric rotation period of 0.5409~d found by \cite{kiraga2012} for the remainder of this work, as this is in excellent agreement with the $0.54\pm0.01$~d period of \cite{lawson2005}, and the $0.54090\pm0.00005$~d period of \cite{skelly2008}. For \tb, we adopt the photometric period of 4.638~d \citep{kiraga2012}, in excellent agreement with the $4.65\pm0.01$~d period found by \cite{lawson2005}, the $4.66\pm0.06$~d period of \citet{messina2010}, and the 4.639~d period found by applying a Lomb-Scargle periodogram analysis to SuperWASP photometric data \citep{butters2010}. The rotational cycles of \ta and \tb (denoted $E_{1}$ and $E_{2}$ in Equation~\ref{eq:ephemeris}) are computed from Barycentric Julian Dates (BJDs) according to the (arbitrary) ephemerides:

\begin{align}
\text{BJD (d)} &= 2456693.9 + 0.5409E_{1} & \hfill \text{(for TWA 6)} \nonumber \\ 
\text{BJD (d)} &= 2457107.9 + 4.638E_{2} & \hfill \text{(for TWA 8A)}
\label{eq:ephemeris}
\end{align}

\subsection{Stellar properties} 
\label{sec:evolution}
To determine the \teff and \logg of our target stars, we applied our automatic spectral classification tool (discussed in \citealt{donati2012}) to several of the highest S/N spectra for both stars. We fit the observed spectrum using multiple windows in the wavelength ranges 515--520~nm and 600--620~nm (using Kurucz model atmospheres, \citealt{kurucz1993}), in a similar way to the method of \cite{valenti2005}. This process yields estimates of \teff and \logg, where the optimum parameters are those that minimize $\chi^{2}$, with errors bars determined from the curvature of the $\chi^{2}$ landscape at the derived minimum. 

For \ta, we find that ${\teff = 4425\pm50}$~K and ${\logg = 4.5\pm0.2}$ (with $g$ in cgs units). While two-temperature modelling such as that carried out by \citet{gullysantiago2017} would provide a better estimate of \teff and the fractional spot coverage, for our purposes, a homogeneous model is sufficient. For \ta, we adopt the $V$ and $B$ magnitudes of $10.88\pm0.05$ and $12.19\pm0.05$ from \citep{messina2010}, and assuming a spot coverage of the visible stellar hemisphere of $\sim30$~per~cent (typical for such active stars, see Section~\ref{sec:tomography}), we derive an unspotted $V$ magnitude of $10.6\pm0.2$. We note that assuming a different spot coverage (such as 0 or 50~per~cent) places our derived parameters within our quoted error bars. Using the relation from \citet{pecaut2013}, the expected visual bolometric correction for \ta is $BC_{\text{v}} = -0.70\pm0.04$, and as there is no evidence of extinction to TWA members \citep[e.g.][]{stelzer2013}, we adopt $A_{\textsc{v}} = 0$. Combining $V$, $BC_{\text{v}}$, $A_{\textsc{v}}$ and the trigonometric parallax distance found by Gaia of $63.9 \pm 1.4$~pc (corresponding to a distance modulus of $4.03\pm0.05$, \citealt{gaia,gaiadr2}, in excellent agreement with the $59.59\pm3.6$~pc of \citealt{donaldson2016}), we obtain an absolute bolometric magnitude of $5.85\pm0.29$, or equivalently, a logarithmic luminosity relative to the Sun of $-0.44\pm0.12$. When combined with the photospheric temperature obtained previously, we obtain a radius of $1.0\pm0.2$~\rsun. 

\begin{figure}
\includegraphics[width=\columnwidth]{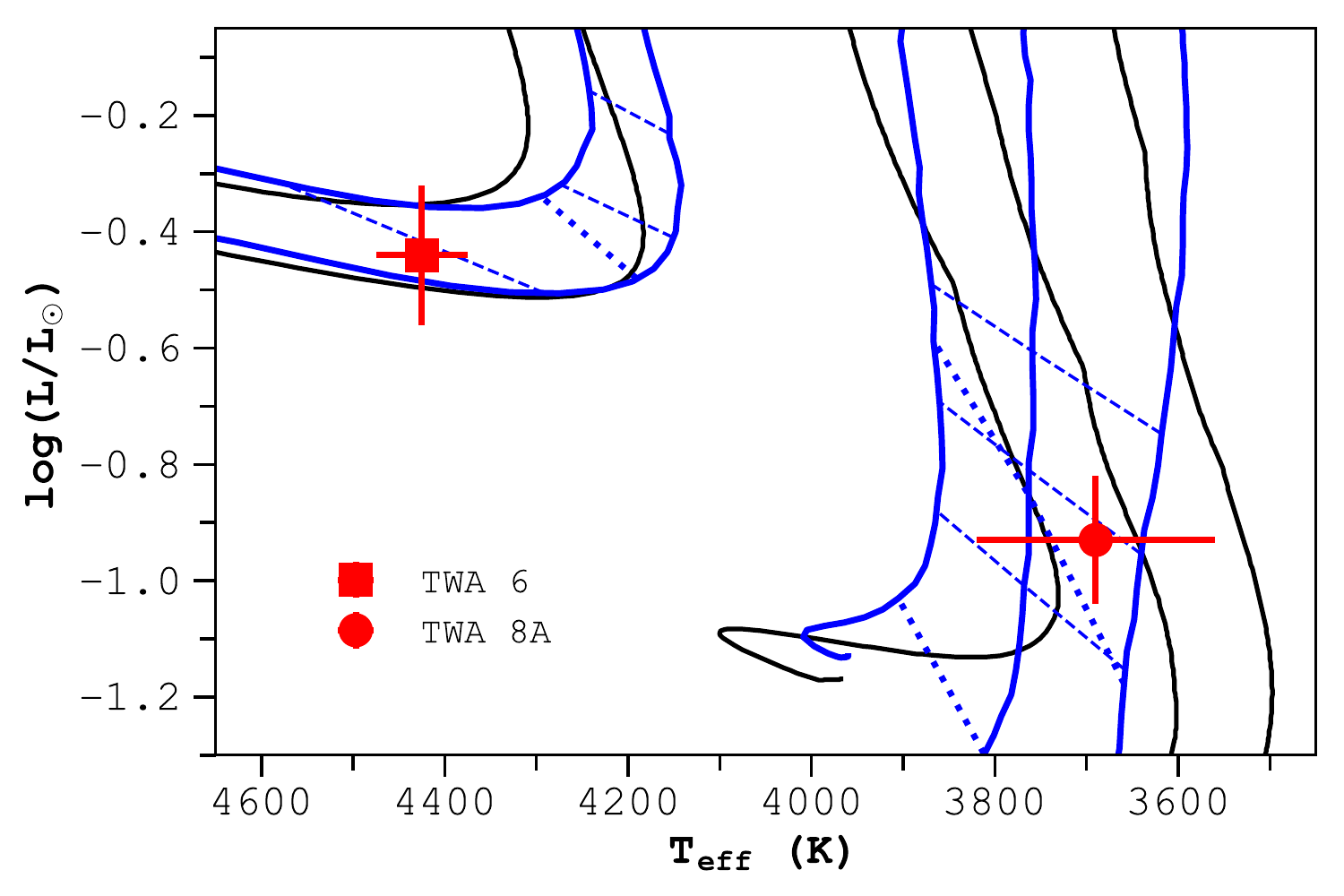}
\caption{H-R diagram showing the stellar evolutionary tracks provided by \citet[blue solid lines]{siess2000} and \citet[black solid lines]{baraffe2015} for masses of 0.3, 0.4, 0.5, 0.9 and 1.0~\msun. Blue dashed lines show the corresponding isochrones for ages 5, 10 \& 20~Myr, and blue dotted lines mark the 0 and 50~\pc fractional radius for the bottom of the convective envelope, both for for \citet{siess2000} models.}
\label{fig:hr}
\end{figure}

Coupling \prot (see Equation~\ref{eq:ephemeris}) with the measured \vsini of $72.6\pm0.5$~\kms (see Section~\ref{sec:tomography}), we can infer that $R_{\star}\sin{i}$ is equal to $0.78\pm0.01$~\rsun, where $R_{\star}$ and $i$ denote the stellar radius and the inclination of its rotation axis to the line of sight. By comparing the luminosity-derived radius to that from the stellar rotation, we derive that $i$ is equal to $49\degr^{+15}_{-8}$, in excellent agreement with that found using our tomographic modelling (see Section~\ref{sec:tomography}). Using the evolutionary models of \cite{siess2000} (assuming solar metallicity and including convective overshooting), we find that \ta has a mass of $0.95\pm0.10$~\msun, with an age of $21\pm9$~Myr (see the H-R diagram in Figure~\ref{fig:hr}, with evolutionary tracks and corresponding isochrones). Similarly, using the evolutionary models of \citet{baraffe2015}, we obtain a mass of $0.95\pm0.10$~\msun and an age of $17\pm7$~Myr.

For \tb, our spectral fitting code yields a best-fit at ${\teff = 3800\pm150}$~K and ${\logg = 4.7\pm0.2}$, however, this \teff is in the regime where the Kurucz synthetic spectra are considered unreliable in terms of temperature. To address this issue, we are currently working on a more advanced spectral classification tool based on PHOENIX model atmospheres and synthetic spectra \citep[see][]{allard2014}. In the mean time for the work presented here, we determined \teff for \tb from the observed $B-V$ value and the relation between \teff and $B-V$ for young stars from \cite{pecaut2013} (and by assuming $A_{\textsc{v}} = 0$). We adopt $V = 12.265\pm0.023$ and $B = 13.70\pm0.03$ from \citet{henden2016}, with $B-V = 1.434\pm0.038$. Using this $B-V$ with the relation between intrinsic colour and \teff for young stars found by \cite{pecaut2013}, and assuming $A_{\textsc{v}} = 0$, we derive $\teff = 3690\pm130$~K. Combining the observed $V$ magnitude with the expected $BC_{\text{v}}$ for \tb of $-1.50\pm0.19$ \citep{pecaut2013} with the trigonometric parallax distance of $46.27\pm0.19$~pc as found by Gaia (\citealt{gaia,gaiadr2}, corresponding to a distance modulus of $3.326\pm0.009$, in excellent agreement with the $47.2\pm2.8$~pc of \citealt{donaldson2016} and $46.9_{-2.9}^{+3.3}$~pc of \citealt{riedel2014}), we obtain an absolute bolometric magnitude of $7.1\pm0.3$, or equivalently, a logarithmic luminosity relative to the Sun of $-0.93\pm0.11$. When combined with the photospheric temperature obtained previously, we obtain a radius of $0.8\pm0.2$~\rsun. Combining this radius with the mass derived below (from \citealt{baraffe2015} evolutionary models), we estimate $\logg = 4.3\pm0.3$.

Combining \prot (see Equation~\ref{eq:ephemeris}) with the \vsini of $4.82\pm0.16$~\kms (see Section~\ref{sec:specfit}), we find $R_{\star}\sin{i} = 0.44\pm0.03$~\rsun, yielding $i = 32\degr^{+13}_{-8}$, in good agreement with our tomographic modelling (see Section~\ref{sec:tomography}). Using \citet{siess2000} models we find M$ = 0.45\pm0.10$~\msun, with an age of $11\pm5$~Myr. Using the evolutionary models of \citet{baraffe2015}, we find M=$0.55\pm0.1$~\msun, with an age of $13\pm6$~Myr.

We note that we do not consider the formal error bars on the derived masses and ages to be representative of the true uncertainties, given the inherent limitations of these evolutionary models. Furthermore, we note that for internal consistency with previous MaPP and MaTYSSE results, the values from the \cite{siess2000} models should be referenced. We note that the ages derived here are consistent with the age of the young TWA moving group (of $10\pm3$~Myr, \citealt{bell2015}), and that both evolutionary models suggest that \ta has a mostly radiative interior, where as \tb is mostly (or fully) convective.

The temperatures measured here are hotter than expected from spectral types estimated from red-optical spectra that cover TiO and other molecular bands  \citep{white2004,stelzer2013,herczeg2014}.  This discrepancy is consistent with past wavelength-dependent differences in photospheric temperatures from young stars, which may be introduced by spots \citep[e.g.][]{bouvier1992,debes2013,gullysantiago2017}.  The interpretation of these differences is not yet understood.  Use of the lower temperatures that are measured at longer wavelengths from molecular bands would lead to lower masses and younger ages. Our temperatures are accurate measurements of the photospheric emission from 5000--6000 \AA\ and are consistent with all temperature measurements for stars in the MaTYSSE program.  

\subsection{Spectral energy distributions}
\label{sec:sed}
\begin{figure}
\centering
\includegraphics[width=\columnwidth]{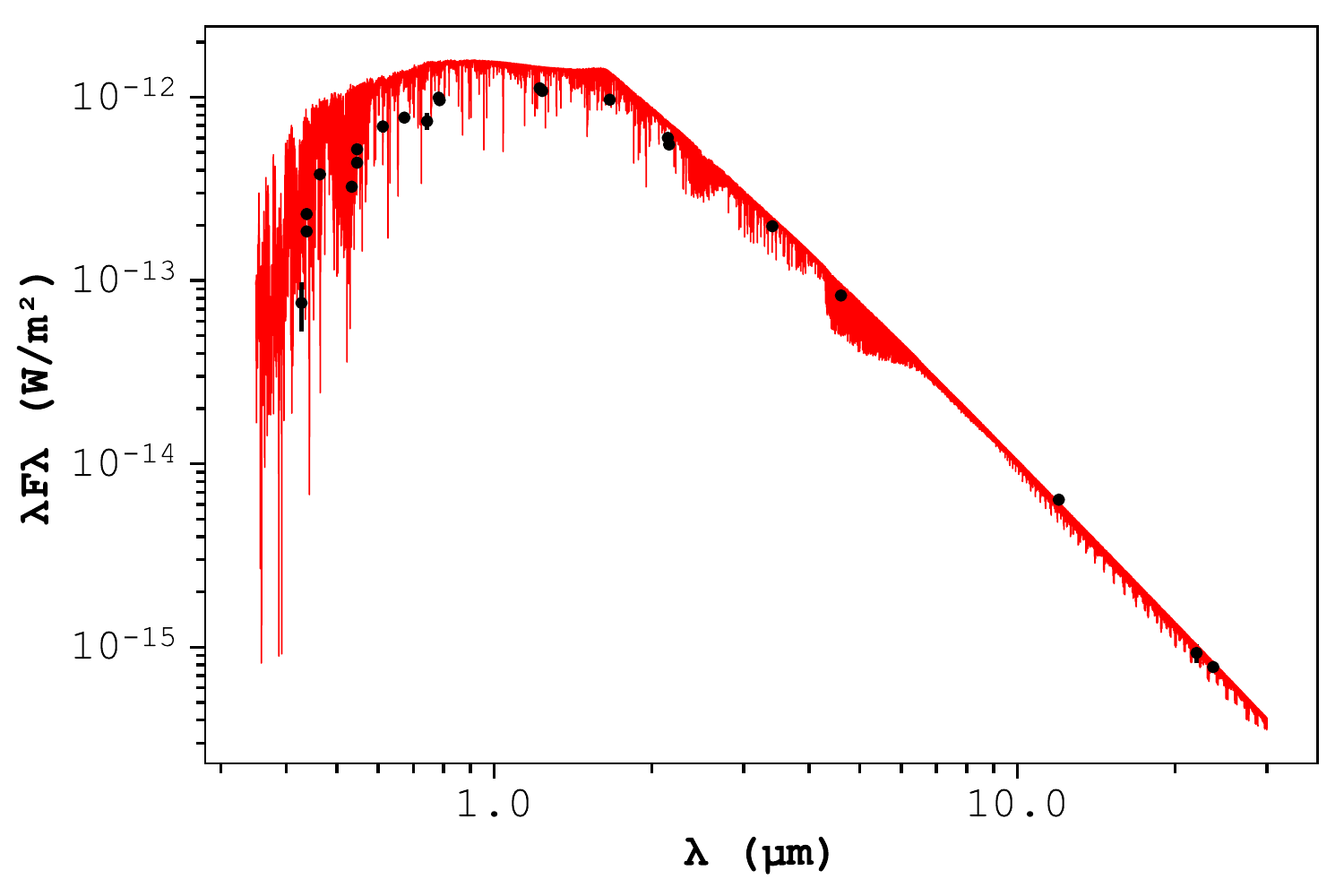}
\includegraphics[width=\columnwidth]{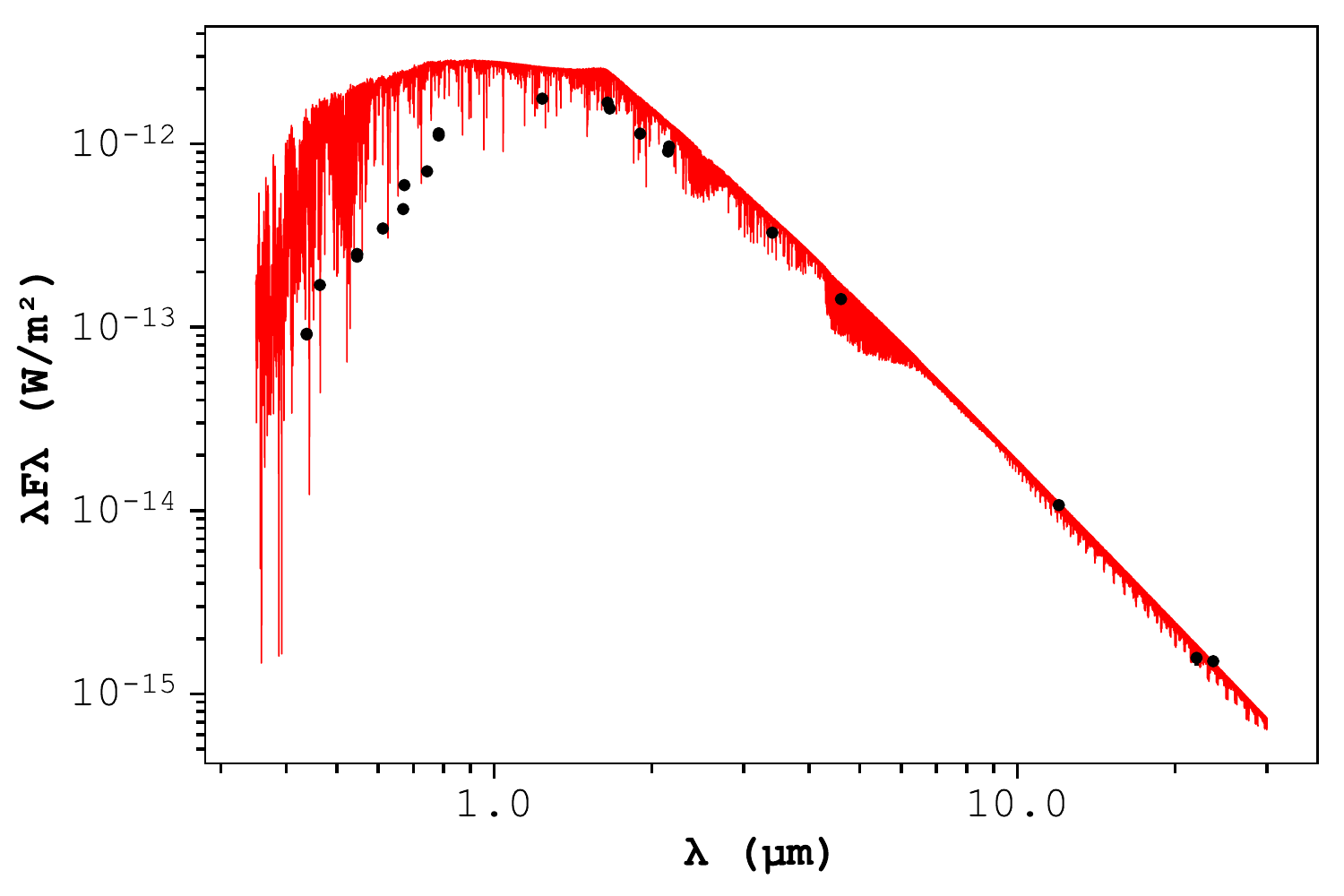}
\caption{Spectral energy distributions (SEDs) of \ta (top) and \tb (bottom), where the photometric data (see text) are shown as black dots, and where PHOENIX-BT-Settl model spectra \protect \citep{allard2014} are shown as a red line. For the model spectra we adopt $\teff = 4400$~K and 3700~K for \ta and \tb respectively, and $\logg = 4.5$ for both stars, as well as the other parameters given in Table~\ref{tab:syspars}, adopting the extinction relation of \citet{cardelli1989}. Furthermore, we assume that both stars have a 30~\pc surface coverage of cool starspots (see Section~\ref{sec:imaging}), and so the displayed spectra have a 30~\pc contribution from a spectrum that is 1000~K cooler.}
\label{fig:sed}
\end{figure}

Spectral Energy Distributions (SEDs) of \ta and \tb were constructed using photometry sourced from the DENIS survey \citep{denis2005}, the AAVSO Photometric All Sky Survey \citep[APASS,][]{henden2015}, the GALEX all-sky imaging survey \citep{bianchi2011}, the TYCHO-2 catalogue \citep{hog2000}, the WISE, Spitzer and Gaia catalogues \citep{wise, spitzer, gaia, gaiadr2}, and \citet{torres2006}. We note that deep, sensitive sub-mm and mm photometry are not currently available for our targets. Comparing the SEDs (shown in Fig.~\ref{fig:sed}) to PHOENIX-BT-Settl synthetic spectra \citep{allard2014}, we find that neither \ta nor \tb have an infrared excess up to 23.675~$\mu$m, indicating that both objects have dissipated their circumstellar discs. Given that the SEDs of \ta and \tb show no evidence of an infra-red excess, both stars are likely disc-less and are not accreting (also see e.g., \citealt{weinberger2004, low2005}). However, for completeness, in Appendix~\ref{sec:acc} we present several metrics that determine the accretion rates from emission lines (if accretion were present), with our analysis showing that chromospheric emission likely dominates the line formation for both targets, confirming their classification as wTTSs.

\subsection{Emission line analysis}
\label{sec:emission}
We find that \ta shows core \ion{Ca}{ii} infrared triplet (IRT) emission (see Fig.~\ref{fig:twa6lines}) with a mean equivalent width (EW) of around 0.3~\AA\ (10.7~\kms), similar to what is expected from chromospheric emission for such PMS stars \citep[e.g.][]{ingleby2011}, and lower than that for accreting cTTSs \citep[e.g.][]{donati2007}. The core \ion{Ca}{ii}~IRT emission is somewhat variable, with both red and blue-shifted peaks (where the red-shifted emission is generally larger), and where the emission is significantly higher at cycles 9.258, 9.359, 14.951 and 15.063. We note that there are some differences in the \sv line profiles of the \ion{Ca}{ii}~IRT, that are likely due to their different atmospheric formation heights. We note that no significant Zeeman signatures are detected in \ion{Ca}{ii}~H\&K, \ion{Ca}{ii}~IRT or \ion{He}{i}~5875.62~\AA\, and so the emission is likely chromospheric rather than from the magnetic footpoints of an accretion funnel. \ta also shows single-peaked \ha and \hb emission that displays relatively little variability over the $\sim30$ rotation cycles (see Fig.~\ref{fig:twa6lines}). For \ha, significantly higher flux is seen in cycles 9.258, 9.359 and 9.439, with the extra emission arising in a predominantly red-shifted component. Moreover, cycle 14.797 displays a significantly higher flux that is symmetric about zero velocity. This higher flux is also seen in \hb, with larger emission for cycles 9.258 and 9.359 (both asymmetric, red-shifted), 14.797 (symmetric) and 14.951 (asymmetric, red-shifted). Given that these emission features occur at similar phases in \ion{Ca}{ii} IRT, \ha and \hb, and are also short lived, they likely stem from the same formation mechanism in the form of stellar prominences that are rotating away from the observer. This conclusion is also supported by the mapped magnetic topology, as we see closed magnetic loops off the stellar limb, along which prominence material may flow. To better determine the nature of the emission and its variability, one can calculate variance profiles and autocorrelation matrices, as described in \citet{johns1995} and given by:

\begin{equation}
V_{\lambda} = \left[\frac{\sum^{n}_{i=1}\left(I_{\lambda,i} - \overline{I_{\lambda}}\right)^{2}}{n-1}\right]^{\frac{1}{2}}
\end{equation}

\noindent Fig.~\ref{fig:twa6var} shows that the \ha emission varies from around -200~\kms to +300~\kms (similar to that found previously for \ta by \citealt{skelly2008}), well beyond the \vsini of 72.6~\kms, and with most of the variability in a red-shifted component. Furthermore, the autocorrelation matrix shows strong correlation of the low-velocity components, indicating a common origin. We find that \hb and \ion{He}{i}~D3 show negligible variability, with a relatively low spectral S/N limiting the analysis. 

In the case of \tb, core \ion{Ca}{ii}~IRT emission is present with a mean EW of around 0.37~\AA\ (11.9~\kms, see Fig.~\ref{fig:twa8alines}). This emission is mostly non-variable, with only cycle 1.958 showing significantly higher (symmetric) emission. Furthermore, the Zeeman signatures in the \sv line profiles (see Fig.~\ref{fig:twa8alinessv}) have the same sign as those of the absorption lines (see Fig.~\ref{fig:lsd}), and so are of photospheric origin. \tb also displays double-peaked \ha and \hb emission, with a peak separation of around 40~\kms. This separation lies well within the co-rotation radius, and is only a few times larger than the \vsini of 4.82~\kms, indicating that the source of the emission is chromospheric. The lines are somewhat variable, with a significant increase in emission (for both \ha and \hb) at cycles 1.958, 2.579 and 3.018. Fig.~\ref{fig:twa8avar} shows the variance profiles and autocorrelation matrices of \ha, \hb and \ion{He}{i}~D3, Here we see that for \ha, the variability concentrates in two peaks centred around -50~\kms and +75~\kms (ranging $\pm150$~\kms), with variability in \hb likewise occurring in two peaks centred around -75~\kms and +65~\kms (ranging $\pm150$~\kms), with both autocorrelation matrices showing the low-velocity components to be highly correlated. For \ion{He}{i}~D3 we find that the variability is single peaked, centred around zero velocity, with only low-velocity components showing significant correlation. We also note that the \ha emission of \tb shows strong Zeeman signatures (see Fig.~\ref{fig:twa8alinessv}) that are opposite in sign to those of the absorption lines (see Fig.~\ref{fig:lsdprofiles}), as expected for chromospheric emission.

\section{Tomographic modelling}
\label{sec:tomography}
In order to map both the surface brightness and magnetic field topology of \ta and \tb, we have applied our dedicated stellar-surface tomographic-imaging package to the data sets described in Section~\ref{sec:observations}. In doing this, we assumed that the observed variability is dominated by rotational modulation (and optionally differential rotation). Our imaging code simultaneously inverts the time series of \si and \sv profiles into brightness maps (featuring both cool spots and warm plages) and magnetic maps (with poloidal and toroidal components, using a spherical harmonic decomposition). For brightness imaging, a copy of a local line profile is assigned to each pixel on a spherical grid, and the total line profile is found by summing over all visible pixels (at a given phase), where the pixel intensities are scaled iteratively to fit the observed data. For magnetic imaging, the Zeeman signatures are fit using a spherical-harmonic decomposition of potential and toroidal field components, where the weighting of the harmonics are scaled iteratively \citep{donati2001}. The data are fit to an aim $\chi^{2}$, with the optimal fit determined using the maximum-entropy routine of \cite{skilling1984}, and where the chosen map is that which contains least information (where entropy is maximized) required to fit the data. For further details about the specific application of our code to wTTSs, we refer the reader to previous papers in the series \citep[e.g.,][]{donati2010b,donati2014,donati2015}.

As with previous studies of wTTSs, we applied the technique of Least-Squares Deconvolution \citep[LSD,][]{donati1997} to all of our spectra. Given that relative noise levels are around $10^{-3}$ in a typical spectrum (for a single line), with Zeeman signatures exhibiting relative amplitudes of $\sim0.1$~\pc, the use of LSD allows us to create a single `mean' line profile with a dramatically enhanced S/N, with accurate error bars for the Zeeman signatures. LSD involves cross-correlating the observed spectrum with a stellar line-list, and for this work, stellar line lists were sourced from the Vienna Atomic Line Database \citep[VALD, ][]{ryabchikova2015}, computed for $\teff = 4500$~K  and $\log{g} = 4.5$ (in cgs units) for \ta, and $\teff = 3750$~K  and $\log{g} = 4.5$ for \tb (the closest available to our derived spectral-types, see Section~\ref{sec:evolution}). Only moderate to strong atomic spectral lines were included (with line-to-continuum core depressions larger than 40~per~cent prior to all non-thermal broadening). Furthermore, spectral regions containing strong lines mostly formed outside the photosphere (e.g. Balmer, He, \ion{Ca}{ii}~H\&K and \ion{Ca}{ii}~IRT lines) and regions heavily crowded with telluric lines were discarded (see e.g. \citealt{donati2010b} for more details), leaving 6088 and 5953 spectral lines for use in LSD, for \ta and \tb, respectively. Expressed in units of the unpolarized continuum level $I_{\text{c}}$ (and per 1.8~\kms velocity bin), the average noise level of the resulting \sv signatures range from 4.1--7.5$\times10^{-4}$ (median of $5.3\times10^{-4}$) for \ta, and 2.4--5.7$\times10^{-4}$ (median of $2.6\times10^{-4}$) for \tb.

\begin{figure}
\includegraphics[width=\columnwidth]{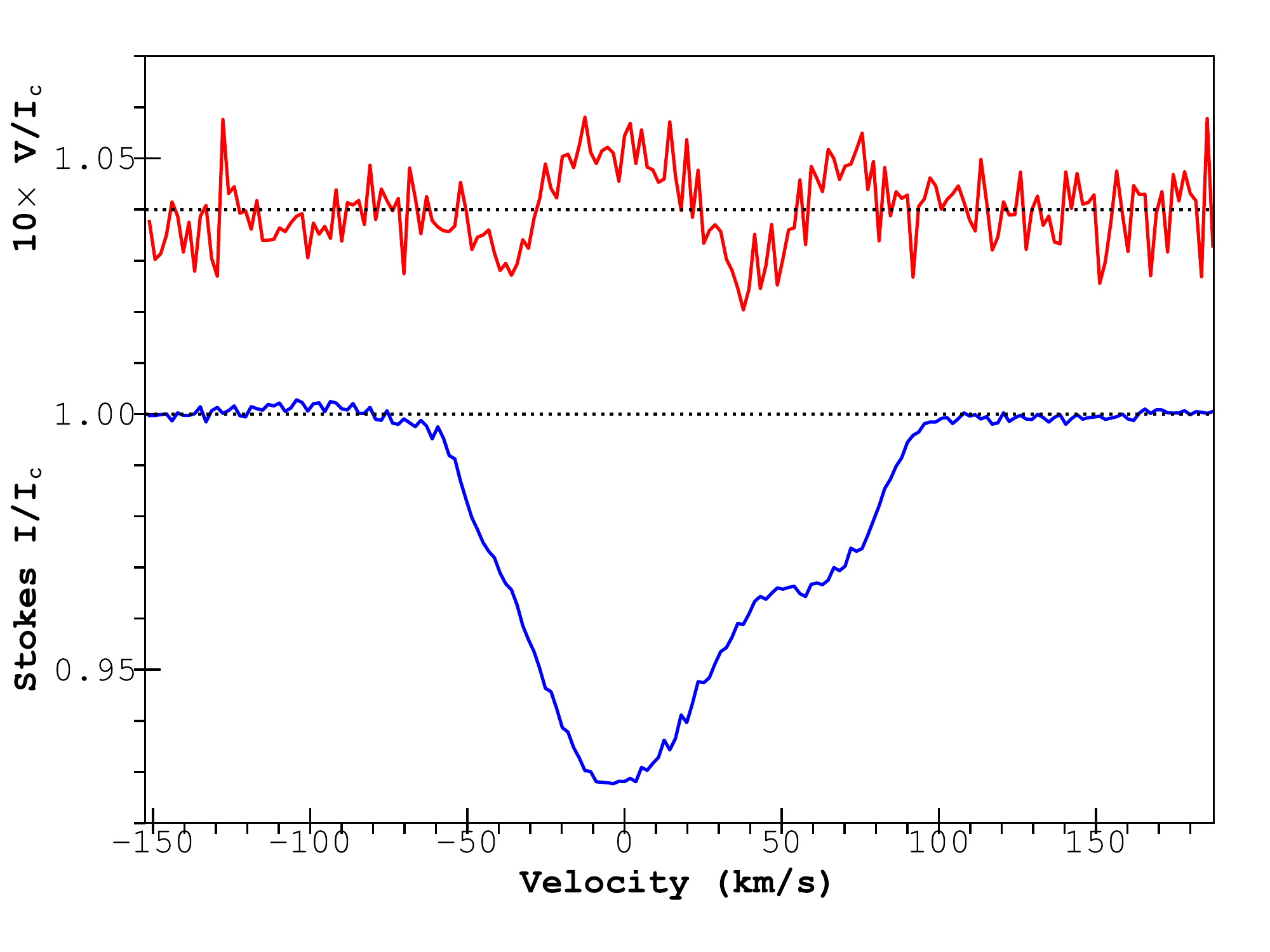}
\includegraphics[width=\columnwidth]{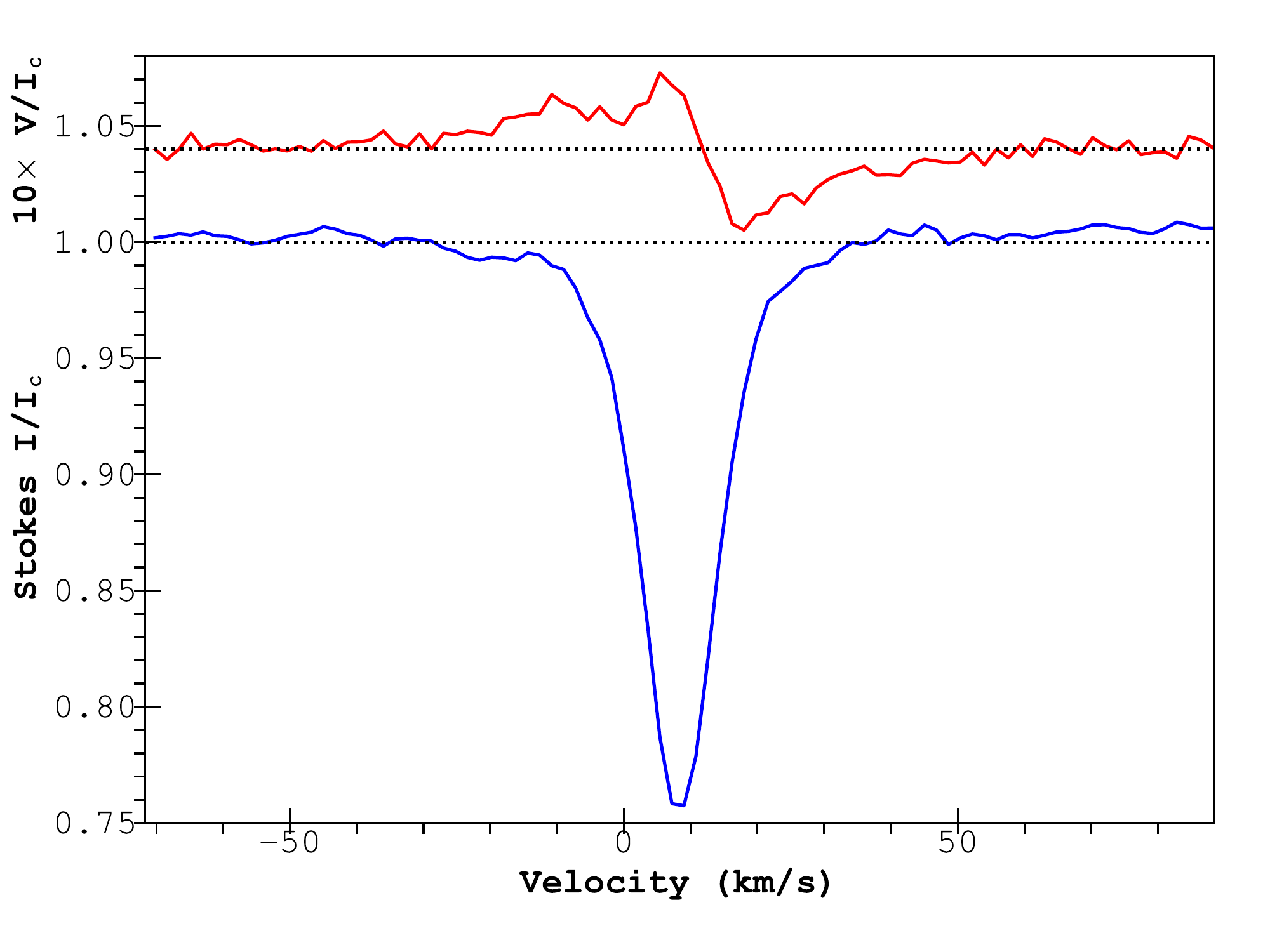}
\caption{LSD circularly-polarized (\sv, top/red curve) and unpolarized (\si, bottom/blue curve) profiles of \ta (top, collected on 19-02-2014, cycle 27.979) and \tb (bottom, collected on 26-03-2015, cycle 0.223). Clear Zeeman signatures are detected in both LSD \sv profiles in conjunction with the unpolarized line profiles. The mean polarization profiles are expanded by a factor of 10 shifted upwards by 0.04 for display purposes.}
\label{fig:lsd}
\end{figure}

The disc-integrated average photospheric LSD profiles are computed by first synthesizing the local Stokes $I$ and $V$ profiles using the Unno-Rachkovsky analytical solution to the polarized radiative transfer equations in a Milne-Eddington model atmosphere, taking into account the local brightness and magnetic field. Then, these local line profiles are integrated over the visible hemisphere (including linear limb darkening, with a coefficient of 0.75, as observed young stars, e.g.~\citealt{donati1997abdor}) to produce synthetic profiles for comparison with observations. This method provides a reliable description of how line profiles are distorted due to magnetic fields (including magneto-optical effects, e.g., \citealt{landi2004}). The main parameters of the local line profiles are similar to those used in our previous studies; the wavelength, Doppler width, equivalent width and Land\'{e} factor being set to 670~nm, 1.8~\kms, 3.9~\kms and 1.2, respectively. 

\begin{figure}
\includegraphics[width=\columnwidth]{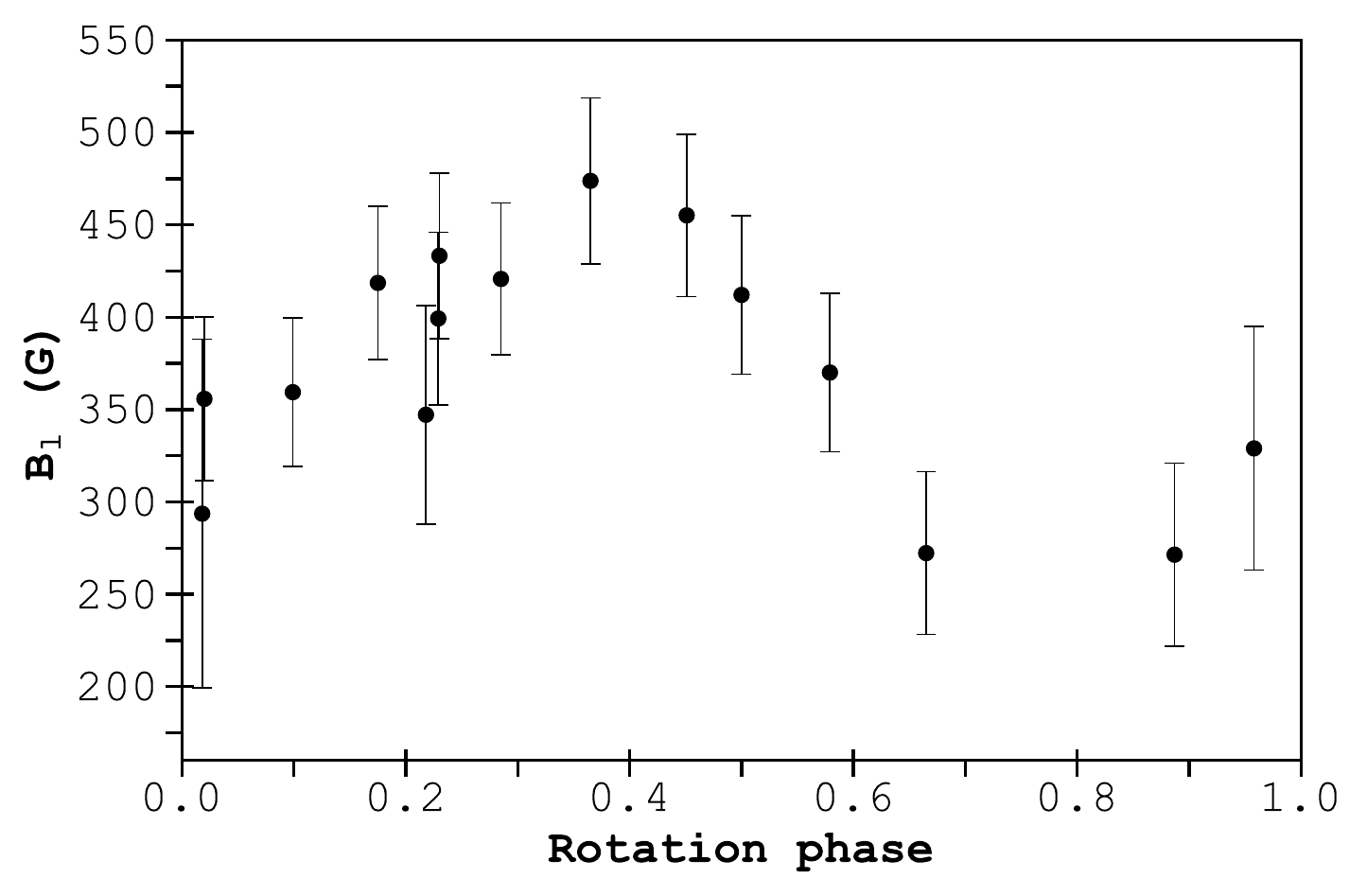}
\caption{The longitudional field strengths $\langle\textbf{B}_{l}\rangle$ for \tb, as measured from the LSD profiles.}
\label{fig:bl}
\end{figure}

We note that while Zeeman signatures are detected at all times in \sv LSD profiles for both stars (see Figure~\ref{fig:lsd} for an example), \tb exhibits much larger longitudinal field strengths ($\textbf{B}_{l}$), similar to those of e.g. mid M dwarfs \citep[see][]{morin2008b}, with values shown in Fig.~\ref{fig:bl}, as calculated from the LSD profiles. Here we clearly see the periodicity in field strength, with the maximum $\textbf{B}_{l}$ around phase 0.37, coincident with the phase of the aligned dipole of the magnetic field (see Fig.~\ref{fig:magmap}) being viewed along the line of sight, with the minimum $\textbf{B}_{l}$ seen around half a rotation later. \tb also exhibits significant Zeeman broadening in the \si profiles that we model in Section~\ref{sec:specfit}, with almost no distortions due to brightness inhomogeneities on the surface. 

As part of the imaging process we obtain accurate estimates for \vrad (the RV the star would have if unspotted), equal to $17.5\pm0.1$~\kms and $8.34\pm0.10$~\kms, the inclination $i$ of the rotation axis to the line of sight, equal to $46\degr\pm10\degr$ and $42\degr\pm10\degr$, for \ta and \tb, respectively, and for \ta the \vsini equal to $72.6\pm0.1$~\kms (see Table~\ref{tab:syspars}, in excellent agreement with the values derived in Section~\ref{sec:evolution}). For \tb, we fixed the \vsini to 4.82~\kms, as this was determined by direct spectral fitting in Section~\ref{sec:specfit} and is more accurate than that derived from ZDI.

\begin{table}
\centering
\caption{Main parameters of \ta and \tb as derived from our study, with \vrad noting the RV that the star would have if unspotted, the equatorial rotation rate $\Omega_{\text{eq}}$ and the difference between equatorial and polar rotation rates $d\Omega$ (as inferred from the modelling of Section~\ref{sec:tomography}). Note, the stellar masses and ages are those determined from \protect \cite{siess2000} models, with values from \protect \cite{baraffe2015} given in parenthesis. The \logg for \tb is estimated from its mass (using \protect \cite{baraffe2015} models) and $R_{\star}$.}
\label{tab:syspars}
\begin{tabular}{lll}
\hline					
	&	TWA 6	&	TWA 8A	\\
\hline					
$M_{\star}$~(\msun)	&	$0.95\pm0.10$ ($0.95\pm0.10$)	&	 $0.45\pm0.10$ ($0.55\pm0.10$)	\\
$R_{\star}$~(\rsun)	&	$1.1\pm0.2$	&	$0.8\pm0.2$	\\
Age (Myr)	&	$21\pm9$ ($17\pm7$)	&	$11\pm5$ ($13\pm6$)	\\
$\log{g}$ (cgs units)	&	$4.5\pm0.2$	&	$4.3\pm0.3$	\\
\teff~(K)	&	$4425\pm50$	&	$3690\pm130$	\\
log($L_{\star}$/\lsun)	&	$-0.40\pm0.12$	&	$-0.93\pm0.11$	\\
\prot~(d)	&	$0.54095\pm0.00003$	&	$4.578\pm0.006$	\\
\vsini~(\kms)	&	$72.6\pm0.1$	&	$4.82\pm0.16$	\\
\vrad~(\kms)	&	$17.5\pm0.1$	&	$8.34\pm0.1$	\\
$i$~(\degr)	&	$46\pm10$	&	$42\pm10$	\\
Distance (pc)	&	$63.9\pm1.4^{\text{a}}$	&	$46.27\pm0.19^{\text{a}}$	\\
$\Omega_{\text{eq}}$ (rad d$^{-1}$)	&	$11.6199 \pm 0.0005$	&	-	\\
$d\Omega$ (rad d$^{-1}$)	&	$0.0098 \pm 0.0014$	&	-	\\
\hline					
\multicolumn{3}{l}{\textbf{References}: (a) \citealt{gaia, gaiadr2}.}
\end{tabular}
\end{table}

\subsection{Brightness and magnetic imaging}
\label{sec:imaging}
\begin{figure*}
\includegraphics[height=0.4\textheight]{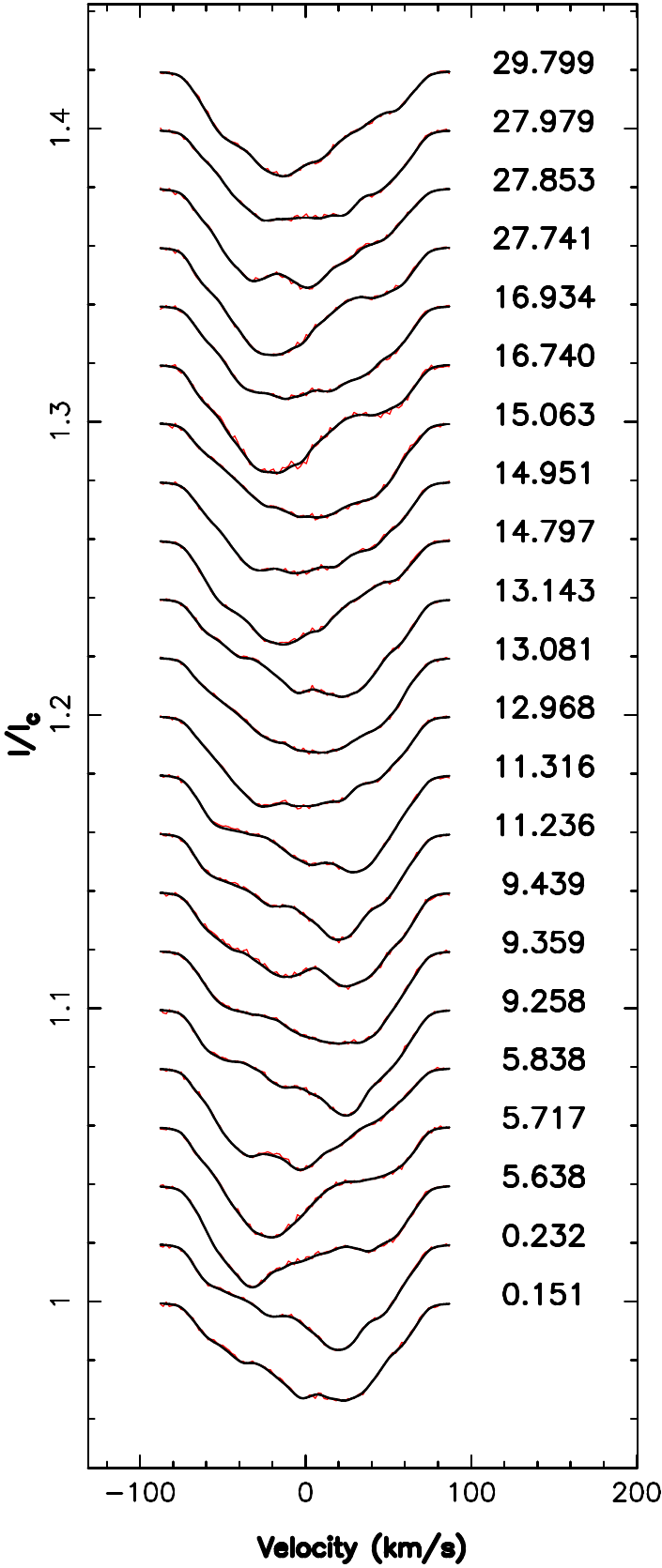}
\includegraphics[height=0.4\textheight]{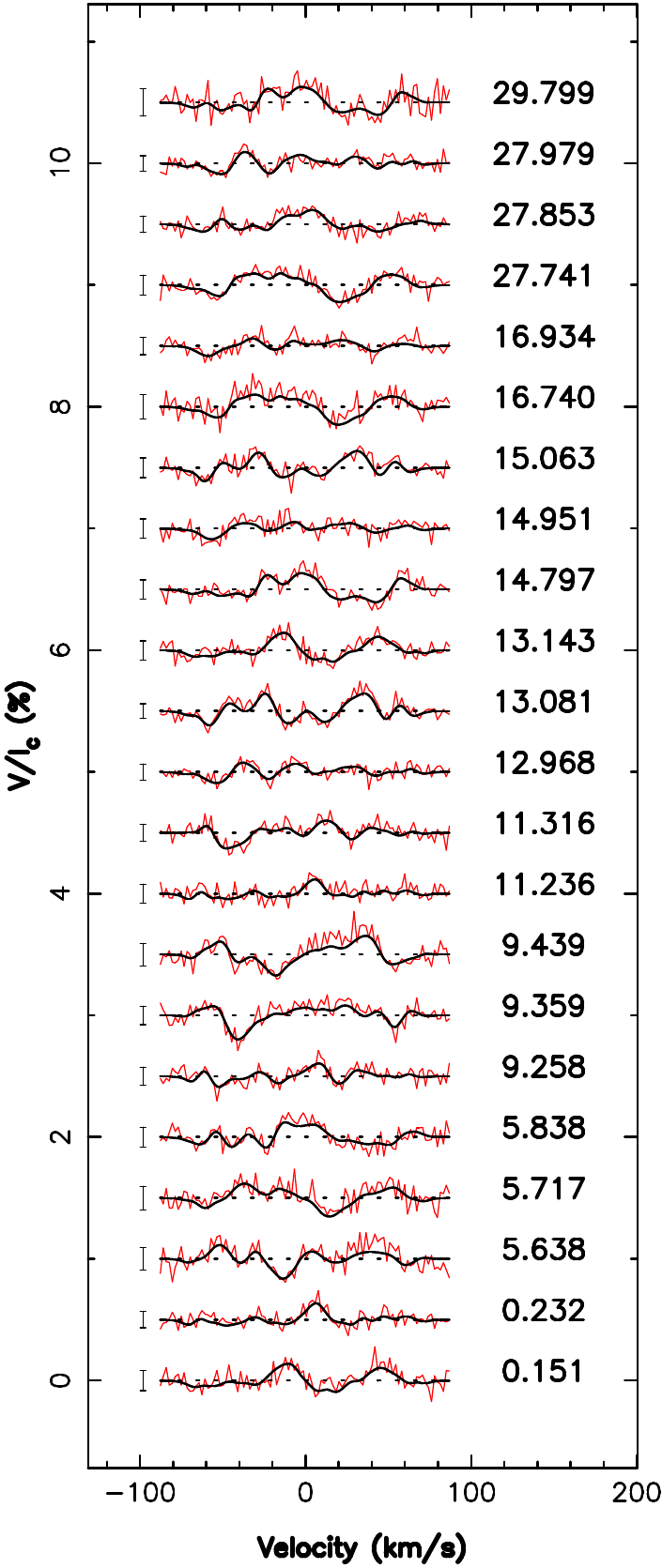}
\hspace{0.04\textwidth}
\includegraphics[height=0.4\textheight]{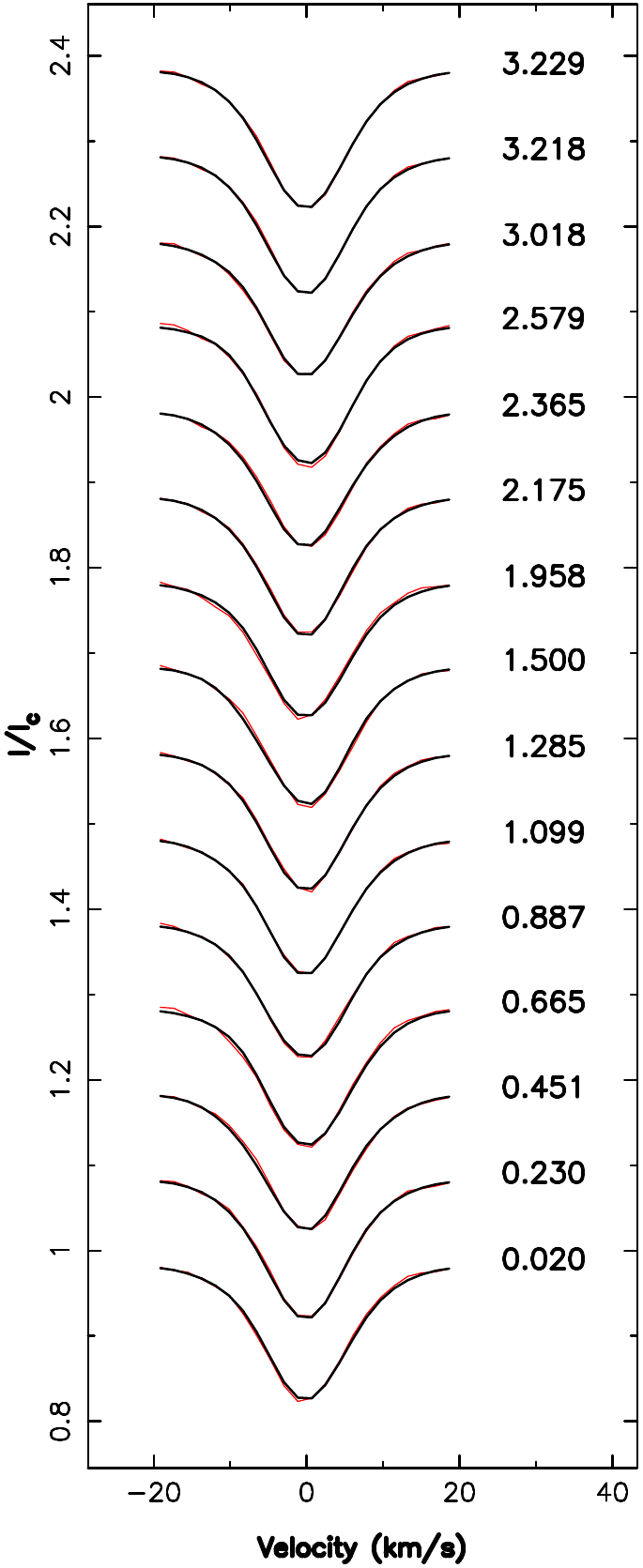}
\includegraphics[height=0.4\textheight]{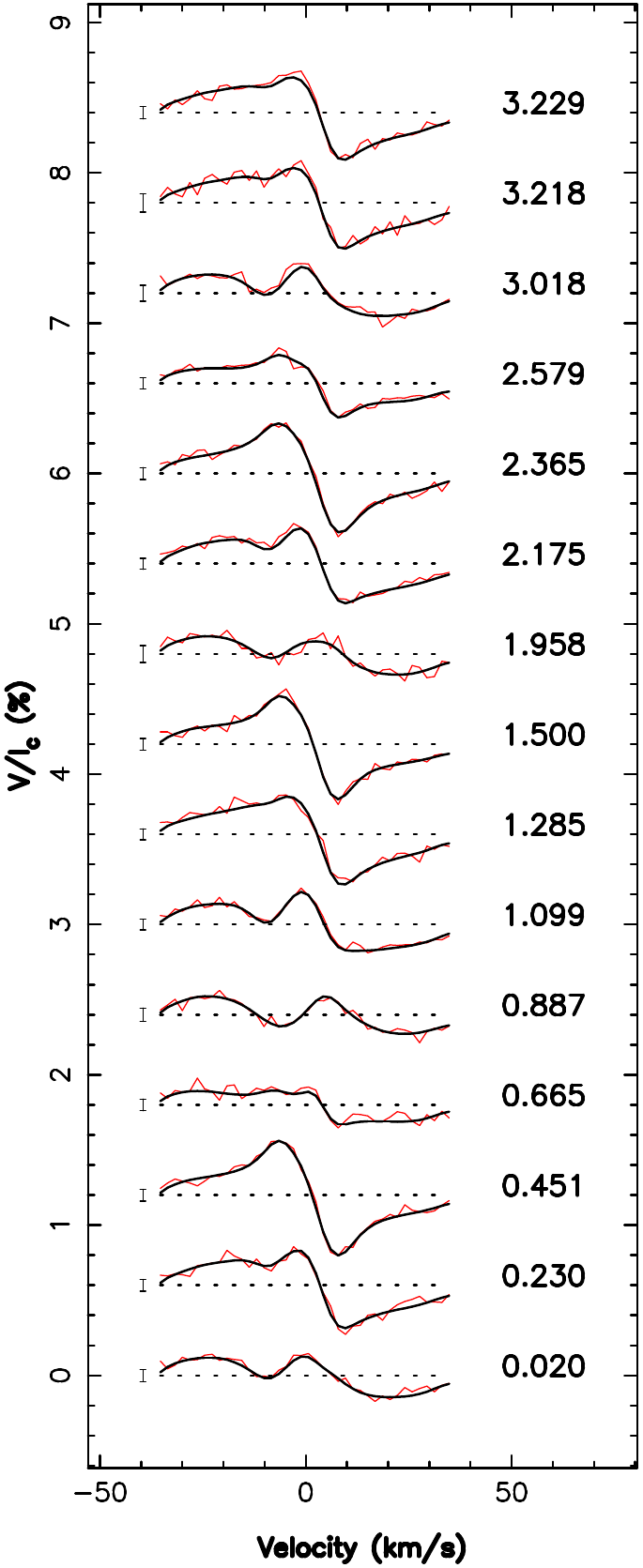}
\caption{Maximum-entropy fit (thin red line) to the observed (thick black line) \si (first and third panels) and \sv (second and fourth panels) LSD photospheric profiles of \ta (first two panels) and \tb (last two panels). Note that for \tb the velocity scales are different. Rotational cycles are shown next to each profile. This figure is best viewed in colour.}
\label{fig:lsdprofiles}
\end{figure*}

The observed LSD profiles for \ta and \tb, as well as our fits to data, are shown in Fig.~\ref{fig:lsdprofiles}. For \ta, we obtain a reduced chi-squared $\chi_{\text{r}}^{2}$ fit equal to 1 (where the number of fitted data points is equal to 4312, with simultaneous fitting of both \si and \sv line profiles). For \tb, the low \vsini means that there is little modulation of the \si line profiles, with the strong magnetic fields causing significant Zeeman broadening of the lines. Indeed, we are able to model the \si line profiles sufficiently well using a stellar model with a homogeneous surface brightness, with our fits to the \sv line profiles yielding $\chi_{\text{r}}^{2} = 1.04$ (for 930 fitted data points). We note that, given the substantially larger \vsini of \ta as compared to \tb, combined with more complete phase coverage, the reconstructed maps of \ta have an effective resolution around 10 times higher.

\begin{figure*}
\includegraphics[width=0.45\textwidth]{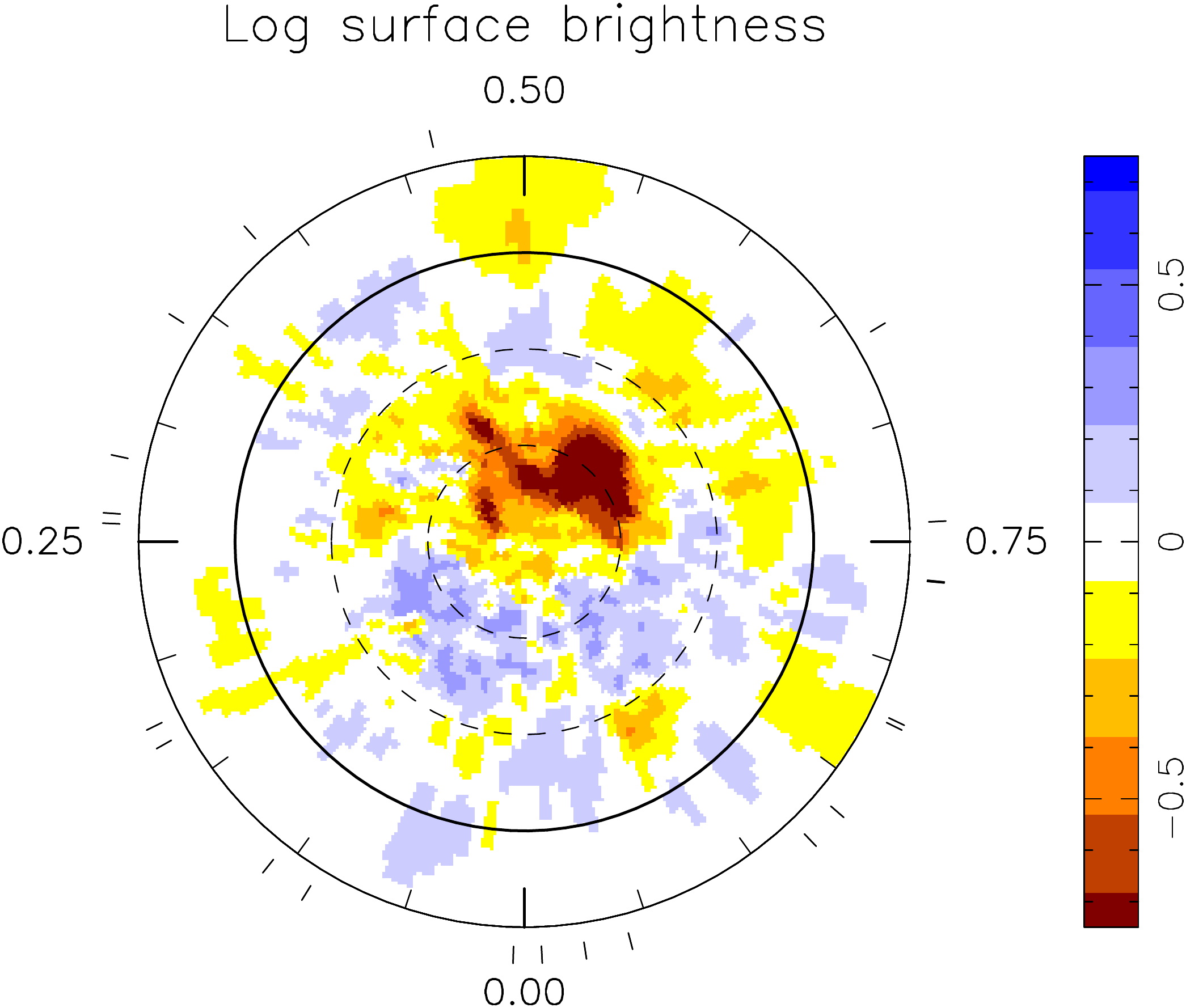}
\caption{Map of the logarithmic brightness (relative to the quiet photosphere) at the surface of \ta. The star is shown in flattened polar projection down to latitudes of $-30\degr$, with the equator depicted as a bold circle, and 30\degr and 60\degr parallels as dashed circles. Radial ticks indicate the phases of observation. This figure is best viewed in colour.}
\label{fig:brightmap}
\end{figure*}

The brightness map of \ta includes both cool spots and warm plages (see Fig.~\ref{fig:brightmap}), with no true polar spot, but rather a large spotted region centred around $60\degr$ latitude (centred around phase 0.6), with the majority of plages at a similar latitude on the opposing hemisphere. These features introduce significant distortions to the \si profiles (see Fig.~\ref{fig:lsdprofiles}), introducing large RV variations (with maximum amplitude 6.0~\kms, see Section~\ref{sec:rv}). Overall, we find a spot and plage coverage of $\simeq17$~\pc (10 and 7~\pc for spots and plages, respectively), similar to that found for V819~Tau, V830~Tau \citep{donati2015}, and Par~2244 \citep{hill2017a}. 

Note that the estimates of spot and plage coverage should be considered as lower limits only, as Doppler imaging is mostly insensitive to small-scale structures that are evenly distributed over the stellar surface (hence the larger minimal spot coverage assumed in Section~\ref{sec:evolution} to derive the location of the stars in the H-R diagram).

\subsection{Magnetic field imaging}
\label{sec:magfield}
Using our imaging code, we have reconstructed the magnetic fields of our target stars using both poloidal and toroidal fields, each expressed using a spherical-harmonic (SH) expansion, with $\ell$ and $m$ denoting the mode and order of the SH \citep{donati2006b}. For a given set of complex coefficients $\alpha_{\ell, m},~\beta_{\ell, m}$ and $\gamma_{\ell,m}$ (where $\alpha_{\ell, m}$ characterizes the radial field component, $\beta_{\ell, m}$ the azimuthal and meridional components of the poloidal field term, and $\gamma_{\ell, m}$ the azimuthal and meridional components of the toroidal field term), one can construct an associated magnetic image at the surface of the star, and thus derive the corresponding \sv data set. Here, we carry out the inverse, where we reconstruct the set of coefficients that fit the observed data.

\begin{figure*}
\includegraphics[width=\textwidth]{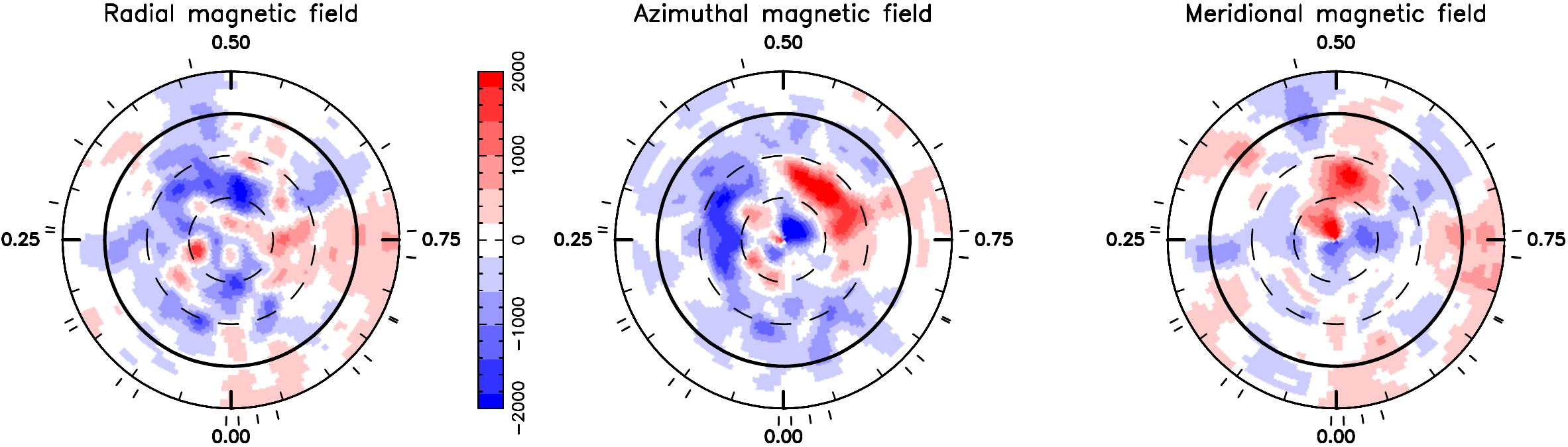}
\includegraphics[width=\textwidth]{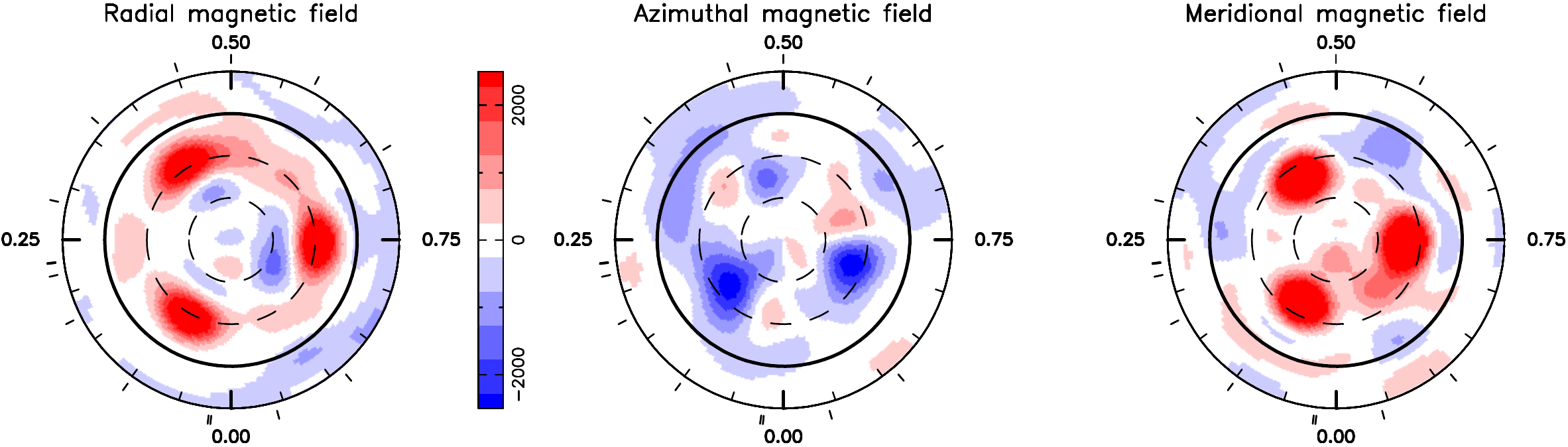}
\caption{Map of the radial (left), azimuthal (middle) and meridional (right) components of the magnetic field \textbf{B} at the surface of \ta (top) and \tb(bottom). Magnetic fluxes in the colourbar are expressed in G. Note that the magnetic filling factor for \tb is $f_{v} = 0.2$. The star is shown in flattened polar projection as in Figure~\ref{fig:brightmap}. This figure is best viewed in colour.}
\label{fig:magmap}
\end{figure*}

For \ta, our reconstructed fields presented in Fig.~\ref{fig:magmap} are limited to SH expansions with terms $\ell \leq 20$. Given the high \vsini of \ta (combined with good phase coverage), we are able to resolve smaller-scale magnetic fields, and indeed such a large number of modes are required to fit the observed \sv signatures. We note, however, that including higher-order terms ($>20$) only marginally improves our fit. Such high-degree modes indicate that the magnetic fields in \ta concentrate on smaller, more compact spatial-scales. In contrast, our fits to the \sv observations of \tb only require terms up to $\ell \leq 10$, with higher order terms providing only a marginal improvement. Hence, the magnetic field of \tb is concentrated at larger spatial scales. 

The reconstructed magnetic field for \ta is split almost evenly between poloidal and toroidal components (53 and 47~\pc, respectively), with a total magnetic energy $\langle B \rangle = 840$~G, where $\langle B \rangle$ is given by

\begin{equation}
\langle B \rangle = \oiint_{\theta, \phi} \left (\mathbf{B}_{\alpha}^{2} + \mathbf{B}_{\beta}^{2} + \mathbf{B}_{\gamma}^{2} \right)^{1/2} d\theta d\phi
\end{equation}

\noindent The poloidal field is mostly axisymmetric (49~\pc), with the largest fraction of energy (58~\pc) in modes with $\ell > 3$, and with 30~\pc of energy in the dipole mode ($\ell = 1$, with a field strength of 550~G). On large scales, the poloidal component is tilted at $35\degr$ from the rotation axis (towards phase 0.34). The toroidal component is also mostly axisymmetric, with the largest fraction of energy (68~\pc) in modes with $\ell >3$, and with 17~\pc of energy in the octupole ($\ell = 3$) mode. These components combine to generate an intense field of $\geq 2$~kG at $45\degr$ latitude around phase 0.50--0.75 and 0.20--0.35, as well as an off-pole 2~kG spot at phase 0.75. We note that the large spotted region reconstructed in the brightness map (around $60\degr$ latitude at phase 0.6, see Fig.~\ref{fig:brightmap}) aligns well with these intense fields, suggesting that they are related.

In the case of \tb, the reconstructed field is 71~\pc poloidal and 29~\pc toroidal, with a total unsigned flux of 1.4~kG, and with a magnetic filling factor of $f_{v} = 0.2$ (where $f_{v}$ is equal to the fraction of the stellar surface that is covered by the mapped magnetic field using \sv data). The poloidal field is mostly axisymmetric (70~\pc), with 16~\pc of the energy in the dipole ($\ell = 1$, with a field strength of 0.72~kG), 21~\pc in the quadrupole ($\ell = 2$), 18~\pc in the octupole ($\ell = 3$), and with the remaining 44~\pc of energy in modes with $\ell >3$. On large scales (several radii from the star), the poloidal component may be approximated by an $B = 0.69$~kG aligned-dipole tilted at $20\degr$ from the rotation axis (towards phase 0.37). The toroidal component is mostly non-axisymmetric, with the majority of energy (55~\pc) in modes with $\ell >3$, and with 21, 6 and 18~\pc in modes with $\ell = 3, 2$ \& $1$. These components combine to generate intense fields in excess of 2~kG in around phases 0.08, 0.42 and 0.75 on the stellar surface, centred around $20\degr$ latitude in the radial field component and around $35\degr$ in the meridional field component. Given the filling factor of $f_{v} = 0.2$, this suggests that surface magnetic fields can locally reach over 10~kG. Moreover, the high fraction of energy in high-order modes suggests that there are a large number of small-scale magnetic features, a conclusion also supported by the direct spectral fitting in Section~\ref{sec:IVspec-fit}.

\begin{figure*}
\includegraphics[width=0.4\textwidth]{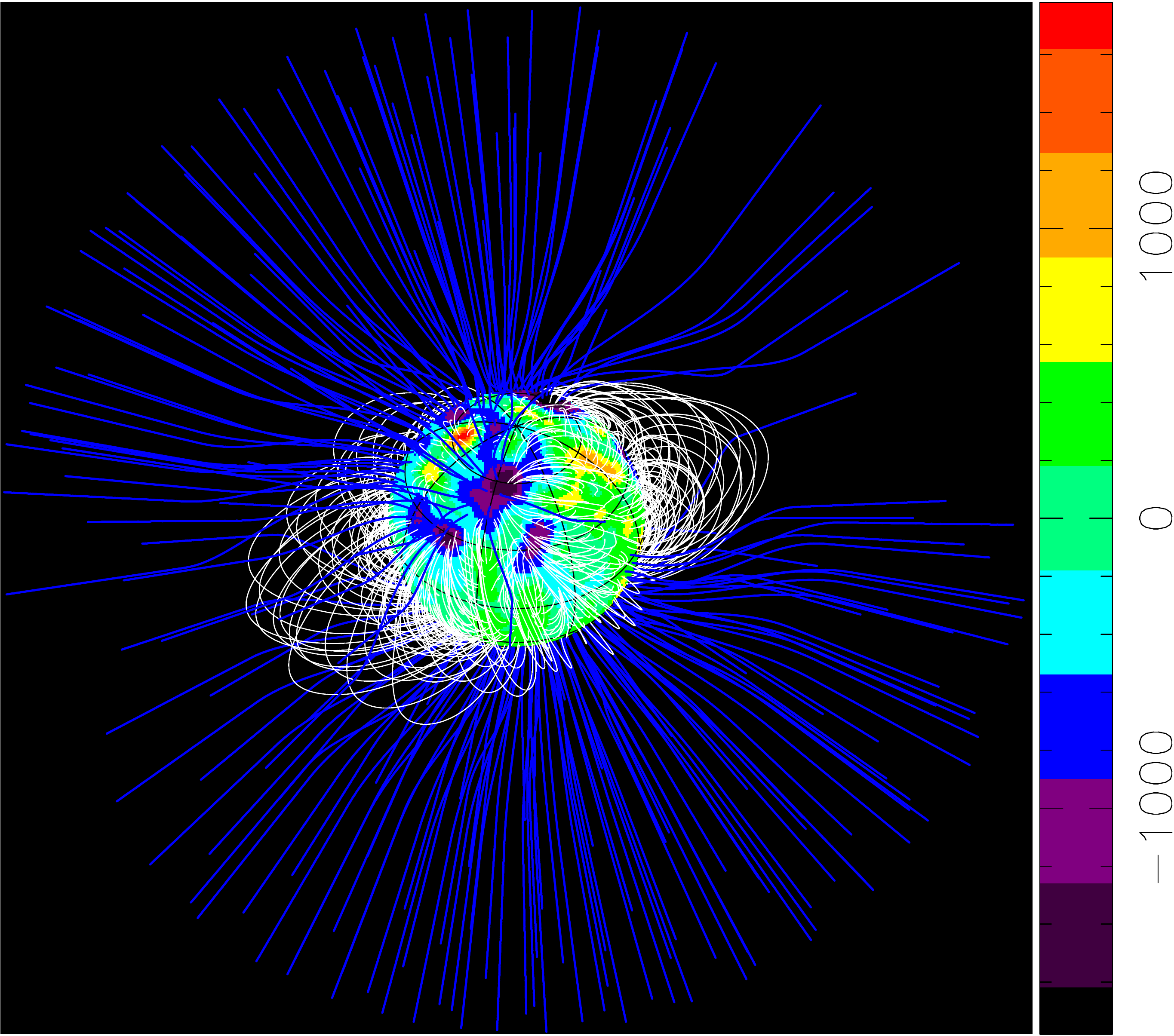}
\hspace{2cm}
\includegraphics[width=0.4\textwidth]{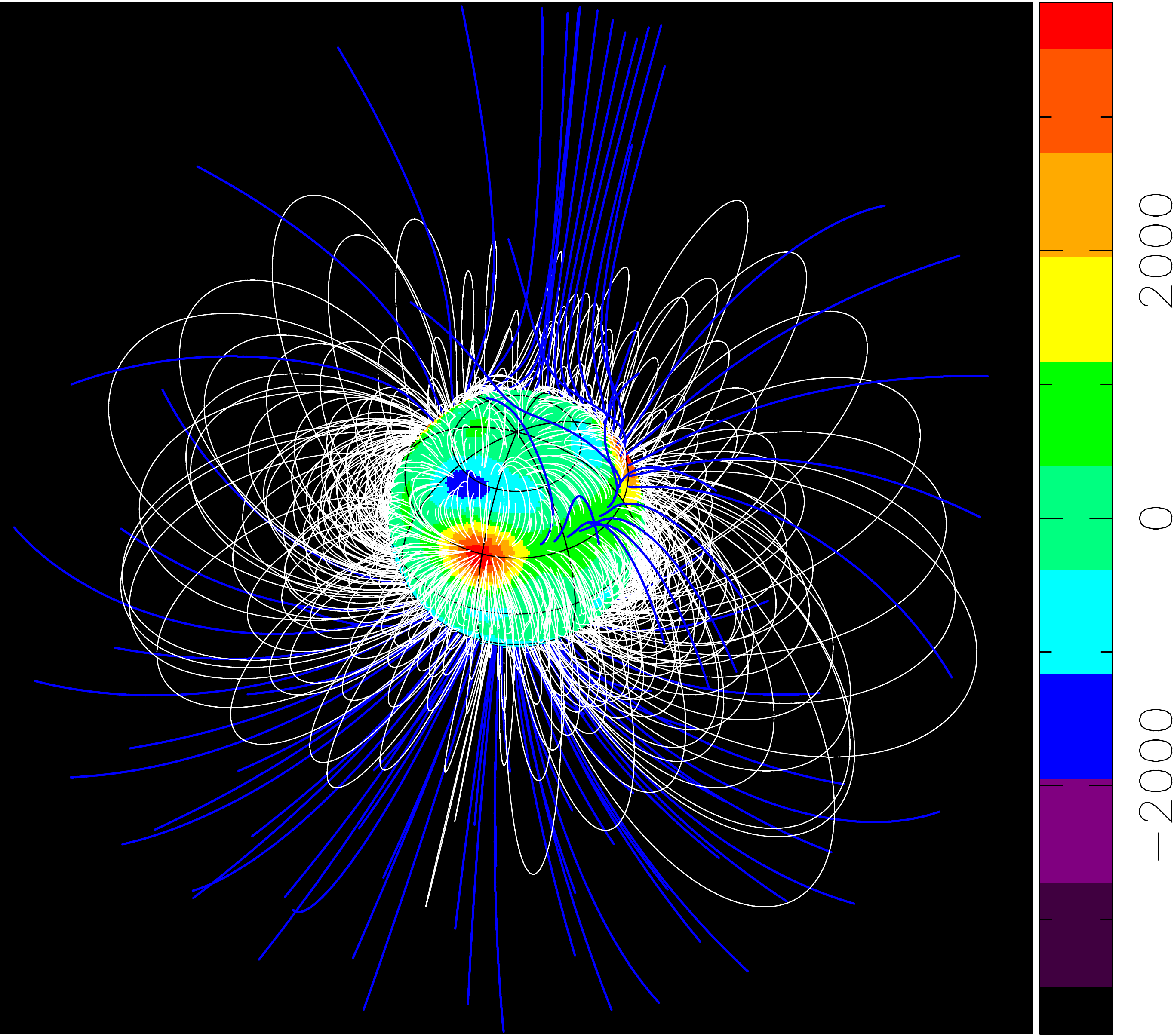}
\caption{Potential field extrapolations of the radial magnetic field reconstructed for \ta (left) and \tb (right), viewed at phases 0.95 and 0.70, with inclinations of $45.6\degr$ and $31\degr$, respectively. Open and closed field lines are shown in blue and white, respectively, whereas colours at the stellar surface depict the local values of the radial field (as shown in the left panels of Figure~\ref{fig:magmap}). The source surfaces at which the field becomes radial are set at distances of 2.6~\rstar for \ta and 10.7~\rstar for \tb, as these are close to the co-rotation radii (where the Keplerian orbital period equals the stellar rotation period, and beyond which the field lines tend to open under the effect of centrifugal forces, \citealt{jardine2004}), and are smaller than or similar to the Alfv\'{e}n radii of $>10$~\rstar \citep{reville2016}. This figure is best viewed in colour. Full animations may be found for both \ta and \tb at https://imgur.com/hSkhYLT and https://imgur.com/AdKptUx.}
\label{fig:extrap}
\end{figure*}

In Figure~\ref{fig:extrap} we use the a potential field approximation (e.g., \citealt{jardine2002}) to extrapolate the large-scale field topologies of \ta and \tb. These topologies are derived solely from the reconstructed radial field components, and represent the lowest possible states of magnetic energy, providing a reliable description of the magnetic field well within the Alfv\'{e}n radius \citep{jardine2013}.

\subsection{Surface differential rotation}
\label{sec:diffrot}
The level of surface differential rotation of \ta was determined in a similar manner as that carried out for other wTTSs \citep[e.g., ][]{skelly2008,skelly2010, donati2014,donati2015}. Assuming that the rotation rate at the surface of the star varies with latitude $\theta$ as $\Omega_{\text{eq}} - d\Omega\sin^2{\theta}$ (where $\Omega_{\text{eq}}$ is the rotation rate at the equator and $d\Omega$ is the difference in rotation rate between the equator and the pole), we reconstruct brightness and magnetic maps at a fixed information content for many pairs of $\Omega_{\text{eq}}$ and $d\Omega$ and determine the corresponding reduced chi-squared $\chi_{\text{r}}^{2}$ of our fit to the observations. The resulting $\chi_{\text{r}}^{2}$ surface usually has a well defined minimum to which we fit a parabola, allowing an estimate of both $\Omega_{\text{eq}}$ and $d\Omega$ (and their corresponding error bars).

\begin{figure}
\includegraphics[width=\columnwidth]{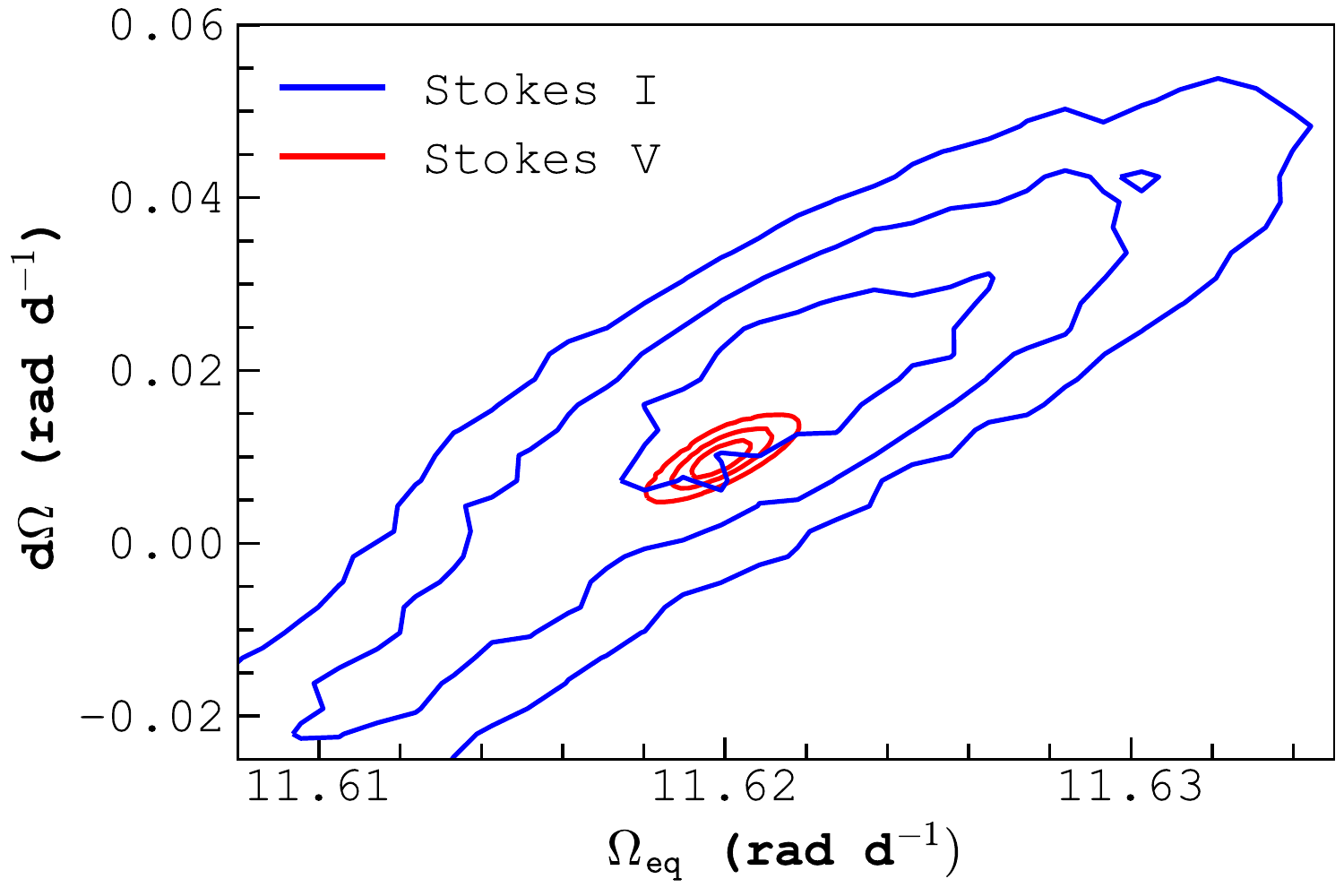}
\caption{Variations of $\chi_{\text{r}}^{2}$ as a function of $\Omega_{\text{eq}}$ and $d\Omega$ for \ta, derived from modelling of our \si (red) and \sv (blue) LSD profiles at a constant information content. For both \si and \sv, a clear and well defined parabola is observed, shown by the 1, 2 and $3\sigma$ ellipses (depicting 68.3, 95.5 and 99.7~\pc confidence levels, respectively), with the $3\sigma$ contour tracing the 5.5~\pc increase in $\chi^{2}_{\text{r}}$ (or equivalently a $\chi^{2}$ increase of 11.8 for 2156 fitted data points). This figure is best viewed in colour.}
\label{fig:twa6difrot}
\end{figure}

Fig.~\ref{fig:twa6difrot} shows the $\chi_{\text{r}}^{2}$ surface we obtain (as a function of $\Omega_{\text{eq}}$ and $d\Omega$) for both Stokes $I$ and $V$, for \ta. We find a clear minimum at $\Omega_{\text{eq}} = 11.6199 \pm 0.0005$~\radd and $d\Omega = 0.0098 \pm 0.0014$~\radd for \si data (corresponding to rotation periods of $0.54073\pm0.00002$~d at the equator and $0.54118\pm0.00002$~d at the poles; see left panel of Figure~\ref{fig:twa6difrot}), with the fits to the \sv data of $\Omega_{\text{eq}} = 11.622\pm 0.004$~\radd and $d\Omega = 0.018 \pm 0.011$~\radd showing consistent estimates, though with larger error bars (right panel of Figure~\ref{fig:twa6difrot}). We note that both these periods are in excellent agreement with those found previously by \citet{skelly2008} and \citet{kiraga2012}.

For \tb, we were able to constrain the rotational period to $4.578\pm0.006$~d (corresponding to $\Omega_{\text{eq}} = 1.3724\pm0.0019$~\radd), in good agreement with the photometric period of 4.638~d found by \citet{kiraga2012}. However, given that the observations span only $\sim 3$ rotation cycles, the recurrence of profile distortions across different latitudes is severely limited, and so we were unable to constrain surface shear. Hence, for our fits with ZDI we have assumed solid body rotation.

\section{Magnetic field strength from individual lines}
\label{sec:specfit}
\begin{figure*}
\centering
\includegraphics[width=0.7\textwidth]{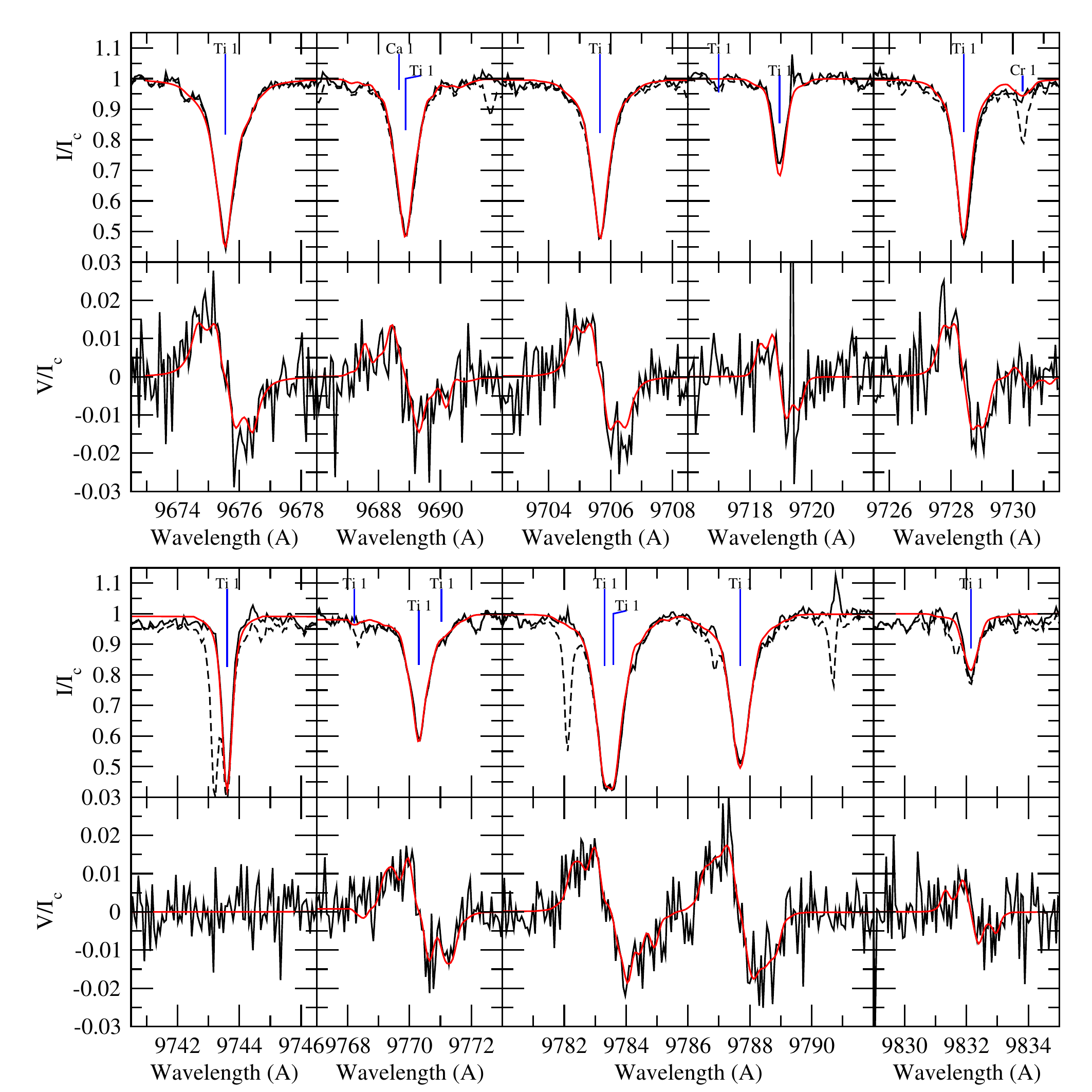}
\caption{Detections of Zeeman broadening in the observation of \tb on March 27. The panels show the \si spectrum at the top and the corresponding \sv spectrum below for the full set of lines used in our fits (see Table~\ref{tab:atomic-data}). Dashed lines show the observation before telluric correction and solid lines show the spectrum after telluric correction. Over-plotted in a red solid line is our best fit using our third model to fit both \si and \sv simultaneously.}
\label{fig:twa8a-IV-fit}
\end{figure*}

\tb has a very strong photospheric magnetic field that can be detected in some individual lines, allowing direct spectral fitting to derive the strength of the magnetic field. As this is not the case for \ta,  it is not included in the following analysis. For \tb, \sv signatures are visible in over 20 lines, mostly redwards of 8000~\AA\ where the S/N is largest. Of particular interest are a set of eleven strong \ion{Ti}{i} lines between 9674 and 9834~\AA\, ten of which are detected in \sv, and one which has a Land\'{e} factor of zero (9743.6~\AA, see Fig.~\ref{fig:twa8a-IV-fit}). These atomic lines have minimal blending from molecular lines, and while there is a some blending from telluric lines, it can be corrected. These lines have the added advantage that all but two of them are from the same multiplet, which mitigates the impact of some systematic errors (e.g., errors in \teff) on our measurements of the magnetic field. A detailed description of these lines is given in Table~\ref{tab:atomic-data}.

\begin{figure}
\includegraphics[width=\columnwidth]{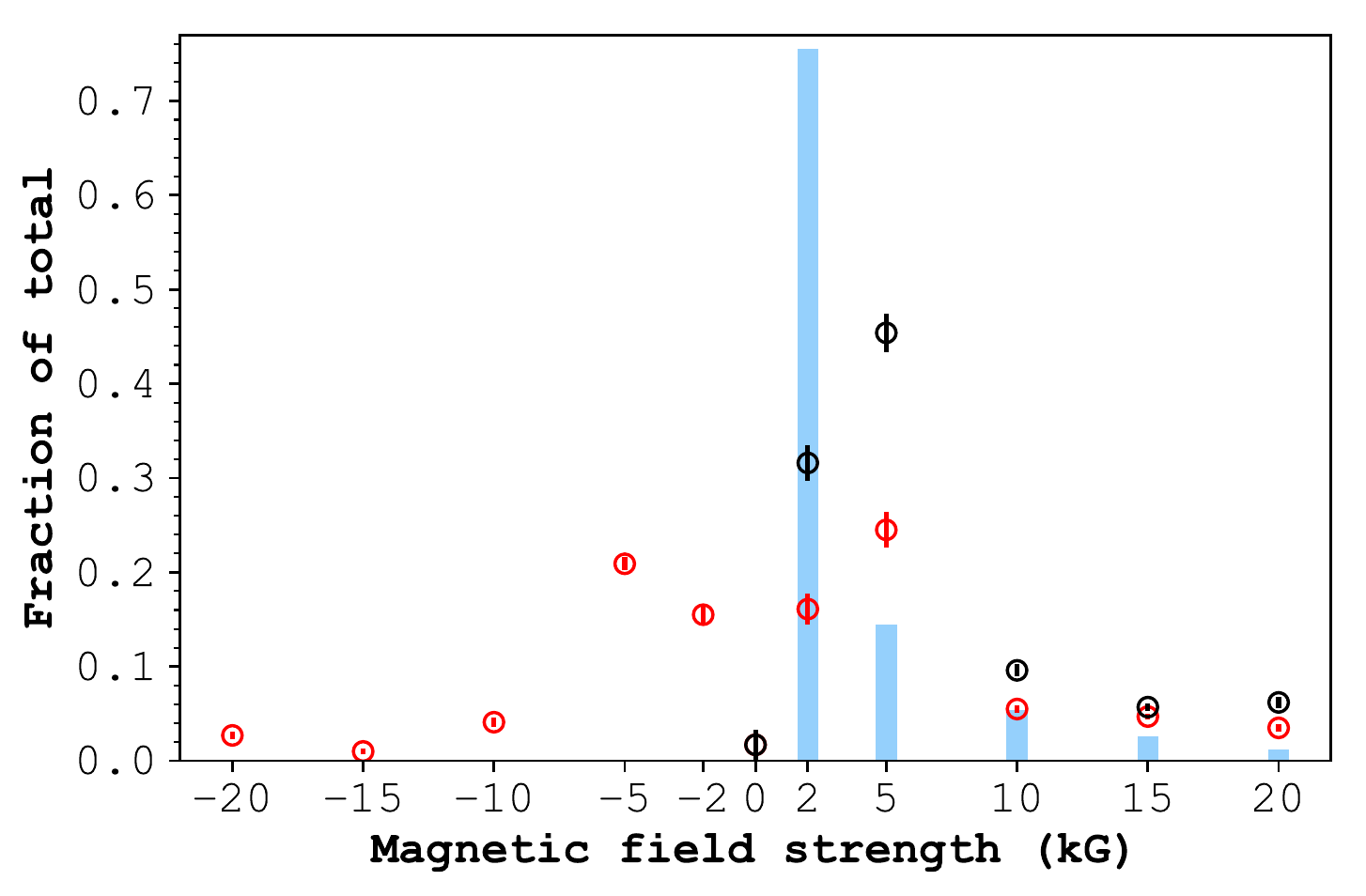}
\includegraphics[width=\columnwidth]{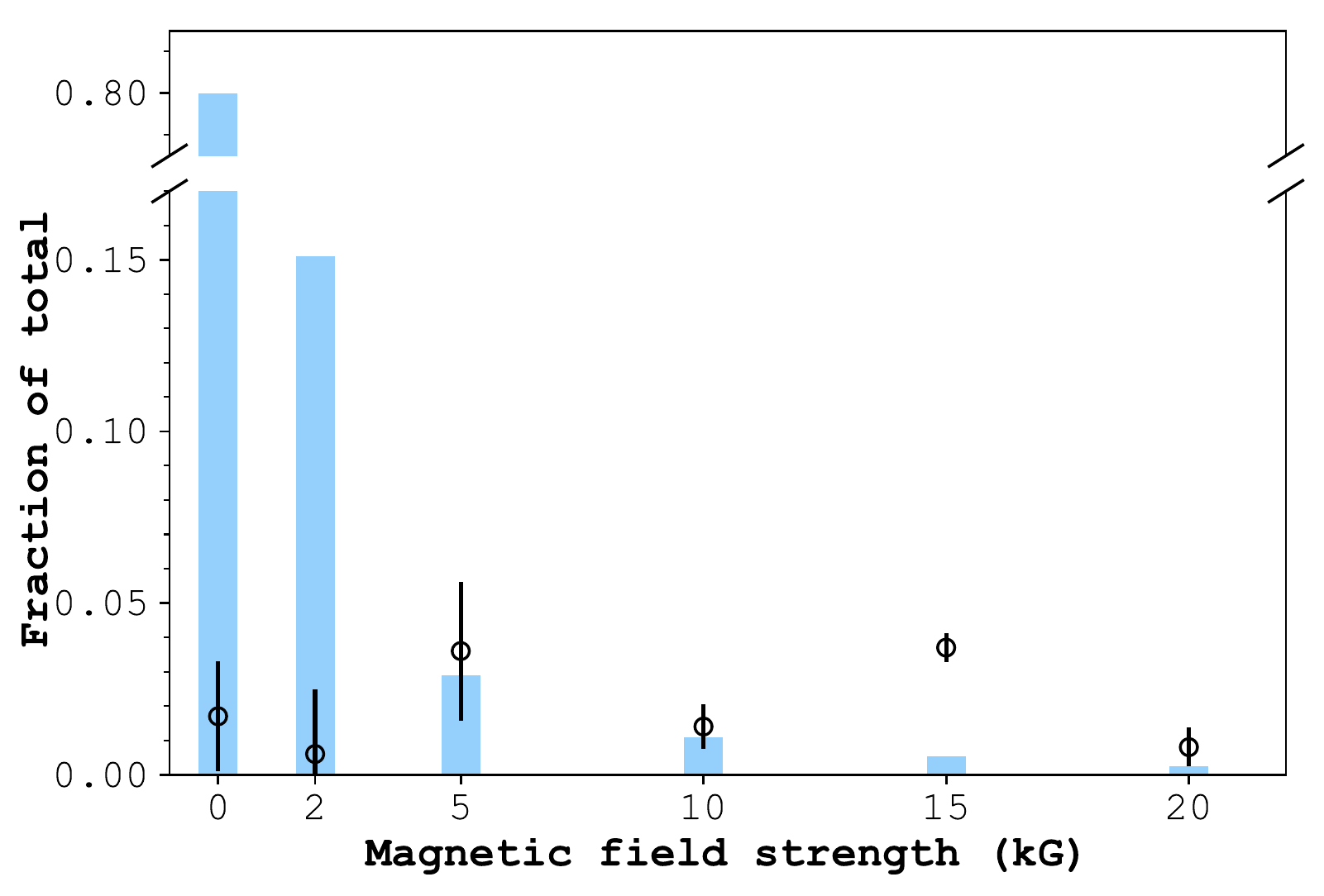}
\caption{The distribution of surface magnetic field strengths for \tb, as determined from ZDI and direct spectral fitting of \ion{Ti}{i} lines. Blue bars show the fraction of the total mapped magnetic field strength from ZDI, for fields of a given bin. 
Top panel: Comparison between the magnetic field strengths determined from fitting \si data. Red circles show the mean filling factors for each field strength using our third model to simultaneously fit \si and $V$ spectra (see Section~\ref{sec:IVspec-fit}). Black circles show the combined filling factors for both the positive and negative fields. Thus, one can directly compare the recovered field strengths for \si data from ZDI and direct spectral fitting by comparing the blue bars and the black circles, respectively. One can see that a significantly larger fraction of higher-strength fields are recovered by direct spectral fitting, as compared to that from ZDI (see discussion in Section~\ref{sec:IVspec-fit}). 
Bottom panel: Comparison between the magnetic field strengths determined from fitting \sv data. Black circles show the resulting filling factors after subtracting the contributions of the negative fields from those of the positive fields. As \sv profiles are sensitive to the sign of the line-of-sight component of $\mathbf{B}$, significant cancellation of fields may occur, and so we must compare our fits with ZDI to \sv profiles, to these black circles. In this case, we see that ZDI recovers a similar fraction of field strengths for the 5 and 10~kG bins, but significantly more for the 2~kG bin, and significantly less for the 15 and 20~kG bins (see discussion in Section~\ref{sec:IVspec-fit}).}
\label{fig:maghist}
\end{figure}

\subsection{Telluric correction}
Before a detailed analysis of the \si spectra may be carried out, we must first correct for the large number of telluric water lines present between 9670--9840~\AA. Telluric lines are not expected to produce circular polarisation, and we see no indication of them in \sv, hence we conclude that their impact on the \sv spectrum is negligible.

As we did not expect to detect magnetic fields in individual telluric blended lines, we did not observe a hot star for telluric calibration. Fortuitously, on some nights, other programs with ESPaDOnS at the CFHT observed the hot stars HD 63401 (PI J.D.\ Landstreet) and HD 121743 (PI G.A.\ Wade).  HD 63401 is a 13500~K Bp star \citep[e.g.,][]{bailey2014} and HD 121743 is a 21000~K B star \citep[e.g.,][]{alecian2014}, with both stars having virtually no photospheric lines in the wavelength range of interest, apart from Paschen lines. Our observations of \tb on the nights of March 25 to April 1, as well as April 5 and 6, had suitable telluric reference observations that were sufficiently close in time and obtained under sufficiently similar conditions.   

The telluric reference spectra were first continuum normalised by fitting low order polynomials through carefully selected continuum regions, then dividing by those polynomials, independently for each spectral order. The telluric reference spectra were then scaled in the form $I^{a}$, where $I$ is the continuum normalised spectrum and $a$ the scaling factor. The scaling factor $a$ and the radial velocity shift for the telluric lines were determined by fitting the modified reference spectrum to telluric lines of the science spectrum through $\chi^2$ minimisation.  Telluric lines around the photospheric lines of interest ($\sim$9650--9850~\AA) were included, as well as some telluric lines in the range 9300--9500~\AA\ where there are fewer strong photospheric lines. The science spectrum was then divided by the scaled shifted telluric spectrum. A example spectrum before and after telluric correction is shown in Fig.~\ref{fig:twa8a-IV-fit}.  

\subsection{Spectrum synthesis}
To constrain the strength of the photospheric magnetic field, we have modelled individual lines in the \si and \sv spectra of \tb. Furthermore, as one of the \ion{Ti}{i} lines has a Land\'e factor of zero, and is narrower in \si as compared to the other \ion{Ti}{i} lines, the magnetic field can also be strongly constrained by the \si spectrum. 

To generate synthetic spectra, we used the {\sc Zeeman} spectrum synthesis program \citep{landstreet1998, wade2001, folsom2012}. This program includes the Zeeman effect and performs polarized radiative transfer in Stokes $IQUV$. The code uses plane-parallel model atmospheres and assumes LTE, and produces disk-integrated spectra.  {\sc Zeeman} includes quadratic Stark, radiative, and van der Waals broadening, as well as optional microturbulence (\vmic) and radial-tangential macroturbulence. A limitation of the code for use in very cool stars is that it does not include molecular lines, or calculations of molecular reactions in the abundances for atomic species.  The \ion{Ti}{i} lines in the 9674--9834 \AA\ region are blended with a few very weak molecular lines, and so {\sc Zeeman} can produce accurate spectra for this region, however most of the spectral region bluewards of this is problematic.

For input to the code we used MARCS model atmospheres \citep{gustafsson2008} and atomic data taken from VALD \citep{ryabchikova2015} (see Table~\ref{tab:atomic-data} for the properties of the atomic lines). VALD data for these particular \ion{Ti}{i} lines were also used by \citet{kochukhov2017} for a similar analysis, and were deemed reliable. Additionally, we can reproduce these Ti lines with near-solar abundances, implying that the oscillator strengths are likely close to correct.

To model the magnetic field of \tb, we adopted a uniform radial magnetic field. While this is an unrealistically simple magnetic geometry, the ZDI analysis found the magnetic geometry to be more complex than a simple dipole. Therefore we leave the geometric analysis to ZDI and adopt the simplest possible geometry here to avoid additional weakly constrained geometric parameters. Furthermore, since this analysis is applied to individual observations, a full magnetic geometry cannot be reliably derived. The model we implement here includes a combination of magnetic field strengths $B$, each with their own filling factor $f$, with the sum of the filling factors (including a region of zero field) equal to unity.

We fit synthetic spectra using a Levenberg-Marquardt $\chi^2$ minimisation routine \citep[similar to][]{folsom2012, folsom2016}, with the radial magnetic field strengths and filling factors as optional additional free parameters. The code was updated to allow fitting observed Stokes $I$ spectra, $V$ spectra, or $I$ and $V$ simultaneously, with wavelength ranges carefully set around the lines of interest. In order to place uncertainties on the fitting parameters, we use the square root of the diagonal of the covariance matrix, as is commonly done.  This is then scaled by the square root of the reduced $\chi^2$, to very approximately account for systematic errors. These formal uncertainties may still be underestimates, and a further consideration of uncertainties is discussed in Sect.~\ref{sec:IVspec-fit}.  

\subsection{Fitting the Stokes $I$ spectrum}
Our initial fits were carried out with the observation on March 27 since the \sv LSD profile for this night has one of the simplest shapes, indicating a more uniform magnetic field in the visible hemisphere.

Measurements of magnetic fields in \si spectra are constrained by both the width and the desaturation of lines with different Land\'e factors. Fitting the \si spectrum to determine magnetic field strengths requires constraints on several other stellar parameters which influence line width and depth. Here, we adopt the \teff and \logg values derived in Section~\ref{sec:prop} (see Table~\ref{tab:syspars}). Since our choice of lines is dominated by one multiplet, adopting these values is a small source of uncertainty. We note that these lines are not well adapted to constraining \teff and \logg spectroscopically. We include \vsini and \vmic as free parameters in the fit, since they can play an important role in line shape and strength, and can only be determined spectroscopically. \vmic is constrained by desaturation of strong (on the curve of growth) lines and, given the lack of weak lines in our spectral range, is determined with only a modest accuracy by different degrees of desaturation of different strong lines. Macroturbulence is assumed to be zero, since it is likely much smaller than the \vsini\ of $\sim$5~\kms. Ti abundance is included as a free parameter, however, we caution the reader that this may not provide reliable results, as the code neglects the fraction of Ti bound in molecules. Nevertheless, this free parameter is necessary to avoid the code fitting line strength entirely by varying magnetic field and \vmic. 

When fitting the spectra of \tb we adopted three main models, each of increasing complexity, to better constrain the nature of the magnetic field. These three models (described below) are used to fit \si spectra only, \sv only, and both \si and $V$ simultaneously.

Our first model consists of fitting the \si spectrum using just one magnetic region with a corresponding filling factor, yielding a best-fit magnetic field strength of $B_1 = 5.65\pm0.10$~kG with a filling factor $f_1 = 0.597\pm0.016$, but at a reduced $\chi^2$ of 19.6. Fits with $f$ fixed to 1 consistently fail to reproduce the line shape, with a core that is far too wide and with wings that are too narrow, implying that only a fraction of the star is covered by very strong magnetic fields. 

Our second model increases the number of free parameters by including two magnetic regions and filling factors, achieving a visibly much better fit with a reduced $\chi^2$ of 12.9, and with field strengths of $B_1 = 4.71\pm0.08$~kG with $f_1 = 0.648 \pm 0.015$, and $B_2 = 15.61 \pm 0.25$~kG with $f_2 = 0.133 \pm 0.007$. This second model does a better job of simultaneously reproducing the narrow core and broad wings of the magnetically sensitive lines, although the high field strength region produces a sharper change in the shape of the wings than seen in the observation, implying that the star has a more continuous distribution of magnetic field strengths than our model.

Our third model again increases the number of free parameters to improve the fit. However, rather than add additional sets of magnetic field strengths and filling factors, which may become more poorly conditioned or not converge well, we instead adopt a grid of fixed magnetic field strengths with filling factors as free parameters (in a similar way to e.g. \citealt{johnskrull1999,johnskrull2004}). This provides an approximate distribution of magnetic field strengths on the visible hemisphere of the star. Using our third model for fitting \si only, we use bins of 0, 2, 5, 10, 15 and 20~kG. Bins of $\sim5$~kG allow for smooth model line profiles, and so smaller bins (that would be less well constrained) are not necessary. Adding bins above 20~kG improves the $\chi^{2}$ fit by a small but formally significant amount. However, the impact on the synthetic line is small and only affects the far wings of the line in \si. Small changes in the far wings of the line are most vulnerable to systematic errors, such as weak lines that are not accounted for, errors in the telluric correction, errors in continuum normalisation, or very weak fringing, all of which could approach the strength of the line this far into the wing. Thus we limit the magnetic field to 20 kG, and caution that even for this bin the filling factor may be overestimated. The resulting best fit parameters for \si only for March 27 using our third model is presented in Table~\ref{tab:specfit}, with a reduced $\chi^2$ of 10.6.

\citet{yang2008} studied \tb and derived some magnetic quantities based on \si observations in the IR. They adopted literature values for the stellar parameters of $\teff = 3400$~K, $\logg = 4.0$ and $\vsini = 4.0$\kms.  Their ``Model 1'' corresponds to our first model with one filling factor and magnetic field strength. They report only the product of their filling factor and magnetic field strength as 2.3 kG, which is close to our value for March 27 of $3.37 \pm 0.11$~kG, although not within uncertainty.  Their ``Model 2'' corresponds to our second model with two filling factors and magnetic field strengths. They report the quantity $\langle |B_f| \rangle = 2.7$~kG, which is comparable but again not consistent with our value of $5.13 \pm 0.14$~kG. The ``Model 3'' of \citet{yang2008} is closest to our third model with a grid of filling factors, although they only fit filling factors for field strengths of 2, 4 and 6 kG. They report $\langle |B_f| \rangle = \sum_i B_i f_i$ of 3.3~kG.  The equivalent value from our fit is $\langle |B_f| \rangle = 5.90 \pm 0.44$~kG, which is again inconsistent. We note that, if we perform our fit using the three bins of 2, 4, and 6 kG used by \citet{yang2008}, we find $\langle |B_f| \rangle = 3.96 \pm 0.38$ kG. While this is much closer to their ``Model 3'' results, we find that the fit to our data is much worse in the wings of the lines, so we consider this model to be less accurate for our spectra. The IR spectra of \citet{yang2008} had a much lower S/N than our observations, and so the wings of the lines may not have been detected as clearly as in our spectra. Indeed, the very strong magnetic field with a very small filling factor necessary to fit the wings of our magnetically sensitive lines is likely the cause of the difference between our results, as well as intrinsic variability of the field.

\subsection{Fitting the Stokes $V$ spectrum}
In order to fit the \sv spectrum we adopt the best fit \vsini, \vmic and Ti abundance from fitting \si with our third model, since these parameters cannot be well constrained from $V$ spectra (see Table \ref{tab:specfit}).  

When directly fitting the \sv spectrum, it becomes immediately apparent that a filling factor (much less than unity) is necessary. To produce \sv profiles with the widths of the observed lines, a very strong magnetic field is necessary. However, to reproduce the amplitudes of the \sv profiles, a weaker field is necessary, or a very strong field covering a small portion of the star.  This can be easily seen by comparing the widths of the observed \si and $V$ profiles (see Fig.~\ref{fig:lsdprofiles}) and noting that the $V$ profiles remain stronger in the far wings compared to the $I$ profiles.  

Fitting the \sv profiles with our first model yields a best fit of $B_1 = 7.09 \pm 0.19$~kG and $f_1 = 0.081 \pm 0.004$, with a reduced $\chi^2$ of 2.27. However, this provides a poor fit to the line profiles, in particular the outer and inner parts of the line cannot be well fit simultaneously. We find a much better fit when using our second model, with a reduced $\chi^2$ of 1.58, and field strengths and filling factors of $B_1 = 4.70\pm0.19$~kG with $f_1 = 0.078 \pm 0.004$, and $B_2 = 14.94\pm 0.26$~kG with $f_2 = 0.051 \pm 0.003$, implying $\langle |B_f| \rangle = 1.13 \pm 0.05$~kG.  The filling factors and $\langle |B_f| \rangle$ derived here are much smaller than those derived from \si. \sv is sensitive to the sign of the line-of-sight component of $\mathbf{B}$, while \si is sensitive to the magnitude of $\mathbf{B}$. The difference in filling factors is likely due to cancelation in $V$ of nearby regions with opposite sign.

We also fit the \sv spectra with our third model, where our use of positive fields is still appropriate as the disc integrated field is positive for March 27, and indeed at all other phases. Our fit yields a reduced $\chi^2$ of 1.56, where the parameters are summarised in Table~\ref{tab:specfit}. The improvement in the fit using our third model is modest compared to the first and second models, but it is clearly better visually, with a formally significant improvement of nearly $3\sigma$. We note that the distribution of filling factors is quite different from that of the \si fit, with most of the surface having no magnetic field detected in \sv, and the remaining field lying more in the 5 and 15 kG bins.

Using our fits to approximate the longitudinal magnetic field ($B_l$), we have taken the line of sight component of the model magnetic field, averaged over the stellar disc and weighted by the brightness of the continuum, i.e.
\begin{equation}
B_{l,\rm{syn}} = \sum_i  \frac {\int  I_c f_i B_i \cos(\theta) {\rm d}\Omega}{\int I_c {\rm d}\Omega}
\label{eqn:Blsyn}
\end{equation}
\noindent where $f_i$ is the filling factor for component $i$, $B_i$ is the purely radial magnetic field for that component, $\theta$ is the angle between the line of sight and the radial field. $I_c$ is the continuum brightness at for a point on the disc (accounting for limb darkening), and the integral of d$\Omega$ is over the visible disk.

From Eqn.~\ref{eqn:Blsyn} we derive $B_{l,\rm{syn}} = 0.78\pm0.04$~kG and $0.82\pm0.08$~kG for our second and third models, respectively. These values agree to within their uncertainties, and are comparable to (but roughly 1.7 times larger than) the actual observed $B_l$ values for this phase, as calculated from the LSD profiles (see Fig.~\ref{fig:bl}). Indeed, if we calculate an observed $B_l$ from just the \ion{Ti}{i} 9705.66~\AA\ line (using the telluric corrected spectrum), rather than an LSD profile, we find $0.88\pm0.13$~kG for March 27. Moreover, the behaviour of this \ion{Ti}{i} line with rotational phase is consistent with the LSD profile, except that it shows a higher field strength. This implies that the signal in the \sv LSD profiles may not be adding perfectly coherently, producing a lower amplitude $V$ profile. This is not surprising as, due to the very large field strength, Zeeman splitting patterns of individual lines begin to matter for the line profile shapes. Thus, simply scaling amplitudes by effective Land\'e factors is a less effective approximation for such strong fields.

\subsection{Simultaneous fitting of Stokes $I$ and $V$}
\label{sec:IVspec-fit}
As we detect magnetic fields in both \si and $V$ observations, our model should be able to reproduce these signatures simultaneously. This requires us to allow a combination of positive and negative magnetic fields, resulting in a cancellation of much of the signal in \sv while allowing for a large unsigned magnetic flux in \si. This is evident from the much smaller filling factor in our fits of \sv compared to our fits to \si.

Firstly, we performed simultaneous fits to \si and $V$ using a simple model with three magnetic regions - two with positive fields and one with a negative field. A model with one positive field and one negative field is insufficient to reproduce the shapes of the \si or $V$ line profiles. For this simple model, the best fit magnetic parameters are $B_{1} = +4.76 \pm 0.07$~kG with $f_{1} = 0.360 \pm 0.007$, $B_{2} = -5.05\pm0.09$~kG with $f_{2} = 0.282 \pm 0.007$, and $B_{3} = +15.92\pm 0.20$~kG with $f_{3} = 0.098\pm0.004$ (with $\vsini = 5.26\pm0.17$~\kms, $\vmic = 1.00\pm0.06$~\kms and [Ti/H]$ = -6.947 \pm 0.013$). This fit gives a reduced $\chi^2$ of 7.94, and fits the $I$ spectrum similarly well to our best model from fitting \si only (see above), although it is too strong in the wings of $V$, implying that there should be additional cancellation. This model implies a total $\langle |B_f| \rangle$ of 4.70~kG, and a synthetic $B_{l,\rm{syn}}$ (allowing for cancellation) of 1.28~kG, although (as noted) this is likely too large.

Using our third model (with a grid of magnetic field strengths and filling factors, see above), we again require both negative and positive magnetic fields. As with fitting only \si or \sv, we use bins of 0~G, $\pm2$~kG, $\pm5$~kG, $\pm10$~kG, $\pm15$~kG, and $\pm20$~kG, for a total of 11 bins. The results of our fit with this model, with 11 filling factors as well as \vsini, \vmic and [Ti/H], are presented in Table~\ref{tab:specfitall}, with a reduced $\chi^2$ of 6.33 - clearly an improvement over the simple three magnetic-region model. Our fit to the observation taken on March 27 is shown in Fig.~\ref{fig:twa8a-IV-fit}, showing a good fit to both \si and $V$ spectra, including matching the width of the magnetically-insensitive line with a Land\'e factor of zero.

A summation of the filling factors for bins with the same $|B|$ yields a very similar distribution to that for the fit to \si only, with differences much smaller than the formal uncertainties. This can be understood as \si is sensitive to the total magnetic field strength but not the orientation of the magnetic field. Similarly, the difference between filling factors for bins with the same $|B|$ but opposite sign produces a distribution very similar to that of the fit to \sv only.  This can also be understood since \sv is sensitive to the line-of-sight component of the magnetic field only, with the spatially unresolved (within the same model pixel) components of opposite orientation cancelling out. For our observation on March 27, we find a total $\langle |B_f| \rangle = 5.71\pm0.22$~kG and $B_{l,\rm{syn}} = 0.78\pm0.15$~kG. This $\langle |B_f| \rangle$ is consistent with our fit of only \si with our third model, and $B_{l,\rm{syn}}$ is consistent with our fit of only \sv.

Over the rotation of \tb, this set of results shows $B_{l,\rm{syn}}$ to range from $640 \pm 150$ to $840 \pm 140$~G, with $\langle |B_f| \rangle$ ranging from $5.71\pm 0.22$ to $6.36\pm 0.22$~kG, and varying coherently with rotation phase.

Given the high S/N of our observations, the results we present here may be limited by systematic errors, and our uncertainties may be underestimated. To investigate the impact of uncertainties in \teff and \logg, we re-fit the observation on March 27 with these two parameters changed by $\pm1\sigma$. The change in \teff produces at most a change of $0.5\sigma$ in the other parameters, and often smaller changes than that, and so we conclude that the uncertainty on \teff has a minor contribution to the total uncertainty. Changing \logg by $1\sigma$ has a large impact on \vsini and [Ti/H] (4--5$\sigma$) and on \vmic ($2\sigma$), although it has a much smaller impact on the magnetic filling factors of only $\sim 1\sigma$, rising to $3\sigma$ for the 2~kG and 5~kG bins when \logg is decreased by $1\sigma$. In that case, the filling factor shifts from the 2~kG bin into the 0 and 5~kG bins, underscoring the uncertainty of the 2~kG bin. The relatively large uncertainty in \logg changes the line broadening, but does so independently of Land\'e factor, and so \vsini and \vmic are more sensitive to \logg than filling factors. It is possible that our \vmic is an over-estimate, since typical \vmic values for PMS M-dwarfs are not well known. To estimate an upper limit on this uncertainty, we re-ran the fit with $\vmic = 0$, finding that the best fitting \vsini decreases by 1~\kms, that [Ti/H] increases by 0.1~dex, and that filling factors generally change by less than $1\sigma$ (except for the 10~kG bin which decreases by $2\sigma$). From these tests we conclude that our formal uncertainties may be underestimated by a factor $\lesssim2$, mostly due to the large uncertainty in \logg and the (potentially) larger systematic errors on the filling factors for the 2 and 20~kG bins.

Having established an analysis method for the observation of March 27 using our third model to fit both \si and $V$, we performed this analysis on all observations for which we could perform reliable telluric correction, providing us with ten sets of results, shown in Table~\ref{tab:specfitall}. Taking an average over all 10 observations, we find a mean magnetic field strength of $\langle |B_f| \rangle = 6.0\pm0.5$~kG, where the amount of magnetic energy in each bin is shown in Table~\ref{tab:specfit}. The standard deviation of these results is close to the mean uncertainty for all parameters, suggesting that our formal uncertainties account well for random errors, with the larger standard deviation likely due to the rotational modulation.

In Fig.~\ref{fig:maghist} we compare the magnetic field strength distribution on \tb as determined by our ZDI maps in Section~\ref{sec:magfield}, to our direct spectral fitting here. As our ZDI map has a continuous distribution of field strengths, we have created histograms using the same bins as that for the direct spectral fitting, allowing for a direct comparison of recovered field strengths. 
For \si, we find that 75~\pc of the field strength recovered by ZDI is in the 2~kG bin, with a 15~\pc in the 5~kG bin, and 9~\pc at higher field strengths. In comparison, direct spectral fitting yields 32~\pc of the magnetic field to be 2~kG, with almost 46~\pc in the 5~kG bin, and with 22~\pc of fields in the 10, 15 and 20~kG bins. For \sv, the line profiles are sensitive to the sign of the line-of-sight component of $\mathbf{B}$, and so there is likely significant cancellation of fields of opposite polarity. Hence, our fits to \sv LSD profiles with ZDI recover only the uncanceled magnetic fields. Therefore, for comparison to direct spectra fitting, we must subtract the filling factors determined for the negative fields from the positive fields, yielding the fraction of uncanceled fields that could be fit with ZDI. For ZDI we find that 80~\pc of the surface has a 0~G field, with 15~\pc of the field in the 2~kG bin, 3~\pc in the 5~kG bin, 1~\pc in the 10~kG bin, and with 1~\pc at higher field strengths. In comparison, for direct spectral fitting we find that less than 1~\pc of the field is 2~kG, with 3.6~\pc of the field at 5~kG, 1.4~\pc at 10~kG, and with 4.5~\pc at higher field strengths. Thus, with ZDI we recover most of the magnetic flux up to 10~kG, but are not as sensitive to fields higher than this. Moreover, our results demonstrate that we underestimate the fraction of high field strengths using the ZDI technique with LSD profiles \sv spectra. As mentioned previously, this may be due to the signal in the \sv LSD profiles not adding perfectly coherently, as variations in line splitting patterns cause variations in line shapes, and so scaling amplitudes by effective Land\'e factors is less accurate. Moreover, there may be significant cancellation in \sv profiles as it is sensitive to the sign of the line-of-sight component of $\mathbf{B}$. The recovery of small-scale, high-field-strength features would likely be improved if linear polarization spectra (Stokes $Q$ and $U$) were included in the ZDI modelling, and would likely increase the recovered total magnetic field energy \citep[see][]{rosen2015}.

\begin{table}
\setlength\tabcolsep{5pt} % default value: 6pt
\centering
\caption{Best fit parameters from direct spectral fitting of \tb. The first and second columns respectively give the results of fitting Stokes $I$ and $V$ separately (using our third model) for the spectrum taken on 27 Mar 2015. Parameters with no error bars for the $V$ fit were held fixed. The third column shows the results of fitting Stokes $I$ and $V$ simultaneously, where we present the mean over the 10 nights that could be reliably telluric-corrected, with error bars given as the standard deviations. Values for fits to individual nights are presented in Table~\ref{tab:specfitall}.}
\begin{tabular}{cccc}
\hline
&	Stokes $I$ only	&	Stokes $V$ only	&	Stokes $I$ and $V$ \\
&	27 Mar 2015		&	27 Mar 2015 		& 	mean \\
\hline                                							
\vsini (\kms)	 & 	$4.77  \pm 0.23$ 	 & 	4.77	 & 	$4.82 \pm 0.16$	\\
\vmic (\kms)	 & 	$1.15  \pm 0.08$	 & 	1.15	 & 	$1.08 \pm 0.05$	\\
$[$Ti/H$]$	 & 	$-7.006\pm 0.017$	 & 	-7.01	 & 	$-6.976\pm 0.022$ 	\\
0 G	 & 	$0.002 \pm 0.084$	 & 	$0.859 \pm 0.014$	 & 	$0.017 \pm 0.016$	\\
+2 kG	 & 	$0.312 \pm 0.067$	 & 	$0.023 \pm 0.010$	 & 	$0.161 \pm 0.016$	\\
+5 kG	 & 	$0.483 \pm 0.044$	 & 	$0.057 \pm 0.007$	 & 	$0.245 \pm 0.019$	\\
+10 kG	 & 	$0.090 \pm 0.020$	 & 	$0.016 \pm 0.005$	 & 	$0.055 \pm 0.004$	\\
+15 kG	 & 	$0.060 \pm 0.013$	 & 	$0.041 \pm 0.005$	 & 	$0.047 \pm 0.003$	\\
+20 kG	 & 	$0.053 \pm 0.011$	 & 	$0.004 \pm 0.004$	 & 	$0.035 \pm 0.004$	\\
-2 kG	 & 	-	 & 	-	 & 	$0.155 \pm 0.010$	\\
-5 kG	 & 	-	 & 	-	 & 	$0.209 \pm 0.007$	\\
-10 kG	 & 	-	 & 	-	 & 	$0.041 \pm 0.005$	\\
-15 kG	 & 	-	 & 	-	 & 	$0.010 \pm 0.003$	\\
-20 kG	 & 	-	 & 	-	 & 	$0.027 \pm 0.004$	\\
$\langle |B_f| \rangle$ (kG)	&	$5.9 \pm 1.0$	 & 	$1.2\pm0.3$	 & 	$6.0\pm0.5$	\\
\hline
\end{tabular} 
\label{tab:specfit} 
\end{table}

\section{Filtering the activity jitter}
\label{sec:rv}
As well as characterizing magnetic fields of wTTSs, the MaTYSSE program also aims to detect close-in giant planets (called hot Jupiters, hJs) to test planetary formation and migration mechanisms. In particular, characterizing the number and position of hJs will allow us to quantiatively assess the likelihood of the disc migration scenario, where giant planets form in the outer accretion disc and then migrate inward until they reach the central magnetospheric gaps of cTTSs (see e.g., \citealt{lin1996,romanova2006}). Given that we map the surface brightness of the host star, we are able to use our fits to the observed data to filter out the activity-related jitter from the RV curves (where the RV is measured as the first-order moment of the LSD profile; see \citealt{donati2014,donati2015}). After subtraction of the RV jitter, we may look for periodic signals in the RV residuals to reveal the presence of hJs. Indeed, this method has so far yielded two detections of hJs in the MaTYSSE sample, around both V830~Tau \citep{donati2015,donati2016a,donati2017} and TAP~26 \citep{yu2017}.

\begin{figure*}
\includegraphics[width=\textwidth]{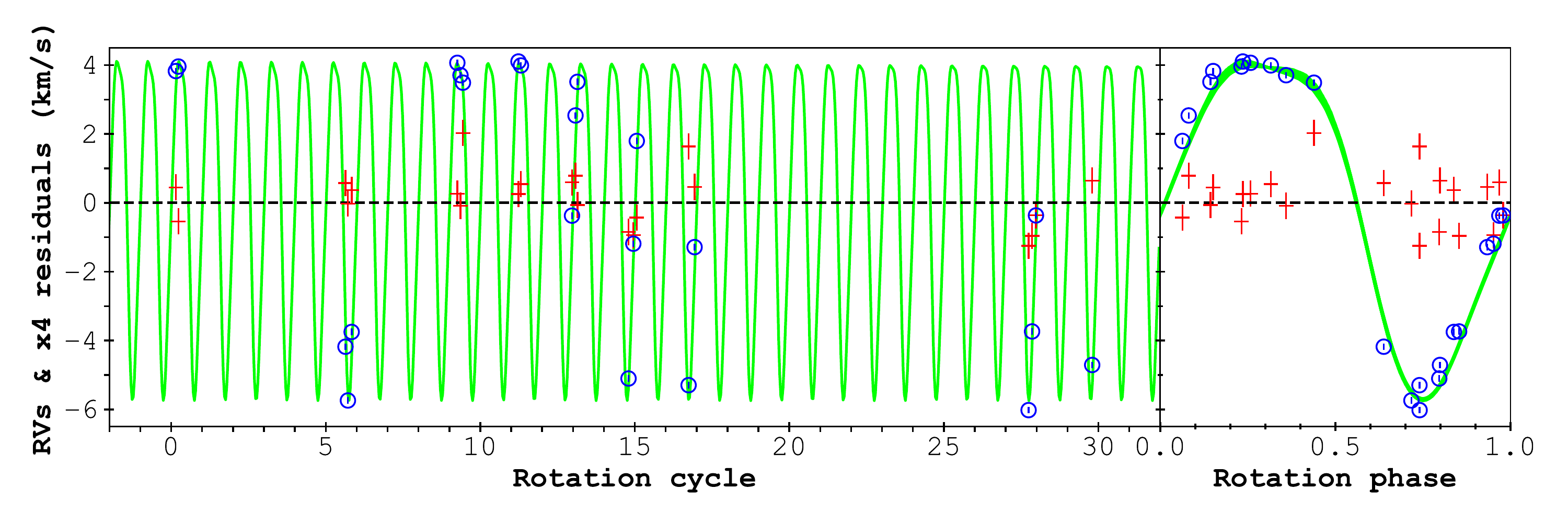}
\includegraphics[width=0.8\textwidth]{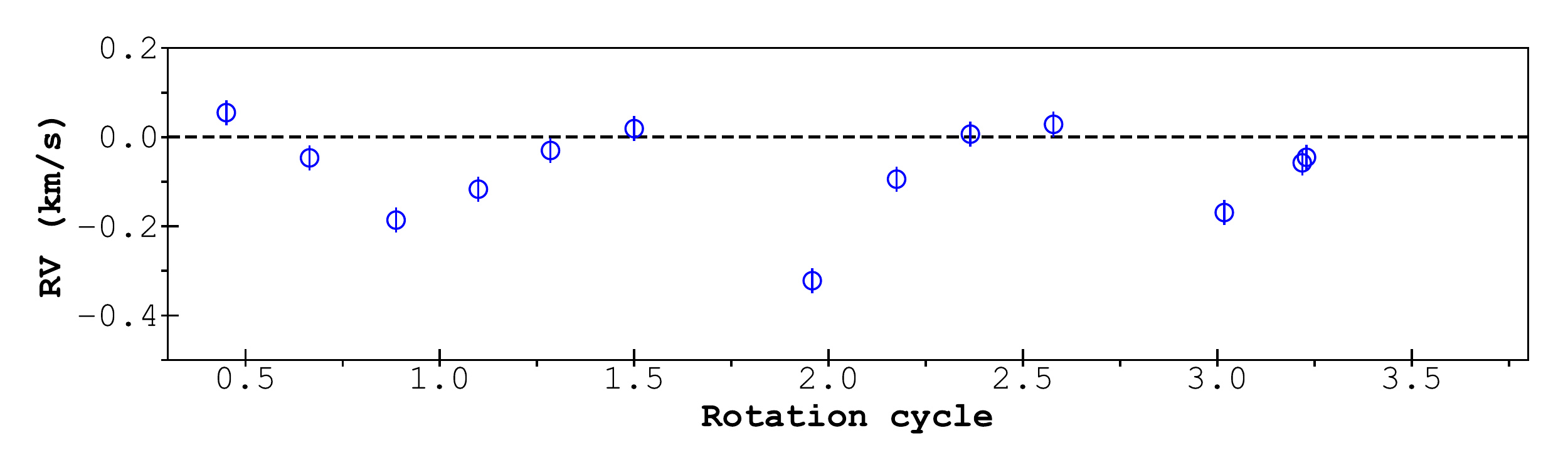}
\caption{Top left panel: RV variations (in the stellar rest frame) of \ta a function of rotation phase, as measured from our observations (open blue circles) and predicted by the tomographic brightness map of Figure~\ref{fig:brightmap} (green line). RV residuals are also shown (red crosses, with values and error bars scaled by a factor of 4 for clarity), and exhibit a rms dispersion equal to 0.20~\kms. RVs are estimated as the first order moment of the \si LSD profiles rather than through Gaussian fits, due to their asymmetric and often irregular shape. Top right panel: The same as the top left panel after phase-folding the data and model. Note that the model shows little variation over the $\sim30$ rotation cycles, showing the very low level of differential rotation. Bottom panel: The measured RVs of \tb as a function of rotation phase. Note that the filtered RVs are not shown for \tb as the line profiles are compatible with a star of uniform brightness. The unfiltered RVs show a period signal that is equal to the stellar rotation period. This figure is best viewed in colour.}
\label{fig:rv}
\end{figure*}

\begin{figure}
\includegraphics[width=\columnwidth]{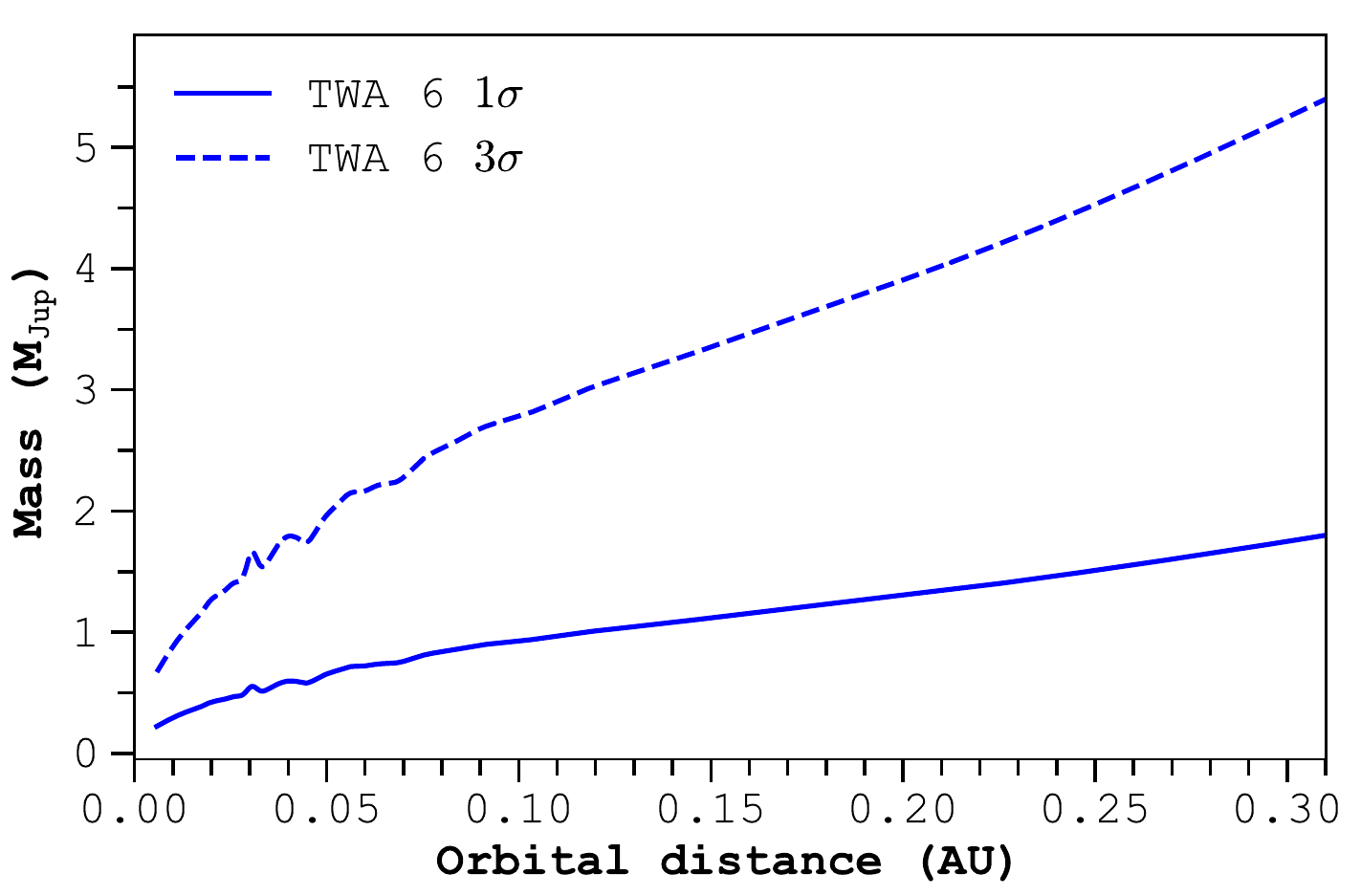} \\
\caption{The $1\sigma$ and $3\sigma$ upper limits (solid and dashed lines, respectively) on the recovered planet mass as a function of orbital distance, using the RVs shown in Fig.~\ref{fig:rv} for \ta. This figure is best viewed in colour.}
\label{fig:planetmass}
\end{figure}

For \ta, the unfiltered RVs have an rms dispersion of 3.8~\kms. The predicted RV due to stellar activity and the filtered RVs are shown in Fig.~\ref{fig:rv}. We find that RV residuals exhibit an rms dispersion of $\sim0.20$~\kms, with a maximum amplitude of 0.51~\kms. This is well above the intrinsic RV precision of ESPaDOnS (around 0.03~\kms, e.g. \citealt{moutou2007,donati2008a}), however, given the high \vsini, the accuracy of the filtering process is somewhat reduced, with an intrinsic uncertainty of around 0.1~\kms. Indeed, we find no significant peaks in a periodogram analysis, and so we find that \ta is unlikely to host a hJ with an orbital period in the range of what we can detect (i.e. not too close to the stellar rotation period or its first harmonics; see \citealt{donati2014}). We find a $3\sigma$ error bar on the semi-amplitude of the RV residuals equal to 0.19~\kms, translating into a planet mass of $\simeq3.1$~\mjup orbiting at $\simeq0.1$~au (assuming a circular orbit in the equatorial plane of the star; see Figure~\ref{fig:planetmass}).

For \tb, the unfiltered RVs have an rms dispersion of 0.13~\kms. Given that the surface brightness of \tb is compatible with that of a homogeneous star, we were unable to filter the RVs in the same manner. However, the measured RVs (shown in Fig.~\ref{fig:rv}) do display a clear periodic signal that is equal to the stellar rotation period, implying that there are starspots on the surface, even though the modulation of the line profiles is minimal.

\section{Summary and discussion}
\label{sec:discussion}

\begin{figure*}
\begin{center}
\includegraphics[width=\textwidth]{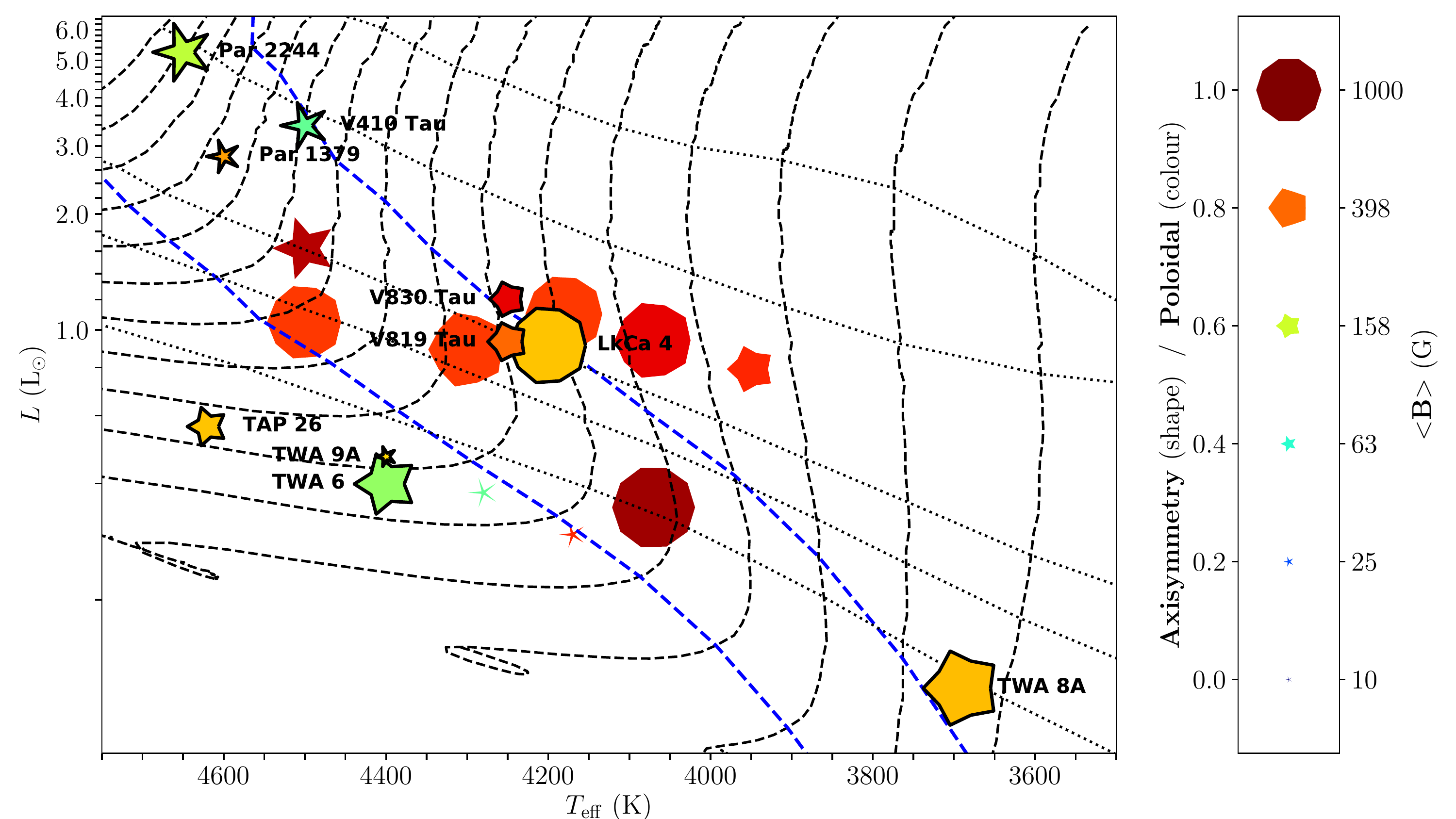} 
\caption{H-R diagram showing the MaTYSSE wTTSs (black line border and labelled) and the MaPP cTTSs (no border). The size of the symbols represents the surface-averaged magnetic field strength (with a larger symbol meaning a stronger field), the colour of the symbol represents the fraction of the field that is poloidal (with red being completely poloidal), and the shape of the symbols represents the axisymmetry of the poloidal field component (with higher axisymmetry shown as a more circular symbol). Also shown are evolutionary tracks from \protect \cite{siess2000} (black dashed lines, ranging from 0.3--1.9\msun), with corresponding isochrones (black dotted lines, for ages of 0.5, 1, 3, 5 \& 10 Myr), and lines showing 100\% and 50\% convective interior by radius (blue dashed).}
\label{fig:confuse}
\end{center}
\end{figure*}

We report the results of our spectropolarimetric observations collected with ESPaDOnS at CFHT of two wTTSs, namely \ta and \tb, in the framework of the international MaTYSSE Large Program. Our spectral analysis reveals that the two stars have quite different atmospheric properties, with photospheric temperatures of $4425\pm50$~K and $3690\pm130$~K and logarithmic gravities (in cgs units) of $4.5\pm0.2$ and $4.3\pm0.3$. The stars are significantly different in mass, with \ta being $1.0\pm0.10$~\msun and \tb being around half that at $0.45\pm0.10$~\msun. Likewise, the radii are also different with $1.0\pm0.2$~\rsun for \ta and $0.8\pm0.2$~\rsun for \tb, viewed at inclinations of $46\degr\pm10\degr$ and $31\degr\pm10\degr$. Using the \citet{siess2000} evolutionary models (for direct comparison to other MaTYSSE and MaPP results), we estimate their ages to be $21\pm9$~Myr and $11\pm5$~Myr, with \ta being mostly radiative, and \tb being fully convective. We note that these masses, ages and internal structures depend strongly on the adopted temperatures.

With a rotation period of $0.54095\pm0.00003$~d, \ta is the most rapidly rotating wTTS yet mapped with ZDI, and one of the fastest rotators in TWA (see \citealt{delareza2004}). By contrast, \tb has a much slower period of $4.578\pm0.006$~d, which is very similar to the median period of 4.7~d of the TWA 1--13 group \citep{lawson2005}, and also more similar to that of other wTTSs such as V819~Tau ($\prot = 5.53113$~d, \citealt{donati2015}), as well as Par~1379 ($\prot = 5.585$~d, \citealt{hill2017a}). 

We find that neither \ta nor \tb have an infrared excess up to 23.675~$\mu$m. Hence, both stars have likely dissipated their circumstellar accretion discs, with either no accretion taking place, or with accretion occurring at an undetectable level, given that standard accretion-rate metrics based on the equivalent widths of \ha, \hb and \ion{He}{i}~$D_{3}$ are strongly affected by chromospheric emission. 

The \ha, \hb and \ion{Ca}{ii}~IRT emission for both stars is mostly non-variable, with only a few spectra showing excess emission that is attributable to flaring events or prominences. In particular, \ta shows excess red-shifted emission in the \ha, \hb and \ion{Ca}{ii}~IRT lines in three spectra, however, these features are not long lasting and are not periodic. Indeed, the magnetic topology at these phases is such that excess emission could be due to off-limb prominence material that is rotating away from the observer in closed magnetic loops.

Using Zeeman Doppler Imaging, we have derived a surface brightness map of \ta, and the magnetic topologies of both stars. We find that \ta has many cool spots and warm plages on its surface, with a total coverage of around 17~\pc. We detect no significant modulation of the \si lines profiles for \tb, and so find its surface to be compatible with a uniformly bright star. The reconstructed magnetic fields for \ta and \tb are somewhat different in strength, and dramatically different in topology. \ta has a field that is split equally between poloidal and toroidal components, with the largest fraction of energy in higher order modes (with $\ell > 3$), with a total unsigned flux of $\langle B \rangle = 840$~G and where the large-scale magnetosphere is tilted at $35\degr$ from the rotation axis. On the other hand, \tb has a highly poloidal field, with most of the energy in the high order modes with $\ell>3$. The field strength is sufficiently large that the \si lines profiles are significantly Zeeman broadened, with Zeeman signatures clearly detected in individual \sv spectral lines. We derive a total unsigned flux of $\langle B \rangle = 1.4$~kG, using a magnetic filling factor $f$ equal to 0.2 (meaning that 20~\pc of the surface was covered with the mapped magnetic features), where on large scales the magnetosphere is tilted at $20\degr$ from the rotation axis. 

For \tb, our simultaneous fits to both \si and $V$ spectra yields a mean magnetic field strength of $\langle |B_f| \rangle = 6.0\pm0.5$~kG, with a significant fraction of energy in high-strength fields ($>5$~kG). Given that we recover a larger fraction of high magnetic field strengths from our direct modelling of \si profiles, with those fields having small filling factors, a significant proportion of magnetic energy likely lies in small-scale fields that are unresolved by ZDI. The difference between direct spectral fitting and ZDI is likely due to several factors; Firstly, by the cancellation of near-by regions of different sign in \sv (providing most of the difference between \si and $V$ in single lines); Secondly, by the signal in \sv LSD profiles not adding perfectly coherently due to the non-self similarity of different lines in \sv, with scaling amplitudes by effective Land\'{e} factors yielding a less accurate line profile (most of the difference between single lines and LSD profiles). Hence, small-scale high-strength magnetic fields are not recovered with LSD, and are thus not reconstructed with ZDI.

Compared to Tap~26, another wTTS that has a similar mass, age and rotation rate \citep{yu2017}, \ta has a larger toroidal field component (50~\pc for \ta versus 30~\pc for Tap~26), with a total field strength that is around twice as large. Likewise, the field of \ta is also around twice as strong as those of the slower rotating (but similarly massive) wTTSs, V819~Tau and V830~Tau \citep{donati2015}. In the case of \tb, we find that is has a  weaker (poloidal) dipole field (of $B = 0.72$~kG) compared to LkCa~4 (with $B = 1.6$~kG), a wTTSs with a similar rotation rate and a slightly higher mass ($\prot = 3.374$~d, 0.8~\msun). Moreover, compared to main-sequence M dwarfs with a similar mass and period, namely EV~Lac ($\langle B \rangle =0.57$~kG) and GJ~182 ($\langle B \rangle=172$~G), we see that \tb has a slightly stronger magnetic field.

In Fig.~\ref{fig:confuse} we compare the magnetic field topologies of all cTTSs and wTTSs so far mapped with ZDI in an H-R diagram. Fig.~\ref{fig:confuse} also indicates the fraction of the field that is poloidal, the axisymmetry of the poloidal component, and shows PMS evolutionary tracks from \cite{siess2000}. In contrast to cTTSs of the MaPP project, the wTTSs that have been analysed (so far) in the MaTYSSE sample do not appear to show many obvious trends with internal structure. The magnetic field strength does not appear to change significantly after the star becomes mostly radiative, with the largely convective V830~Tau, V819~Tau and V410~Tau hosting a similarly strong dipole field to the mostly radiative TAP~26, and with the largely convective Par 2244 hosting a similarly strong field mostly radiative TWA~6. Moreover, the percentage of poloidal field does not appear to change from when the star is fully convective to when it is mostly radiative (e.g., V410 Tau and TWA 6 are both around 50~\pc poloidal). However, the degree of axisymmetry of the poloidal field appears to correlate with the strength of the magnetic field, given that LkCa~4 and \tb (two stars with significantly stronger fields of 1.2~kG and 1.4~kG, respectively) are mostly axisymmetric ($\gtrsim 70$~\pc). Considering both cTTSs and wTTSs as a whole, it appears that stars are mostly poloidal and axisymmetric when they are mostly convective and cooler than $\sim4300$~K. Moreover, stars hotter than $\sim4300$~K appear to be less axisymmetric and less poloidal, regardless of their internal structure. We note that the wTTSs studied thus far clearly show a wider range of field topologies compared to those of cTTSs, with large scale fields that can be more toroidal and non-axisymmetric, consistent with the fact that most of them are largely radiative or are higher mass. We also note that a more complete analysis will be possible once the remainder of the MaTYSSE sample has been analysed.

Through our tomographic modelling, we were able to determine that \ta has a non-zero surface latitudinal-shear at a confidence level of over 99.99~\pc for the brightness map, and 90~\pc for the magnetic map, as measured over the 16 nights of observation. Its shear rate is around 56 times smaller than the Sun, with an equator-pole lap time of $640^{+110}_{-80}$~d. Given the lack of variability in the lines profiles and the small number of observed rotations ($\sim3$ cycles), we were unable to measure the shear rate for \tb. Out measured shear rate for \ta is similar to that found for V410~Tau, V819~Tau, V830~Tau and LkCa~4 \citep{skelly2010,donati2014,donati2015}, which are all of similar mass.

Finally, the brightness map of \ta was used to predict the activity related RV jitter due to stellar activity, allowing us to filter the measured RVs in the search for potential hJs (in the same manner as \citealt{donati2014,donati2015}). Here, the activity jitter was filtered down to a rms RV precision of $\sim0.20$~\kms (from an initial unfiltered rms of 3.8~\kms). While this is well above the RV precision of ESPaDOnS, the high \vsini decreases the accuracy of the filtering process, with an intrinsic uncertainty of around 0.1~\kms. We find no significant peaks in a periodogram analysis, and find that \ta is unlikely to host a hJ with an orbital period in the range of what we can detect, with a $3\sigma$ error bar on the semi-amplitude of the RV residuals equal to 0.19~\kms, translating into a planet mass of $\simeq3.1$~\mjup orbiting at $\simeq0.1$~au.

\section*{Acknowledgements}
This paper is based on observations obtained at the CFHT, operated by the National Research Council of Canada, the Institut National des Sciences de l'Univers of the Centre National de la Recherche Scientifique (INSU/CNRS) of France and the University of Hawaii. We thank the CFHT QSO team for its great work and effort at collecting the high-quality MaTYSSE data presented in this paper. MaTYSSE is an international collaborative research programme involving experts from more than 10 different countries (France, Canada, Brazil, Taiwan, UK, Russia, Chile, USA, Switzerland, Portugal, China and Italy). Observations of \tb are supported by the contribution to the MaTYSSE Large Project on CFHT obtained through the Telescope Access Program (TAP), which has been funded by the``the Strategic Priority Research Program - The Emergence of Cosmological Structures" of the Chinese Academy of Sciences (Grant No.11 XDB09000000) and the Special Fund for Astronomy from the Ministry of Finance. GJH is supported by general grants 11473005 and 11773002 awarded by the National Science Foundation of China. We also thank the IDEX initiative at Universit\'{e} F\'{e}d\'{e}rale Toulouse Midi-Pyr\'{e}n\'{e}es (UFTMiP) for funding the STEPS collaboration program between IRAP/OMP and ESO and for allocating a `Chaire d'Attractivit\'{e}' to GAJH. 
JFD acknowledges funding from the European Research Council (ERC) under the H2020 research \& innovation programme (grant agreement \#740651 NewWorlds). SHPA acknowledges financial support from CNPq, CAPES and Fapemig. This work has made use of the VALD database, operated at Uppsala University, the Institute of Astronomy RAS in Moscow, and the University of Vienna.
%%%%%%%%%%%%%%%%%%%%%%%%%%%%%%%%%%%%%%%%%%%%%%%%%%

\bibliographystyle{mnras}
\bibliography{matysse}

%%%%%%%%%%%%%%%%% APPENDICES %%%%%%%%%%%%%%%%%%%%%
\appendix

\section{Line profiles of \ion{Ca}{ii} infrared triplet, \ha, and \hb, for \ta and \tb}
\label{app:trails}
\begin{figure*}
\begin{center}
\includegraphics[height=0.65\textheight]{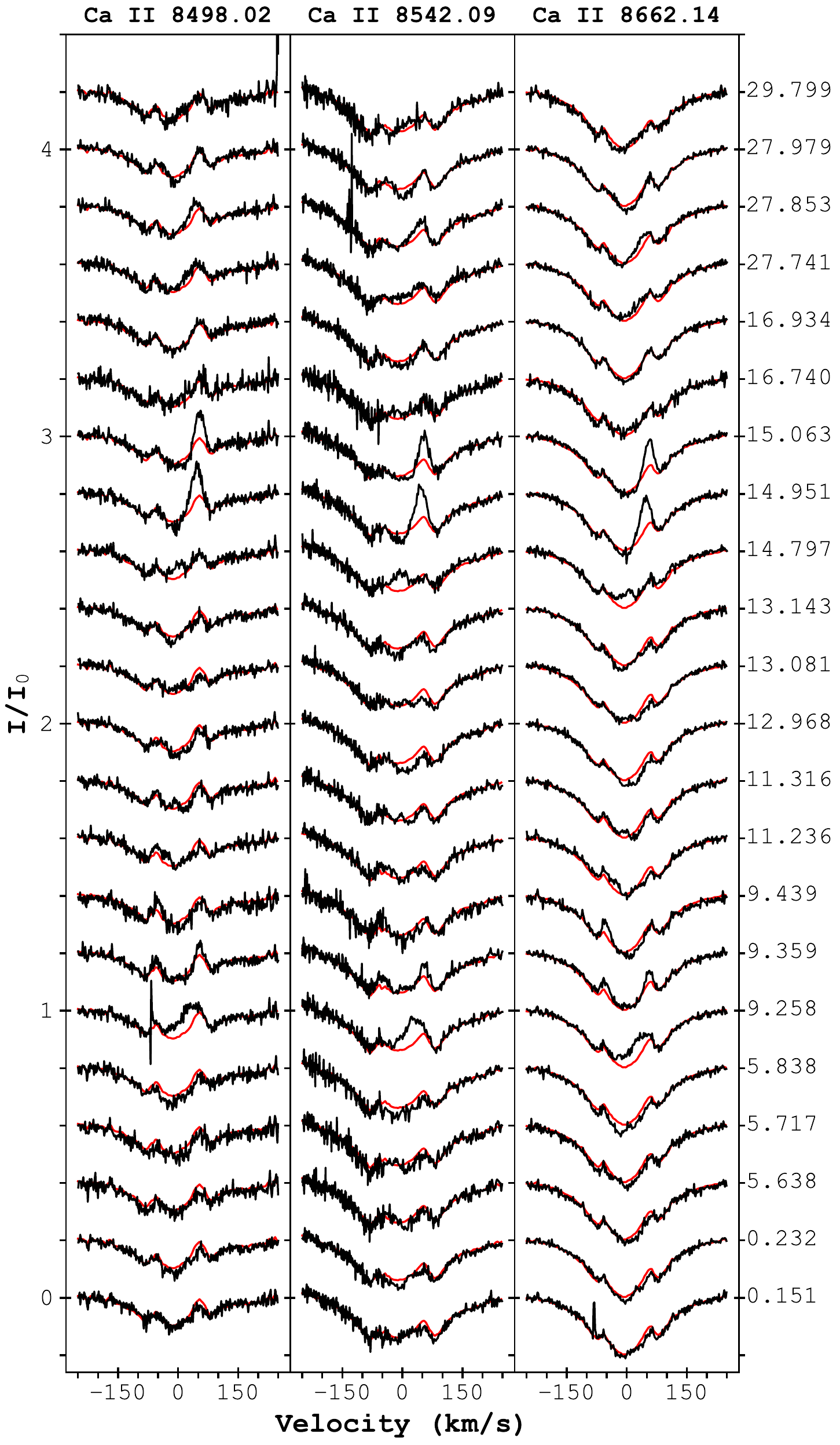}
\hspace{0.5cm}
\includegraphics[height=0.65\textheight]{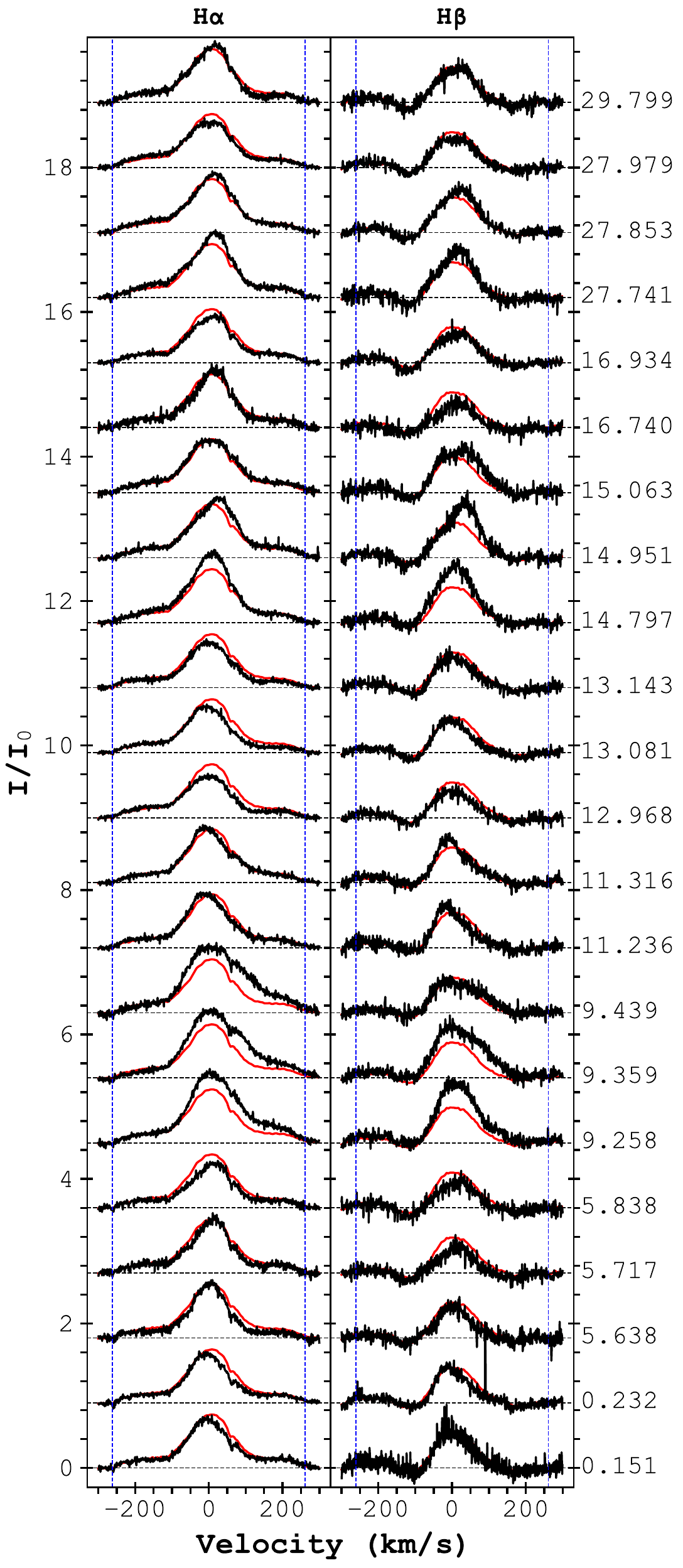}
\caption{For \ta. Left panel: The \ion{Ca}{ii} infrared triplet, with line profiles of the 8498.0.2~\AA\, 8542.09~\AA\ and 8662.14~\AA\ components shown (left to right) as black solid lines, where the mean line profile is shown in red, with the cycle number displayed on the right of the profiles. Right panel: \ha and \hb line profiles, shown in the same manner, additionally showing the co-rotation radius as a dashed blue line.}
\label{fig:twa6lines}
\end{center}
\end{figure*}

\begin{figure*}
\begin{center}
\includegraphics[height=0.65\textheight]{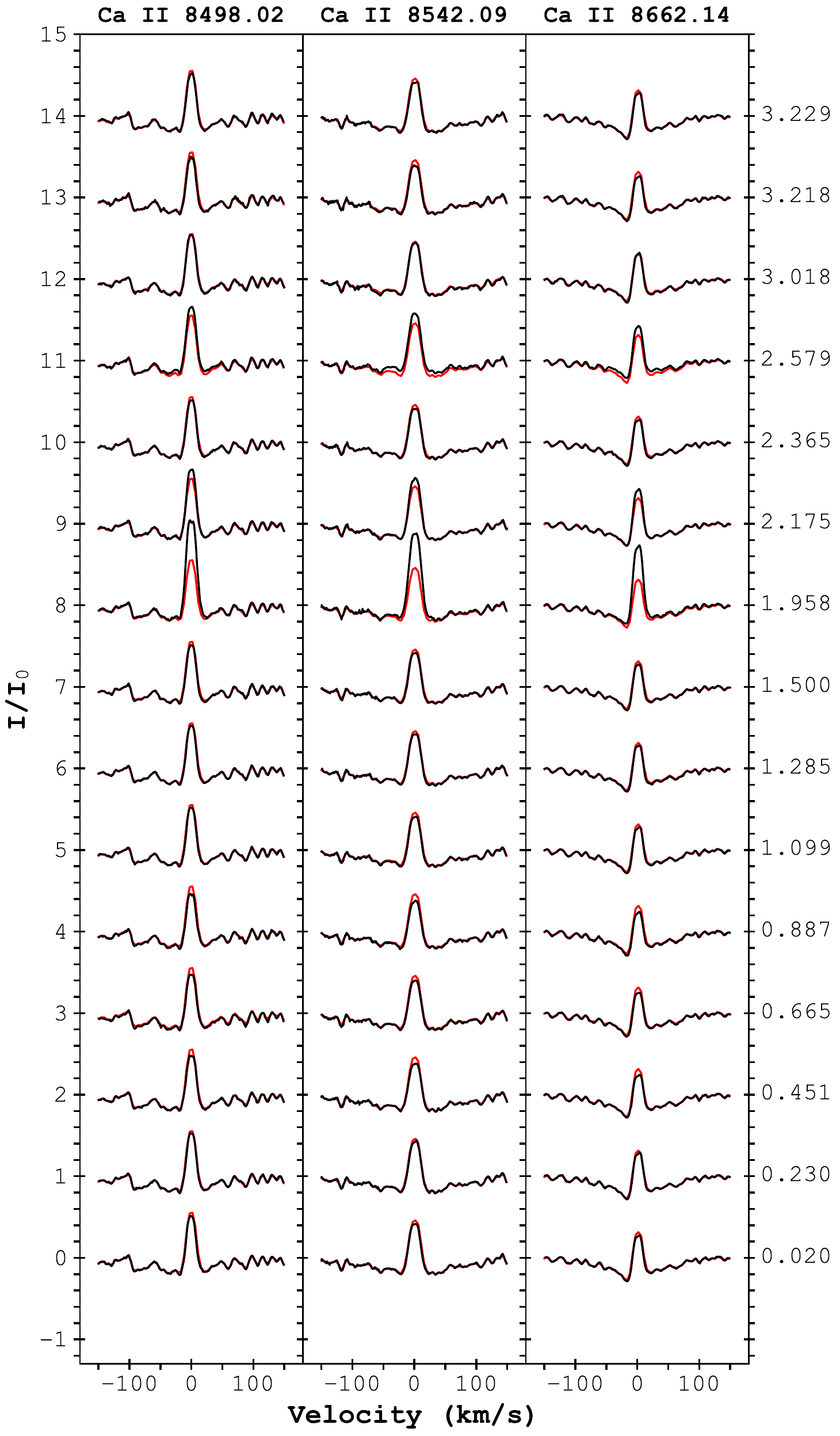}
\hspace{0.5cm}
\includegraphics[height=0.65\textheight]{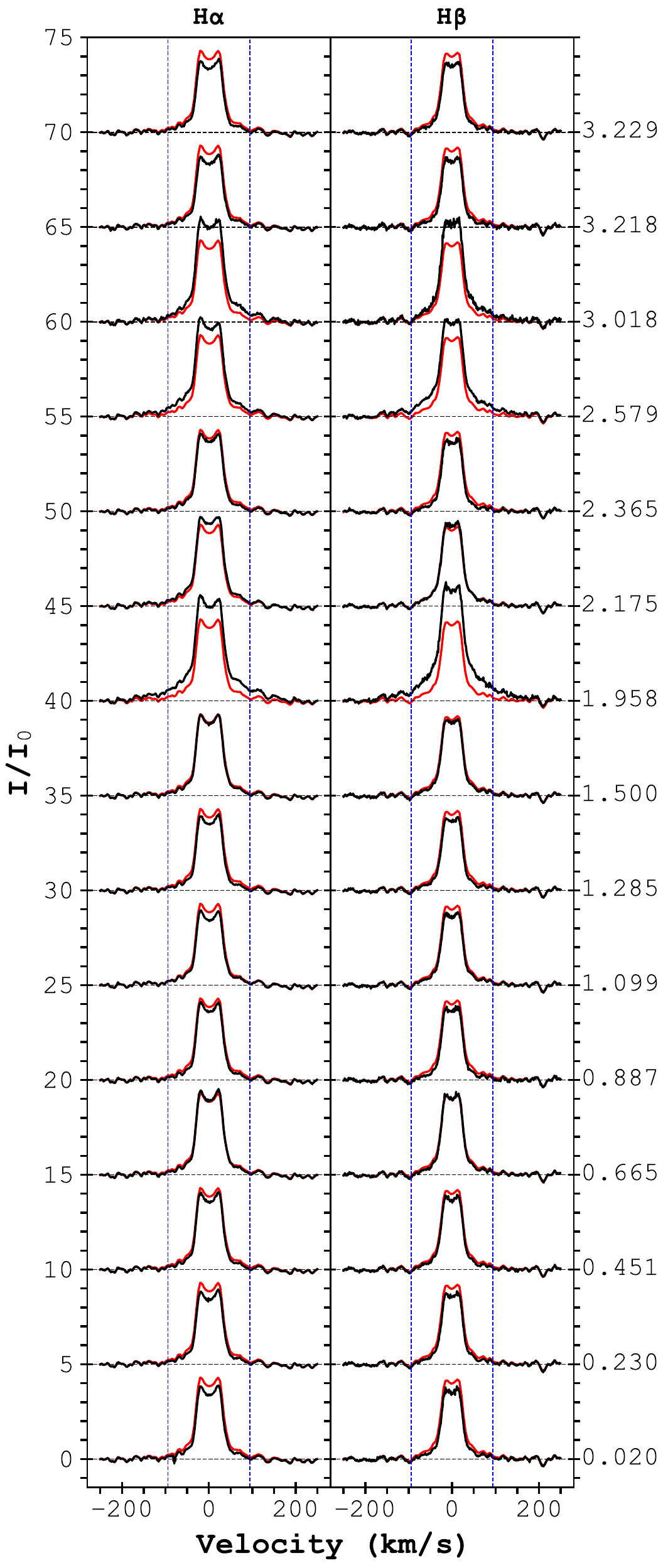}
\caption{Same as Fig.~\ref{fig:twa6lines} but for \tb, with the left panel showing the \ion{Ca}{ii} infrared triplet, and the right panel showing \ha and \hb line profiles.}
\label{fig:twa8alines}
\end{center}

\end{figure*}

Line profiles of the \ion{Ca}{ii} infrared triplet, \ha and \hb are shown in Fig.~\ref{fig:twa6lines} and Fig.~\ref{fig:twa8alines} for \ta and \tb, respectively.

\begin{figure*}
\begin{center}
\includegraphics[height=0.65\textheight]{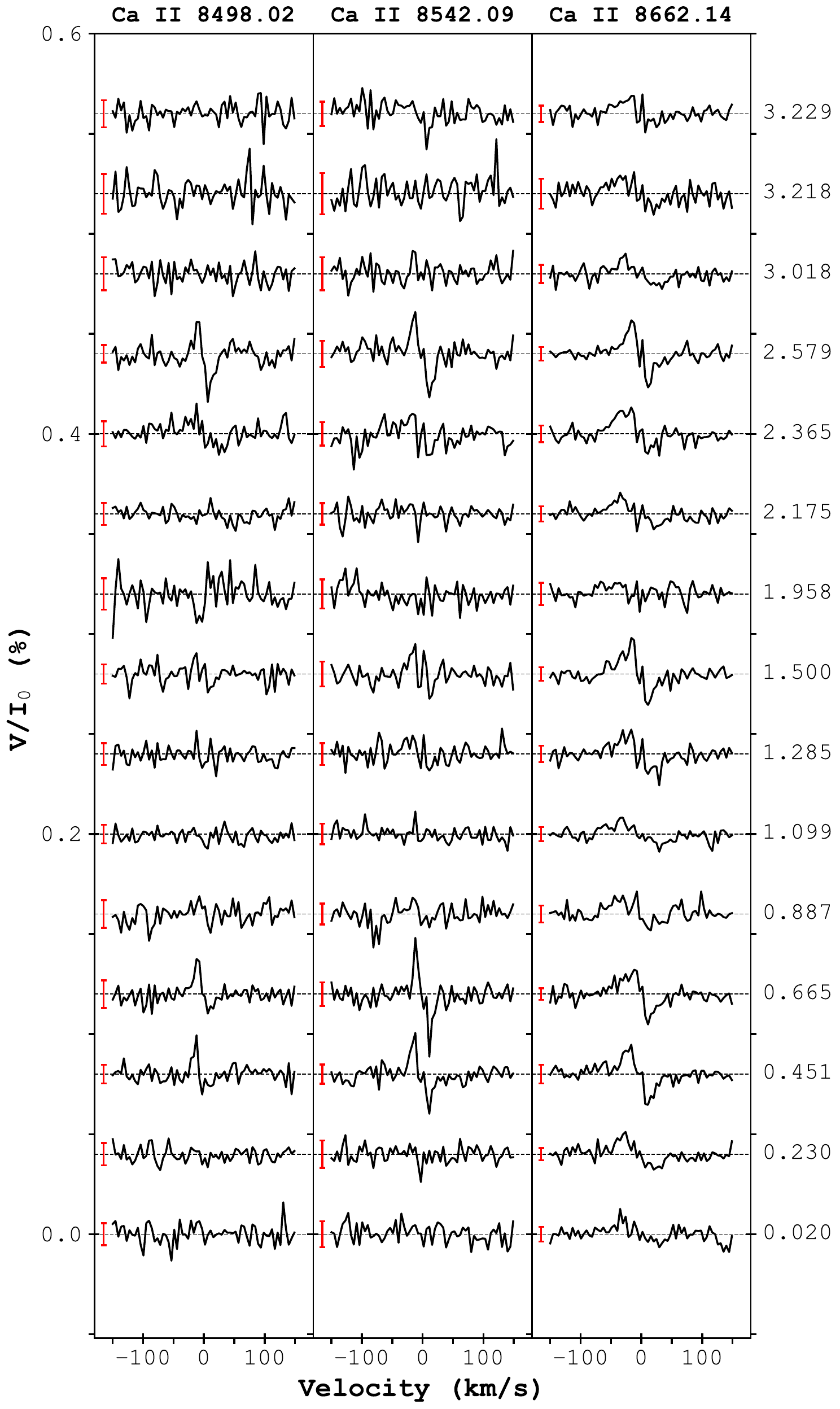}
\hspace{0.5cm}
\includegraphics[height=0.65\textheight]{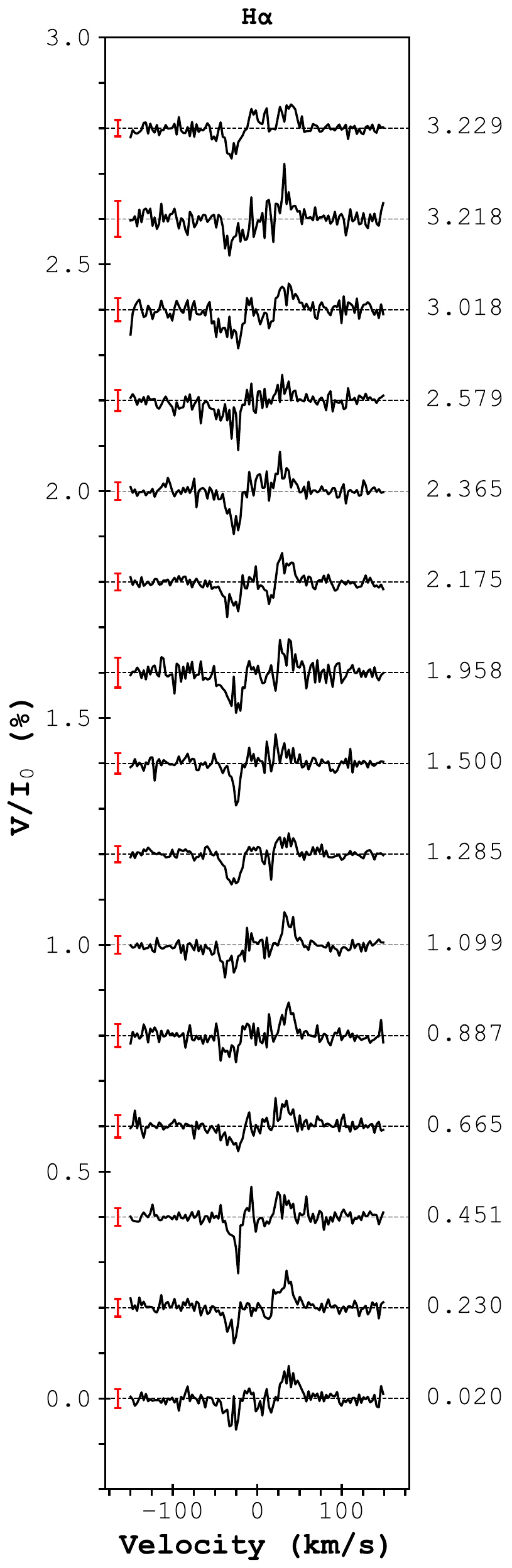}
\caption{\sv line profiles of \tb with the \ion{Ca}{ii} IRT shown in the left panel, and \ha shown in the right panel. $3\sigma$ errorbars are shown in red on the left side of the line profiles.}
\label{fig:twa8alinessv}
\end{center}
\end{figure*}

\begin{figure*}
\centering
\includegraphics[width=0.33\textwidth]{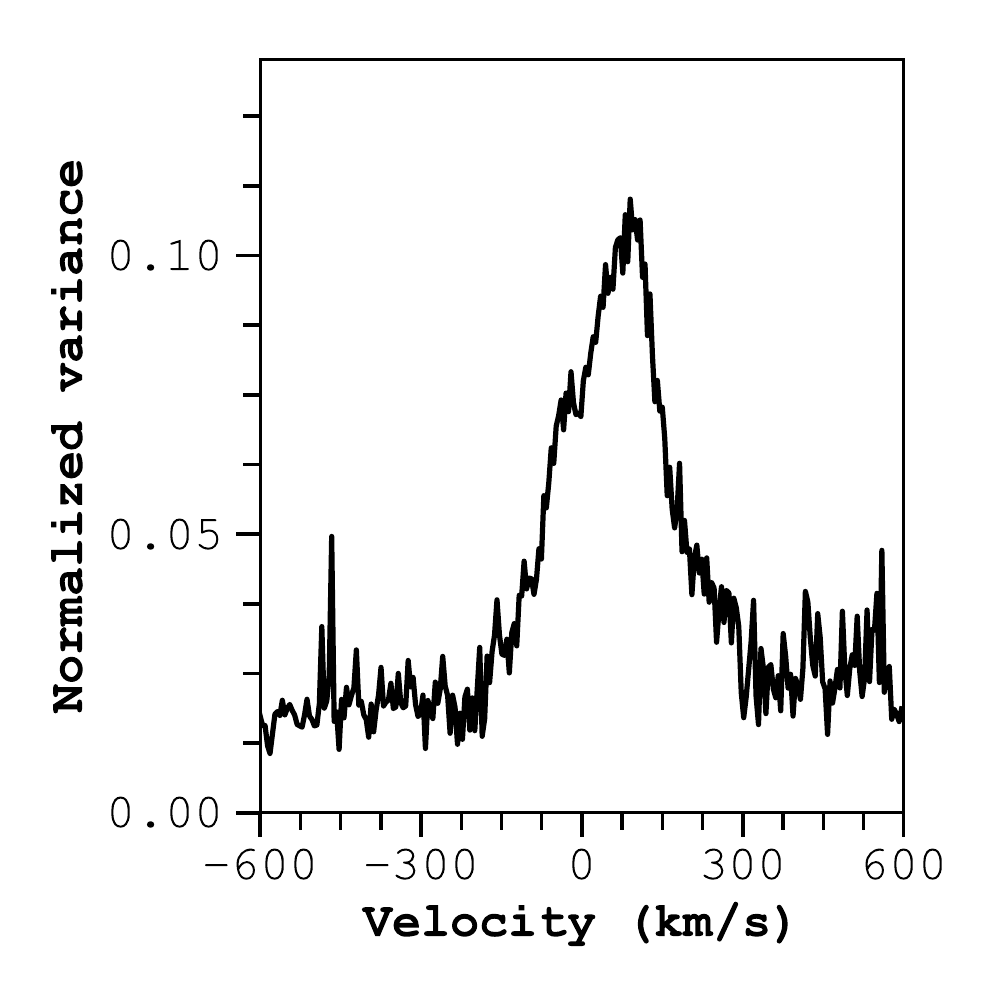}
\includegraphics[width=0.33\textwidth]{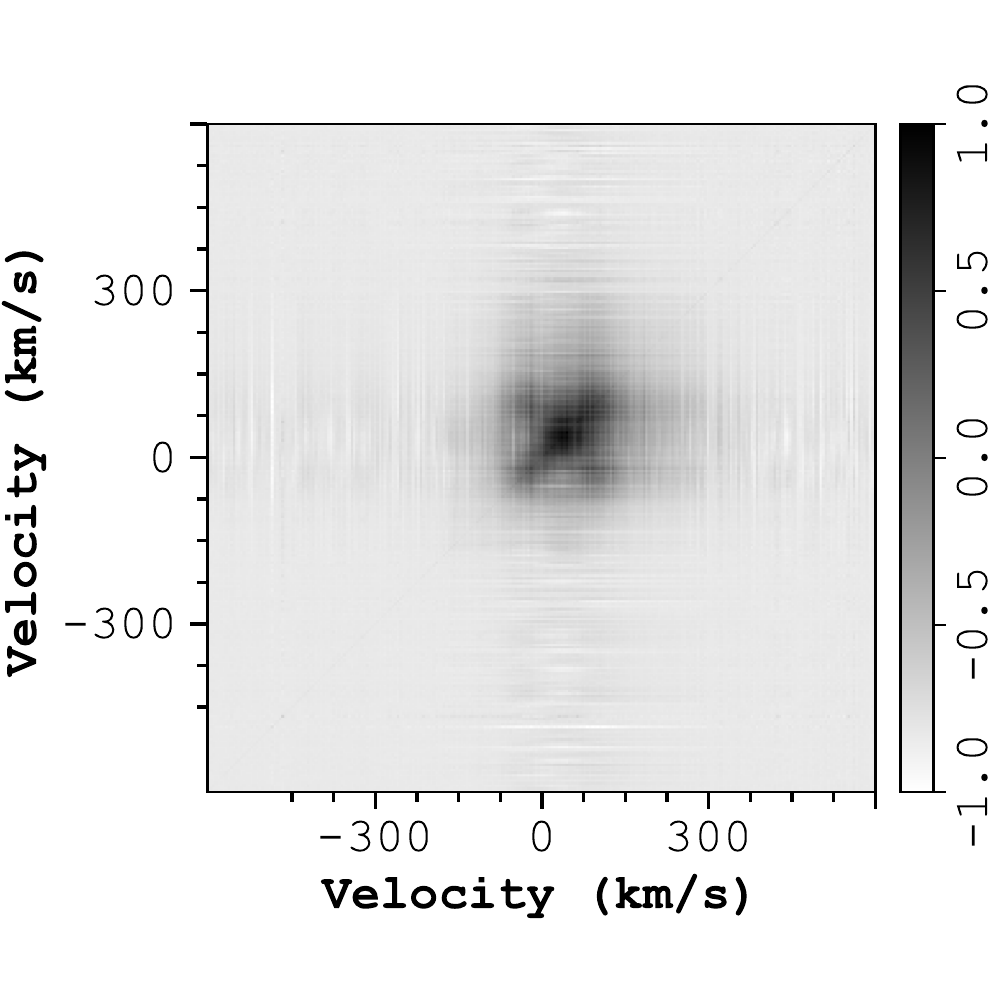}
\caption{Left panel: The normalized variance profile of \ha for \ta. There is variance from around -200~\kms up to around +300~\kms. Right panel: The autocorrelation matrix for \ha, where black means perfect correlation and white means perfect anticorrelation.}
\label{fig:twa6var}
\end{figure*}

\begin{figure*}
\centering
\includegraphics[width=0.33\textwidth]{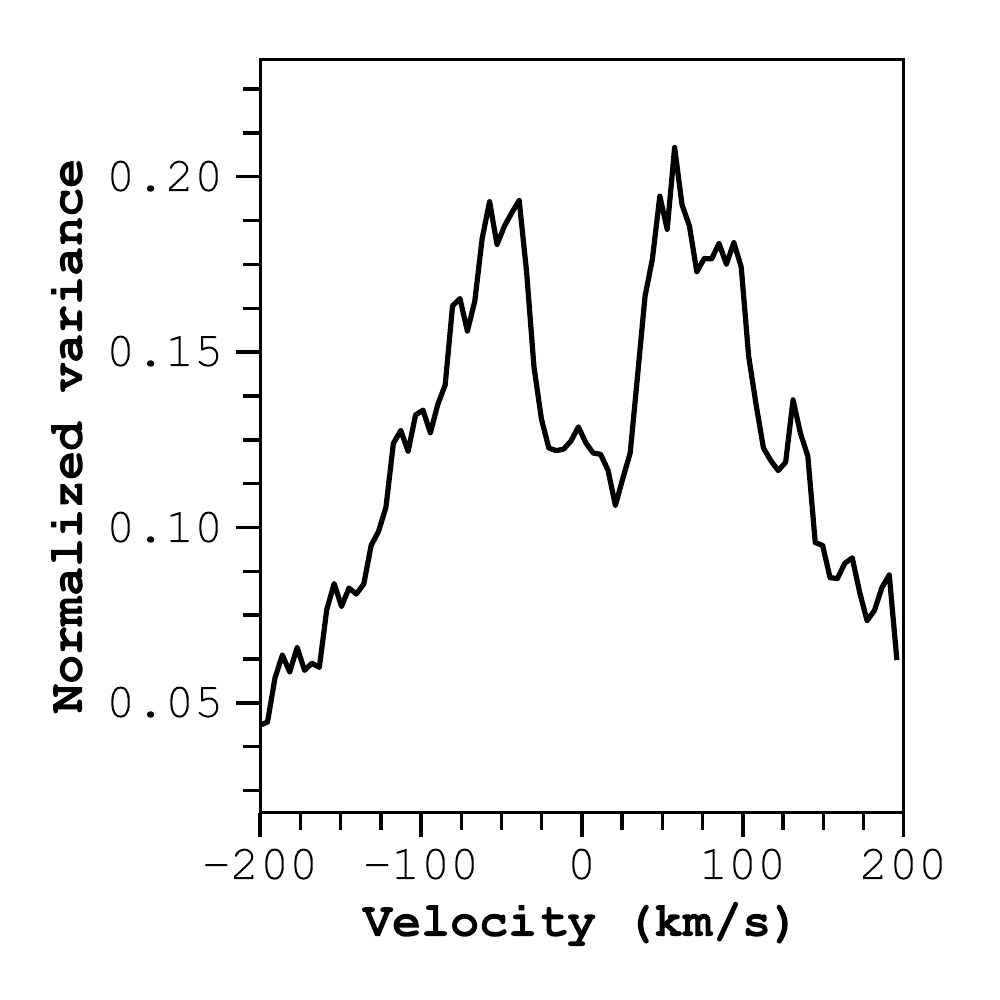}
\includegraphics[width=0.33\textwidth]{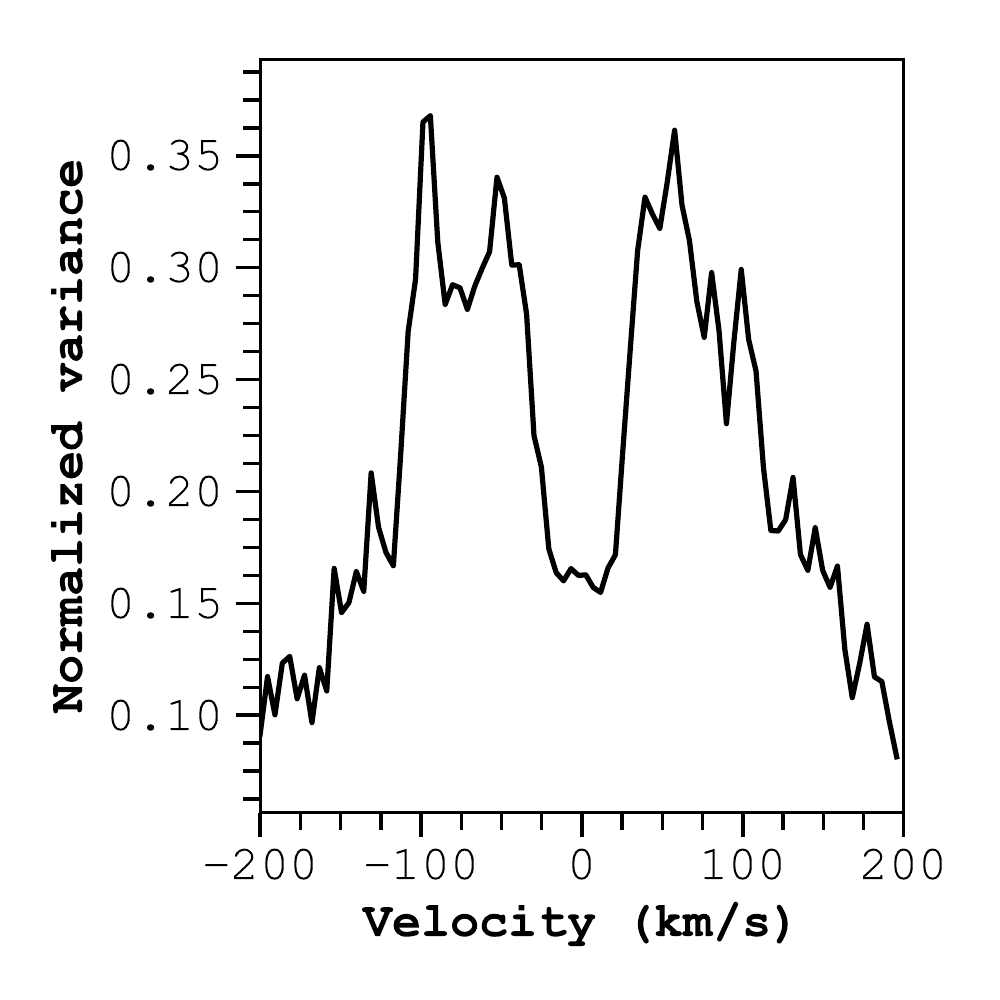}
\includegraphics[width=0.33\textwidth]{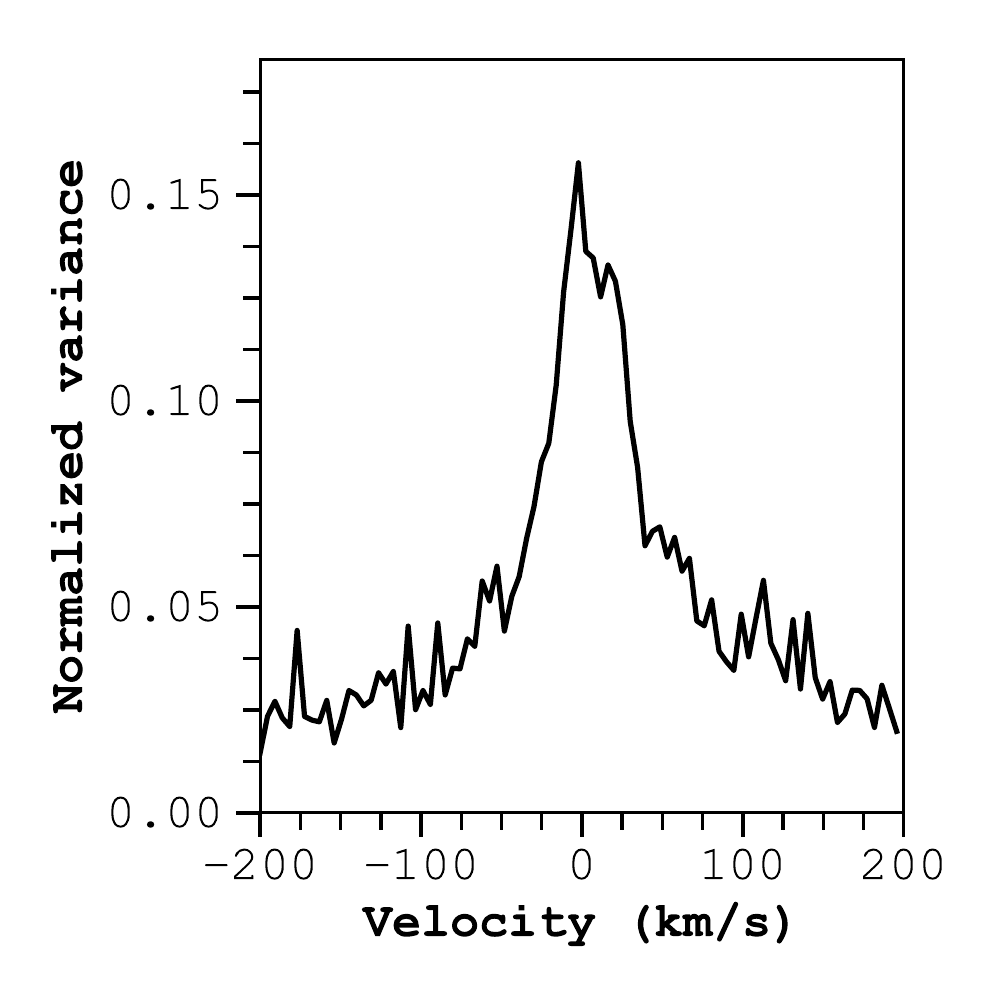}\\
\includegraphics[width=0.33\textwidth]{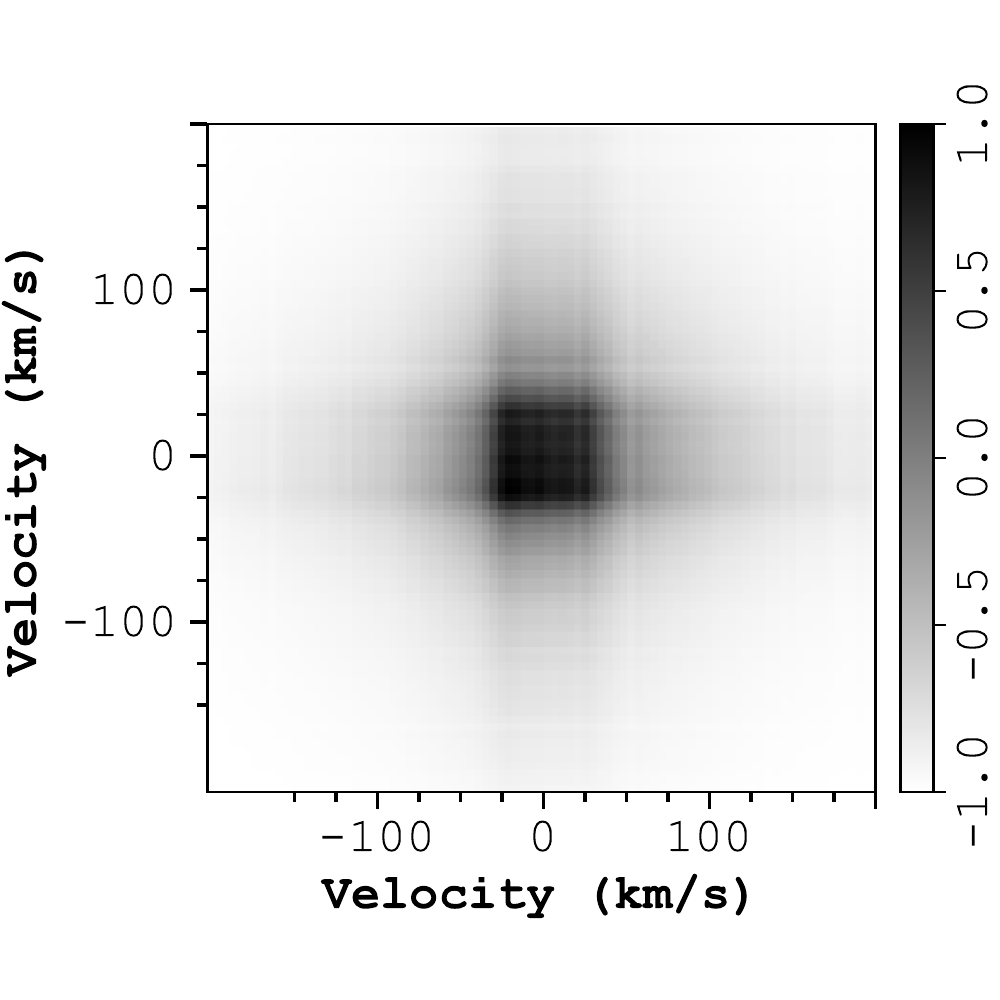}
\includegraphics[width=0.33\textwidth]{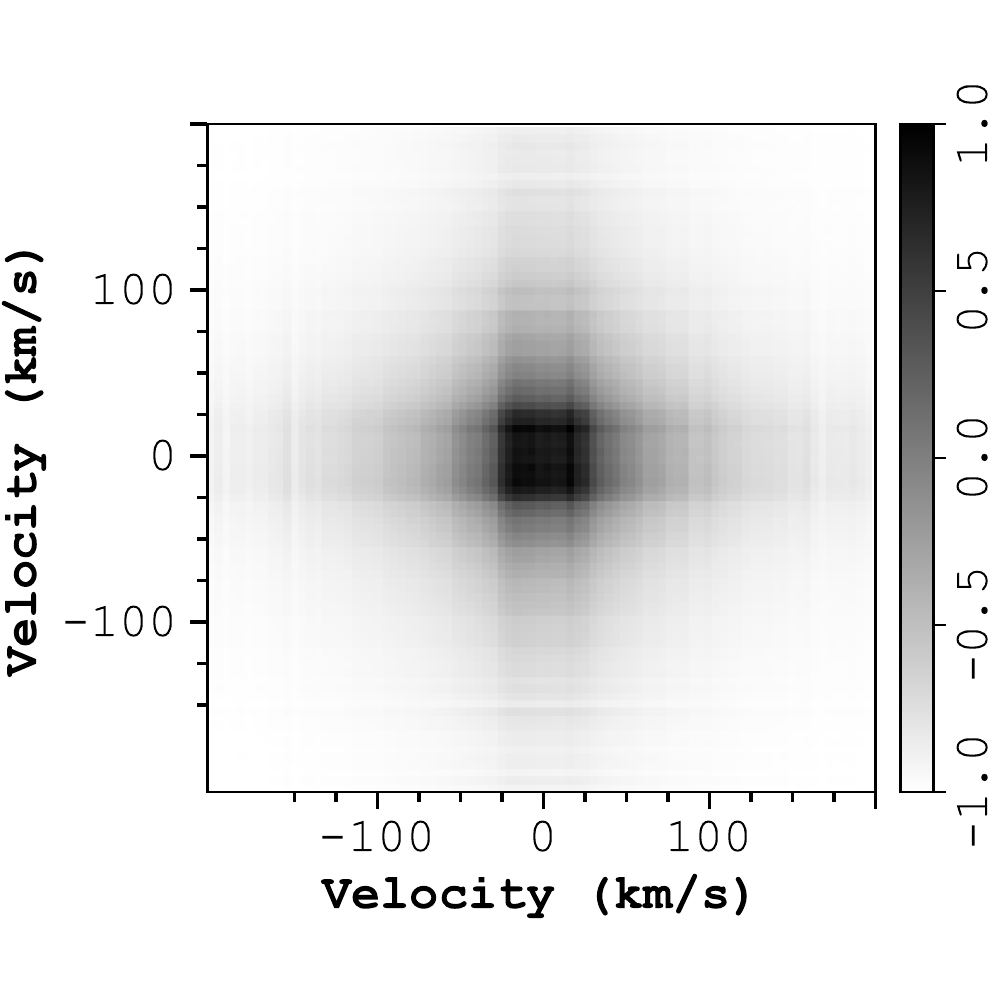}
\includegraphics[width=0.33\textwidth]{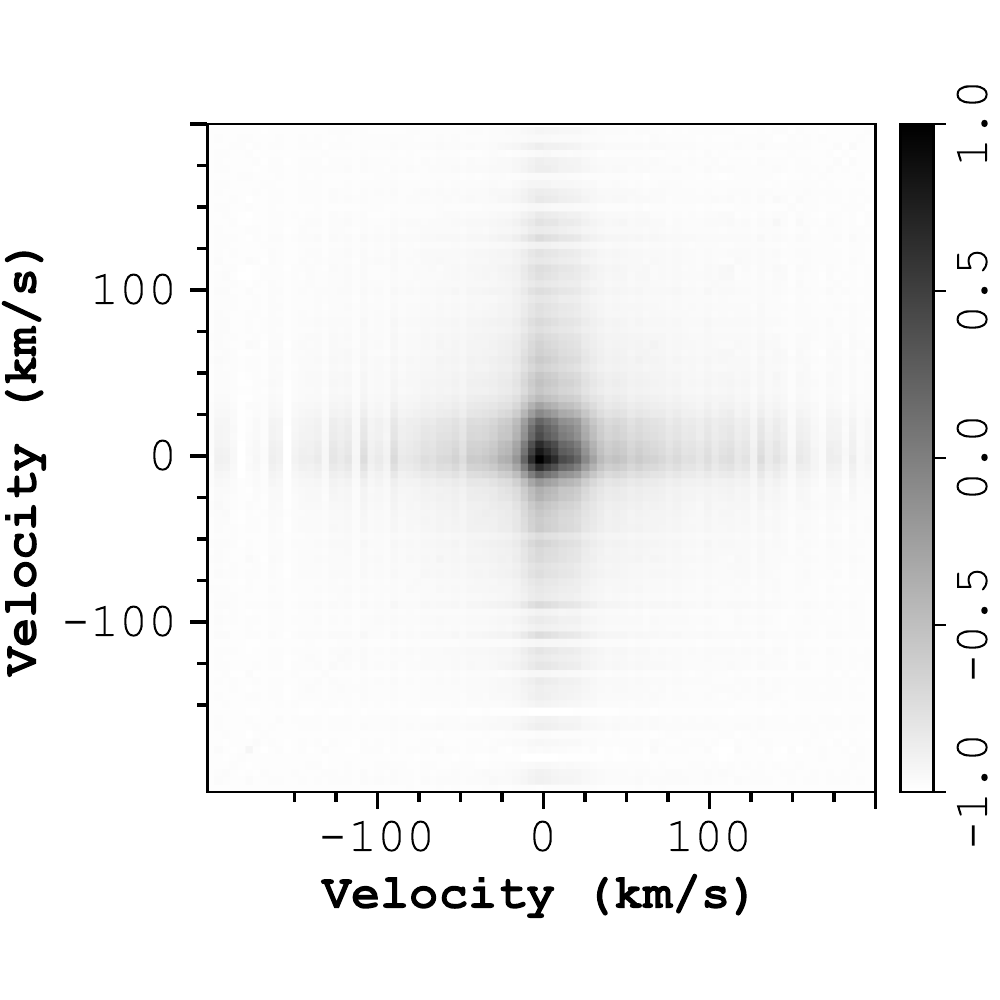}
\caption{Same as Fig.~\ref{fig:twa6var} but for \tb. The top row shows normalized variance profiles for \ha, \hb and \ion{He}{i}~D3 (left to right), with the bottom row showing the corresponding autocorrelation matrices.}
\label{fig:twa8avar}
\end{figure*}

\section{Accretion status of TWA 6 and TWA 8A}
\label{sec:acc}
The SEDs of \ta and \tb show no evidence of an infra-red excess (see Fig.~\ref{fig:sed}), suggesting that both stars are disc-less. Nevertheless, we may use our high-quality spectra of both targets to determine their accretion status using several metrics.

Following our previous studies \citep[e.g.,][]{hill2017a}, one may estimate the level of surface accretion in TTSs by adopting the relations between line luminosity $L_{\text{line}}$ and the accretion luminosity $L_{\mathrm{acc}}$ of \cite{alcala2017}. For this purpose we determined $L_{\text{line}}$ by assuming blackbody scaling using the stellar radius $R_{\star}$ and \teff given in Table~\ref{tab:syspars}. Then, the mass accretion rate ${\dot{M}_{\mathrm{acc}}}$ was calculated using the relationship
\begin{equation}
\dot{M}_{\mathrm{acc}} = \frac{L_{\mathrm{acc}} R_{\star}}{GM_{\star}(1-\frac{R_{\star}}{R_{\text{in}}})}
\end{equation}

\noindent where $R_{\text{in}}$ denotes the truncation radius of the disc, and is taken to be $5R_{\star}$ \citep{gullbring1998}. 

For \ta, we detect weak \ion{He}{i}~$D_{3}$ emission with an EW of around 0.03~\AA\ (1.6~\kms), corresponding to ${\log{\dot{M}_{\mathrm{acc}}} \simeq -10.6}$~\msunyr. We find the H$\beta$ emission (see Fig.~\ref{fig:twa6lines}) to have an EW ranging between 0.69--1.44~\AA\ (average of 0.96~\AA\/, equivalent to 59.5~\kms, corresponding to ${\log{\dot{M}_{\mathrm{acc}}} \simeq -10.1}$~\msunyr), and the H$\alpha$ emission to have an EW ranging between 2.26--4.03~\AA\ (average of 2.85~\AA\/, equivalent to 130.2~\kms, corresponding to ${\log{\dot{M}_{\mathrm{acc}}} \simeq  -10.2}$~\msunyr).

For \tb, we detect moderate \ion{He}{i}~$D_{3}$ emission with an EW of around 0.3~\AA\ (17.4~\kms), corresponding to ${\log{\dot{M}_{\mathrm{acc}}} \simeq -9.7}$~\msunyr. We find the H$\beta$ emission (see Fig.~\ref{fig:twa8alines}) to have an EW ranging between 3.1--6.8~\AA\ (average of 3.9~\AA\/, equivalent to 238.0~\kms, corresponding to ${\log{\dot{M}_{\mathrm{acc}}} \simeq -9.7}$~\msunyr), and the H$\alpha$ emission to have an EW ranging between 5.9--10.9~\AA\ (average of 7.2~\AA\/, equivalent to 326.9~\kms, corresponding to ${\log{\dot{M}_{\mathrm{acc}}} \simeq -9.8}$~\msunyr). 

These accretion rates would suggest that both stars are weakly accreting, however, as discussed in \citet{hill2017a}, chromospheric activity in TTS becomes a significant influence on the strength and width of emission lines in the low accretion regime. Pertinently, the large convective turnover times of TTSs \citep{gilliland1986} combined with their rapid rotation means they possess a low Rossby number, placing them well within the saturated activity regime (e.g., \citealt{reiners2014}). Indeed, the H$\alpha$ line luminosity is observed to saturate in young stars at around $\log[L(H\alpha)/L_{\text{bol}}] = -3.3$ or lower \citep{barrado2003}, and as both our target stars show line luminosities that are similar to (or below) this level, with $\log[L(H\alpha)/L_{\text{bol}}]$ equal to $-3.19$ for \ta and $-3.78$ for \tb, any estimate of accretion rates based on line luminosities (especially H$\alpha$, and to a lesser extent H$\beta$) must be considered to be significantly influenced or even dominated by chromospheric activity.

The distinction between emission due to accretion and that due to chromospheric activity has been characterized by several authors, yielding a distinct threshold between these regimes. Using the empirical spectral-type-dependant relationship between the EW(H$\alpha$) and the accretion rate of \citet{barrado2003}, the defining threshold of an accreting TTS is EW(H$\alpha$) equal to 5.1~\AA\ and 12.2~\AA\ for K5 and M3 spectral types, respectively (appropriate for \ta and \tb, see Section~\ref{sec:evolution}). Given that the maximum EW(H$\alpha$) of \ta and \tb are equal to 4.0~\AA\ and 10.9~\AA\ , both stars lie below these limits and fall into the non-accreting regime (where line broadening is dominated by chromospheric activity).  

Elsewhere, \citet{manara2017b} derived an empirical relationship between a star's spectral-type and the point at which line emission may be dominated by chromospheric activity (termed chromospheric accretion `noise'). In the case of \ta, this threshold is at $\log{(L_{\text{acc,noise}}/L_{\text{star}})} = -2.3\pm0.1$. Given that the average line luminosities $\log{(L_{\text{acc}}/L_{\text{star}}})$ for H$\alpha$, H$\beta$ and \ion{He}{i}~$D_{3}$ are respectively equal to $-3.19\pm0.01$, $-3.04\pm0.02$ and $-3.63\pm0.03$, the luminosity of all three emission lines are significantly below the threshold of chromospheric noise. Likewise for \tb, this threshold is estimated as $\log{(L_{\text{acc,noise}}/L_{\text{star}})} = -2.59\pm0.13$. Here, the average line luminosities for H$\alpha$, H$\beta$ and \ion{He}{i}~$D_{3}$ are respectively equal to $-3.78\pm0.02$, $-3.65\pm0.03$ and $-3.62\pm0.02$, where again, all emission is well below the threshold where one can distinguish between accretion and chromospheric emission. 

Thus, the accretion rates determined above for \ta and \tb must be taken to be upper limits, given that chromospheric emission is likely the dominant broadening mechanism. Hence, our target stars are likely not accreting (or are doing so at an undetectable level), thus confirming their classification as wTTSs - a result consistent with past work by \citet[e.g.][]{white2004, kastner2016}.

\section{Magnetic fields from direct spectral fitting}
\label{app:specfit}

\begin{landscape}
\begin{table}
\centering
\caption{Atomic data used in the direct spectrum fitting, from VALD, for the major lines. Additional much weaker lines were included in the spectrum synthesis for completeness, but are omitted here for brevity. The quantities $_{\rm low}$ and $_{\rm high}$ refer to the lower and upper level of the transition, respectively.  Term symbols are provided to identify lines of the same multiplet. }
\begin{tabular}{ccccccccc}
\hline
Species   & Wavelength (\AA) & $\log gf$ & $E_{\rm low}$ (Ev) & $J_{\rm low}$ & $J_{\rm high}$ & Land\'e $g_{\rm low}$ & Land\'e $g_{\rm high}$ & Multiplet terms \\
  \hline
 Ti \sc{i} & 9675.54 & -0.804 & 0.8360 & 4 &  4 & 1.34 & 1.35 & a\,$^5$F -- z\,$^5$F$^\circ$ \\  
 Ti \sc{i} & 9688.87 & -1.610 & 0.8129 & 1 &  2 & 1.00 & 1.50 & a\,$^5$F -- z\,$^5$F$^\circ$ \\  
 Ti \sc{i} & 9705.66 & -1.009 & 0.8259 & 3 &  3 & 1.26 & 1.26 & a\,$^5$F -- z\,$^5$F$^\circ$ \\  
 Ti \sc{i} & 9718.96 & -1.181 & 1.5025 & 4 &  3 & 1.00 & 0.95 & a\,$^1$G -- z\,$^1$F$^\circ$ \\  
 Ti \sc{i} & 9728.41 & -1.206 & 0.8181 & 2 &  2 & 1.00 & 1.00 & a\,$^5$F -- z\,$^5$F$^\circ$ \\  
 Ti \sc{i} & 9743.61 & -1.306 & 0.8129 & 1 &  1 & 0.00 & 0.00 & a\,$^5$F -- z\,$^5$F$^\circ$ \\  
 Ti \sc{i} & 9770.30 & -1.581 & 0.8484 & 5 &  4 & 1.34 & 1.55 & a\,$^5$F -- z\,$^5$F$^\circ$ \\  
 Ti \sc{i} & 9783.31 & -1.428 & 0.8360 & 4 &  3 & 1.26 & 1.48 & a\,$^5$F -- z\,$^5$F$^\circ$ \\  
 Ti \sc{i} & 9783.59 & -1.617 & 0.8181 & 2 &  1 & 0.00 & 1.49 & a\,$^5$F -- z\,$^5$F$^\circ$ \\  
 Ti \sc{i} & 9787.69 & -1.444 & 0.8259 & 3 &  2 & 1.00 & 1.50 & a\,$^5$F -- z\,$^5$F$^\circ$ \\  
 Ti \sc{i} & 9832.14 & -1.130 & 1.8871 & 5 &  4 & 1.21 & 1.21 & a\,$^3$G -- y\,$^3$F$^\circ$ \\
\hline
\end{tabular} 
\label{tab:atomic-data} 
\end{table}

\begin{table}
\centering
\caption{Best fit parameters from direct spectral fitting of \tb, using our third model that fits Stokes $I$ and $V$ simutaneously, for observations that could be adequately telluric-corrected. Each column gives the fitted parameters for the spectrum obtained on the date given at the top. Mean values are presented in Table~\ref{tab:specfit}.}
\begin{tabular}{ccccccccccc}
\hline
       & 2015-03-25          & 2015-03-26          & 2015-03-27          & 2015-03-28          & 2015-03-29          & 2015-03-30          & 2015-03-31          & 2015-04-01          & 2015-04-05          & 2015-04-06          \\
\hline
\vsini\ (\kms)& $ 4.75  \pm 0.19  $ & $ 4.78  \pm 0.17  $ & $ 4.72  \pm 0.18  $ & $ 5.08  \pm 0.17  $ & $ 4.61  \pm 0.18  $ & $ 4.66  \pm 0.16  $ & $ 4.66  \pm 0.16  $ & $ 4.95  \pm 0.17  $ & $ 4.92  \pm 0.17  $ & $ 5.02  \pm 0.17  $ \\
\vmic\ (\kms) & $ 1.18  \pm 0.07  $ & $ 1.07  \pm 0.06  $ & $ 1.14  \pm 0.06  $ & $ 1.13  \pm 0.06  $ & $ 1.08  \pm 0.06  $ & $ 1.06  \pm 0.05  $ & $ 1.06  \pm 0.06  $ & $ 1.08  \pm 0.06  $ & $ 1.01  \pm 0.06  $ & $ 1.02  \pm 0.06  $ \\
   $[$Ti/H$]$ & $-7.007 \pm 0.014 $ & $-6.985 \pm 0.012 $ & $-7.005 \pm 0.013 $ & $-6.974 \pm 0.012 $ & $-6.981 \pm 0.012 $ & $ 6.988 \pm 0.011 $ & $-6.966 \pm 0.011 $ & $-6.963 \pm 0.012 $ & $-6.934 \pm 0.011 $ & $-6.952 \pm 0.012 $ \\
        +2 kG & $ 0.168 \pm 0.023 $ & $ 0.182 \pm 0.017 $ & $ 0.167 \pm 0.018 $ & $ 0.154 \pm 0.018 $ & $ 0.150 \pm 0.018 $ & $ 0.176 \pm 0.016 $ & $ 0.175 \pm 0.016 $ & $ 0.165 \pm 0.018 $ & $ 0.147 \pm 0.019 $ & $ 0.125 \pm 0.019 $ \\
        +5 kG & $ 0.229 \pm 0.010 $ & $ 0.227 \pm 0.009 $ & $ 0.259 \pm 0.010 $ & $ 0.239 \pm 0.010 $ & $ 0.221 \pm 0.010 $ & $ 0.228 \pm 0.009 $ & $ 0.239 \pm 0.009 $ & $ 0.268 \pm 0.010 $ & $ 0.258 \pm 0.010 $ & $ 0.282 \pm 0.010 $ \\
       +10 kG & $ 0.053 \pm 0.008 $ & $ 0.061 \pm 0.006 $ & $ 0.051 \pm 0.007 $ & $ 0.056 \pm 0.007 $ & $ 0.052 \pm 0.006 $ & $ 0.056 \pm 0.006 $ & $ 0.061 \pm 0.006 $ & $ 0.055 \pm 0.006 $ & $ 0.058 \pm 0.006 $ & $ 0.048 \pm 0.007 $ \\
       +15 kG & $ 0.052 \pm 0.007 $ & $ 0.044 \pm 0.006 $ & $ 0.049 \pm 0.006 $ & $ 0.051 \pm 0.006 $ & $ 0.045 \pm 0.006 $ & $ 0.047 \pm 0.005 $ & $ 0.043 \pm 0.005 $ & $ 0.043 \pm 0.006 $ & $ 0.045 \pm 0.006 $ & $ 0.047 \pm 0.006 $ \\
       +20 kG & $ 0.034 \pm 0.005 $ & $ 0.030 \pm 0.004 $ & $ 0.027 \pm 0.005 $ & $ 0.042 \pm 0.005 $ & $ 0.038 \pm 0.005 $ & $ 0.034 \pm 0.004 $ & $ 0.037 \pm 0.004 $ & $ 0.037 \pm 0.005 $ & $ 0.038 \pm 0.004 $ & $ 0.038 \pm 0.005 $ \\
        -2 kG & $ 0.163 \pm 0.023 $ & $ 0.166 \pm 0.017 $ & $ 0.145 \pm 0.018 $ & $ 0.159 \pm 0.018 $ & $ 0.148 \pm 0.018 $ & $ 0.173 \pm 0.016 $ & $ 0.161 \pm 0.016 $ & $ 0.144 \pm 0.018 $ & $ 0.146 \pm 0.019 $ & $ 0.141 \pm 0.019 $ \\
        -5 kG & $ 0.221 \pm 0.010 $ & $ 0.202 \pm 0.009 $ & $ 0.205 \pm 0.010 $ & $ 0.202 \pm 0.010 $ & $ 0.217 \pm 0.010 $ & $ 0.213 \pm 0.009 $ & $ 0.199 \pm 0.009 $ & $ 0.209 \pm 0.010 $ & $ 0.205 \pm 0.010 $ & $ 0.213 \pm 0.010 $ \\
       -10 kG & $ 0.044 \pm 0.008 $ & $ 0.045 \pm 0.006 $ & $ 0.036 \pm 0.007 $ & $ 0.053 \pm 0.007 $ & $ 0.041 \pm 0.006 $ & $ 0.041 \pm 0.006 $ & $ 0.036 \pm 0.006 $ & $ 0.039 \pm 0.006 $ & $ 0.035 \pm 0.006 $ & $ 0.040 \pm 0.007 $ \\
       -15 kG & $ 0.010 \pm 0.006 $ & $ 0.000 \pm 0.006 $ & $ 0.010 \pm 0.006 $ & $ 0.010 \pm 0.006 $ & $ 0.012 \pm 0.006 $ & $ 0.010 \pm 0.005 $ & $ 0.012 \pm 0.005 $ & $ 0.011 \pm 0.006 $ & $ 0.009 \pm 0.006 $ & $ 0.013 \pm 0.006 $ \\
       -20 kG & $ 0.022 \pm 0.005 $ & $ 0.028 \pm 0.004 $ & $ 0.024 \pm 0.005 $ & $ 0.035 \pm 0.005 $ & $ 0.023 \pm 0.005 $ & $ 0.023 \pm 0.004 $ & $ 0.024 \pm 0.004 $ & $ 0.028 \pm 0.005 $ & $ 0.031 \pm 0.004 $ & $ 0.029 \pm 0.005 $ \\
         0 kG & $ 0.004 \pm 0.039 $ & $ 0.017 \pm 0.031 $ & $ 0.028 \pm 0.033 $ & $ 0.000 \pm 0.033 $ & $ 0.051 \pm 0.032 $ & $ 0.000 \pm 0.029 $ & $ 0.014 \pm 0.029 $ & $ 0.000 \pm 0.032 $ & $ 0.028 \pm 0.033 $ & $ 0.023 \pm 0.033 $ \\
\hline
\end{tabular} 
\label{tab:specfitall} 
\end{table}
\end{landscape}

%%%%%%%%%%%%%%%%%%%%%%%%%%%%%%%%%%%%%%%%%%%%%%%%%%

% Don't change these lines
\bsp	% typesetting comment
\label{lastpage}
\end{document}